\documentclass[reprint, aps, nofootinbib, groupedaddress, superscriptaddress]{revtex4-2}
\input{preamble.sty}

\makeatletter
\def\maketitle{
	\@author@finish
	\title@column\titleblock@produce
	\suppressfloats[t]}
\makeatother

\begin{document}
	\begin{abstract}
	Biophysical models describing complex, cellular phenomena typically include systems of nonlinear differential equations with many free parameters. While experimental measurements can fix some parameters, those describing internal cellular processes frequently remain inaccessible. Hence, a proliferation of free parameters risks overfitting the data, limiting the model's predictive power. In this study, we develop robust methods, applying statistical analysis and dynamical-systems theory, to reduce a biophysical model's complexity. We demonstrate our techniques on an elaborate computational model designed to describe active, mechanical motility of auditory hair cells. Specifically, we use two statistical measures, the total-effect and PAWN indices, to rank each free parameter by its influence on selected, core properties of the model. With the resulting ranking, we fix most of the less influential parameters, yielding a low-parameter model with optimal predictive power. We validate the theoretical model with experimental recordings of active hair-bundle motility, specifically by using Akaike and Bayesian information criteria after obtaining maximum-likelihood fits. As a result, we determine the system's most influential parameters, which illuminate its key biophysical elements of the cell's overall features. While we demonstrated our techniques on a concrete example, they provide a general framework, applicable to other biophysical systems.
\end{abstract}

\title{Developing Parameter-Reduction Methods \\ on a Biophysical Model of Auditory Hair Cells}

\author{Joseph M. Marcinik}
\author{Mart\'in A. Toderi}
\affiliation{Department of Physics \& Astronomy, University of California, Los Angeles, California, 90095, USA}
\author{Dolores Bozovic}
\affiliation{Department of Physics \& Astronomy, University of California, Los Angeles, California, 90095, USA}
\affiliation{California NanoSystems Institute, University of California, Los Angeles, California, 90095, USA}

\date{\today}
\maketitle

\section{Introduction}
The auditory system provides humans and other animals with crucial information about the external world. Auditory cues enable communication with conspecifics, detection of prey, avoidance of predators, and they enhance an animal's spatial awareness. Hearing research has accrued extensive progress over the past decades, with many of its biophysical mechanisms, molecular components, and cellular processes now fairly well established \cite{weverTheoryHearing1949,geislerSoundSynapsePhysiology1998,picklesIntroductionPhysiologyHearing2013}. A number of phenomena, nonetheless, still remain elusive and as subjects of ongoing research.

Specifically, the remarkable sensitivity of hearing, which allows sub-nanometer detection in the presence of comparable or higher levels of ambient noise, is not yet fully explained. This sensitivity is controlled by hair cells, sensory cells that transduce mechanical deflections (from incoming sound waves) into electrical signals, which are further propagated down innervating neurons. Many recent studies have searched for an internal amplifier, an element that expends energy to enhance the mechanical response of auditory end organs to low-level sounds. Extensive experimental evidence advocates for its prevalence in hair cells \cite{gopfertActiveAuditoryMechanics2001,hudspethIntegratingActiveProcess2014,bozovicActiveBiomechanicsSensory2019}, which constitute the first active and nonlinear element in sound detection and processing \cite{hudspethHowEarWorks1989,smothermanHearingFrogs2000}.

As one manifestation of this active process, hair cells exhibit spontaneous oscillations. These oscillations, measured experimentally \textit{in vitro}, require an energy-consuming process \cite{martinSpontaneousOscillationHair2003}. They provide an experimental probe for the underlying, biophysical mechanisms of amplification. Consequently, measurements of this active motility relative to various cellular environments (e.g.\ ionic concentrations \cite{beurgCalciumBalanceMechanotransduction2010,pengAdaptationMammalianAuditory2013}, membrane potential \cite{quinonesVoltageandCalciumdependentMotility2015, meenderinkVoltagemediatedControlSpontaneous2015}, mechanical loading \cite{jaramilloDisplacementclampMeasurementForces1993,strimbuCouplingElasticLoading2012}, channel blockers \cite{ramunno-johnsonEffectsSomaticIon2010}, and other pharmacological manipulations \cite{assadTiplinkIntegrityMechanical1991,beurgActionsCalciumHair2008}) have yielded details on the cellular processes governing internal mechanics.

Models of various complexities have been developed to describe hair-cell dynamics. They have reproduced all of the main experimental findings \cite{hudspethPuttingIonChannels2000,martinSpontaneousOscillationHair2003,nadrowskiActiveHairbundleMotility2004,fruthActiveOscillatorModel2014}. However, with each refinement, they introduced additional differential equations to describe the internal cellular mechanisms. The models, hence, suffer from a proliferation of parameters. While measurements can constrain or approximate some parameters, many are not experimentally accessible and must, justly, be treated as free parameters. At best, we can assert a range over which they reside.

In the current work, we apply some standard as well as recent techniques from the field of dynamical systems modeling, to assess and rank the importance of various parameters on hair-cell dynamics. We use these techniques to reduce the space of free parameters \cite{hommaImportanceMeasuresGlobal1996,sobolGlobalSensitivityIndices2001,saltelliVarianceBasedSensitivity2010,pianosiSimpleEfficientMethod2015}, while ensuring that the model adequately reproduces experimental measurements. We develop and test this reduced model by comparing it to empirical data. While we demonstrate these methods on a concrete example, we emphasize that our techniques generalize readily to other species and hearing organs, even to completely different biological systems.

Parameter reduction produces many desirable outcomes. Firstly, by simplifying a model, we diminish its computational demands for the simulation. Secondly, by using well-tuned techniques to fix a subset of the parameters, we both alleviate the risk of overfitting and limit the occurrence of underfitting \cite{mcelreathStatisticalRethinking2015}. Finally, by reducing the parameter count, we illuminate the biophysical processes that constitute cellular dynamics. For each model parameter, there exists a corresponding mechanism (e.g.\ binding or dissociation of an ion, motion of a molecular motor, unfolding of a protein). Hence, by determining which parameters most strongly influence overall dynamics, we illuminate which internal processes shape the cellular response.

Others have executed similar methods to simplify numerical models in other areas, particularly in biological research, where high-complexity systems often arise. For example, a model for the JAK-STAT signal pathway was reduced from about 60 parameters down to 33 influential ones \cite{mortlockDynamicRegulationJAKSTAT2021}, and a model for voltage fluctuations across the AMPAR receptor was reduced from 24 to seven parameters \cite{lindenBayesianParameterEstimation2022}. These examples comprise only a small portion of the models benefiting from this methodology.

Specifically within auditory research, others have made previous efforts to reduce complex hair-cell models. One example includes the normal form equation of the Hopf bifurcation, a simple nonlinear differential equation that explains the compressive nonlinearity, active amplification, and frequency selectivity of the cell response \cite{ospeckEvidenceHopfBifurcation2001,hudspethCritiqueCriticalCochlea2010}. Other efforts include an empirical study, which unfolded the attractor characterizing innate, hair-bundle oscillations and concluded that roughly three to six degrees of freedom describe the measured oscillator sufficiently \cite{faberChaoticDynamicsInner2018}. However, while these general mathematical frameworks provide insight into the global features of the auditory system, they do not illuminate the biophysical mechanisms underlying the signal detection. For direct comparison to experimental data, models that reflect specific internal processes are needed.

In this paper, we construct a reduced, biophysically motivated model for hair-bundle motion. We start with a comprehensive theoretical model of hair-cell dynamics, garnered from prior literature to describe the hair cell's internal processes. Next, we simplify it algebraically to produce a nondimensional version of the full, biophysical model. We select several key features of the simulation and rank the full set of parameters by their influence on these. Because this ranking provides a quantitative assessment of relative influence, justified by well-studied statistical analyses, we fix the less influential ones. Finally, we compare the resulting reduced model to experimental measurements, demonstrating the effectiveness of this technique.

\section{Methods} \label{sec:methods}
\subsection{Materials and Experimental Techniques}
As our biological model system, we used the North American bullfrog (\textit{Rana catesbeiana}). The amphibian sacculus, i.e.\ an end organ specializing in vestibular and low-frequency auditory detection, has been used extensively for experiments on hair-bundle mechanics, as it provides a robust, optically accessible preparation. We imaged hair cells from dissected sacculi \textit{in vitro}, in semi-intact preparations that maintained their physiological integrity. We then used optical imaging to track the motion of the hair bundle, i.e.\ an organelle comprised of 30-50 stereocilia protruding from the apical surface of each cell. These measurements yielded traces of active, hair-bundle oscillations, allowing direct comparison to the numerical simulations.

\subsubsection{Biological Preparation} 
\begin{figure}
	\includegraphics[width=3.25in]{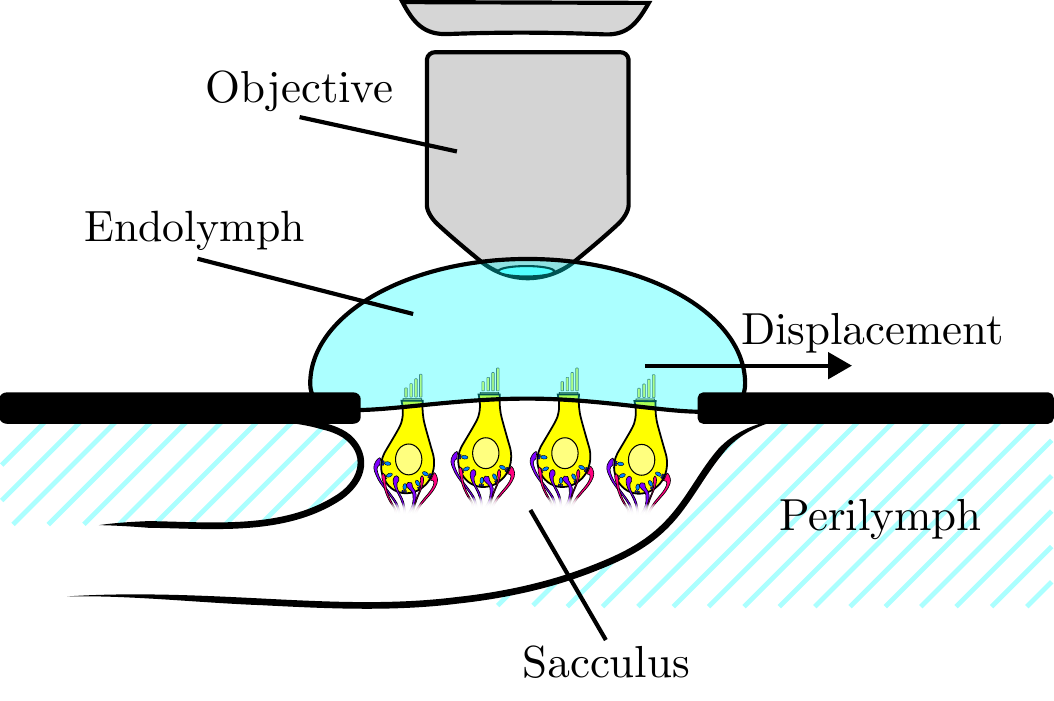}
	\caption{Schematic diagram of the experimental recordings. Hair cells (not drawn to scale), embedded in the supporting tissue of the sensory epithelium, are mounted in a two-compartment chamber, allowing for a separation of fluids on the apical and basal sides of the sacculus. The artificial solutions mimic ionic concentrations of the sacculus's natural fluid environment. Immersed in artificial perilymph (bottom compartment) are cell bodies, supporting cells, and innervating neurons; and immersed in artificial endolymph (top compartment) are hair bundles protruding from the apical side. Stereocilia atop the bundle oscillate horizontally as shown by the arrow, with deflection toward the tallest stereocilium defined as a positive position.} \label{fig:prep}
\end{figure}

Frogs of either gender were anesthetized (pentobarbital: $\SI{150}{\milli\litre\per\kilo\gram}$), pithed, and decapitated following protocols approved by the University of California, Los Angeles Chancellor’s Animals Research Committee. We excised sacculi from the frog inner ears and placed them in oxygenated artificial perilymph solution (in $\si{\milli\molar}$ as follows: 110 Na$^+$ , 2 K$^+$ , 1.5 Ca$^{2+}$ , 113 Cl$^-$ , 3 D-($+$)-glucose, 1 Na$^+$ pyruvate, 1 creatine, 5 HEPES). We mounted the epithelium in a two-compartment chamber, emulating the fluid partitioning of the \textit{in vivo} physiological conditions. In this arrangement, we bathed apical surfaces in artificial endolymph (in $\si{\milli\molar}$ as follows: 2 Na$^+$, 118 K$^+$, 0.25 Ca$^{2+}$, 118 Cl$^-$, 3 D-($+$)-glucose, 5 HEPES) and basolateral membranes in perilymph (as depicted in \cref{fig:prep}). We carefully removed the otolithic membrane from the epithelium after an $\SI{8}{\minute}$ enzymatic dissociation with $\SI{15}{\gram\per\milli\liter}$ collagenase IV (Sigma-Aldrich).

\subsubsection{Optical Recordings}
We collected recordings using an upright optical microscope (Olympus BX51WI) with a water-immersion objective (Olympus LUMPlanFL N 60X, NA:1.00), mounted on an optical table (Technical Manufacturing). We placed the setup inside an acoustically isolated chamber (Industrial Acoustics), so as to avoid introducing external perturbations to the highly sensitive hair cells. We obtained 16-bit TIFF images at a resolution of $\SI{108.3}{\nano\meter\per\pixel}$, with a high speed camera (ORCA-Flash4.0 CMOS) at 1000 frames per second (fps). We observed innate bundle motion, verifying integrity of the biological preparation. 

We processed the collected differential interference contrast (DIC) images, each accounting for $\SI{1}{\milli\second}$ of exposure, using custom-developed MATLAB scripts. Specifically, for each frame of the recording, we determined the mean bundle position weighted by pixel intensity along a line of pixels. Plots of hair-bundle position over time then provided traces of its motion (see \cref{fig:IC-bestfit}). Typical measurements obtained with this procedure yielded noise floors on the order of $\qtyrange{3}{5}{\nano\meter}$. To account for the gradual sag of the biological preparation, we calculated a wide-size (selected manually by visual inspection), Hann-window moving average of each trace and subtracted this long-term drift in the bundle position from the corresponding raw trace.

\subsection{General Procedure for Model Reduction}
\begin{figure}
	\centering
	\includegraphics[width=3.25in]{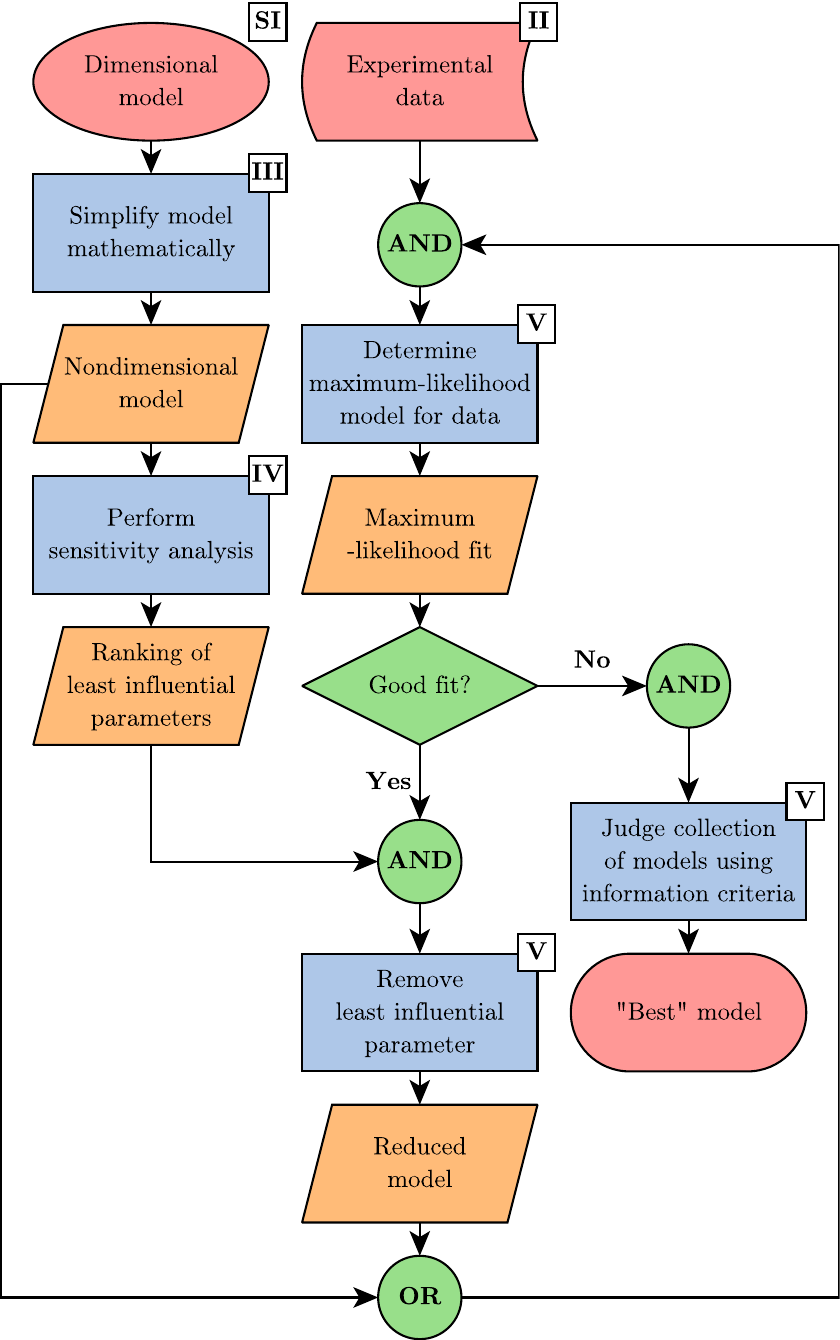}
	\caption{General procedure used for model reduction. Red nodes indicate starting (ellipse and curved parallelogram) or ending (pill shape) nodes. Blue nodes (rectangle) indicate an action to perform. Orange nodes (parallelogram) indicate outputs from the preceding action and inputs for the proceeding action. Green nodes (rhombus and circle) indicate decisions and logical operators, respectively. For relevant input and action nodes, the top-right corner displays the relevant section number in this paper.} \label{fig:flowchart}
\end{figure}

In this section, we outline the general procedure followed in the derivation of a reduced model, which incorporates statistical analyses of the impact of its free parameters (see \cref{fig:flowchart} for a diagram of the overall procedure). The approach consists of three primary components: model derivation, sensitivity analysis, and model selection.
	
We first derived a model for spontaneous hair-bundle motion, based on prior literature (see \cref{sec:model}). The model manifested as a five-variable system of ODEs, which we simplified algebraically by formulating it in a nondimensional form. This mathematical manipulation reduced the number of parameters from 27 to 15, yielding a simpler version, more conducive to our model-selection process.

We next conducted a sensitivity analysis procedure (see \cref{sec:sensitivity}) on our model to rank its parameters by importance. We determined the \enquote{influence} of a parameter by how much it affected five properties characterizing hair-bundle motion produced by the simulation (see \cref{sec:properties}). We applied two definitions to quantify parameter influence, namely total-effect (TE; see \cref{sec:TEind}) and PAWN (see \cref{sec:PAWNind}) indices. Finally, we deemed parameters with larger indices as more influential.

Finally, we applied quantitative metrics to select the best model for a particular dataset. We used two metrics, namely Akaike (see \cref{sec:AIC}) and Bayesian (see \cref{sec:BIC}) information criteria, each of which balance the risk of underfitting and overfitting. A minimum information criterion means that a model poises itself desirably between underfitting and overfitting, yielding an optimal fit, and able to extrapolate outside of the dataset. Hence, we deem models with lesser information criteria (see \cref{sec:IC}) as superior. We started our analysis of the full, 15-parameter system of ODEs, by finding the maximum-likelihood (ML) parameter set using Markov chain Monte Carlo (MCMC). Then, we fixed the least influential parameter and found the ML parameter set for the 14-parameter system of ODEs. We iteratively fixed the remaining least influential parameter until the ML probability decreased sufficiently to yield a poor match to the dataset (illustrated in \cref{fig:IC-bestfit}) We executed this procedure in parallel for three distinct datasets, each obtained from a different hair cell, fixing one parameter value after fitting all three. We determined the best models from these ML parameter sets by comparing their information criteria. Finally, we validated the accuracy of the reduced model on a new, fourth dataset.

\section{Nondimensional Model for Hair-Bundle Dynamics}
Hair cells of the inner ear are comprised of a cell body and a bundle of stereocilia that protrude from the apical surface. Stereocilia are actin-filled, columnar structures, arranged in rows of increasing height and interconnected by tip links, i.e.\ polymers that reach upward from the tip of a shorter stereocilium to the side of a taller neighboring one. These and other linkers between individual stereocilia maintain the integrity of a hair bundle, ensuring that it moves as one object \cite{kacharHighresolutionStructureHaircell2000,furnessDimensionsCompositionStereociliary2008,schwanderCellBiologyHearing2010}. Embedded in the tips of stereocilia and structurally connected to tip links, are mechano-sensitive ion channels. When incoming sound waves deflect the stereocilia, tip-link tension increases, opening channels and subsequently generating an influx of ionic current \cite{hudspethAcousticWavesBrain2019}. This influx of ions adjusts the voltage across the cell membrane, a process known as (mechano-electrical) transduction \cite{hudspethHowEarWorks1989}, starting the process of sound detection.

Coupled to the transduction channels and internal to the stereocilia, myosin motors climb and slip along the actin strands. This energy-consuming process allows active tuning of tip-link tension by the hair bundle, a process known as (myosin-mediated) adaptation, modulating opening probability of the channels and again altering the influx of ions. Ultimately, this interaction between transduction and adaptation repeats, originating a stable limit cycle oscillation. These two processes, namely mechano-electrical transduction \cite{ohmoriMechanoElectricalTransduction1985,eatockAdaptationMechanoelectricalTransduction1987,howardMechanoelectricalTransductionHair1988} and myosin-mediated adaptation \cite{walkerCalmodulinCalmodulinbindingProteins1993,walkerCalmodulinControlsAdaptation1996,cyrMyosin1cInteractsHairCell2002,holtChemicalGeneticStrategyImplicates2002,gillespieMyosin1cHairCell2004}, have been shown to describe the bulk of spontaneous hair bundle motility.

A number of other internal processes have been shown to play a role in shaping the active bundle motion. These include multiple effects of calcium feedback, which modulates the rates of myosin motor activity as well as the mechanical compliance of internal components \cite{cheungCa2ChangesForce2006,beurgCalciumBalanceMechanotransduction2010,quinonesVoltageandCalciumdependentMotility2015,zhangFunctionalCalciumImaging2016}. Please see \cref{sec:model} for additional details on the numerical model for hair-bundle oscillation. Prior work has demonstrated that the full, biophysical model accurately and reliably reproduces the experimental observations.

To simplify our parameter-reduction approach, we first algebraically convert the full system of dimensional ODEs into nondimensional form. Apart from reducing the number of free parameters, this simplified form also clearly elucidates the main dynamics underlying the time evolution of different observables. To distinguish the two models, we signify all nondimensional quantities with a tilde ( $\nondim{}$ ).

The nondimensional system of equations (see \cref{sec:model} for its derivation) is given as follows:
\begin{equation}
	\begin{split} \label{eq:nondim}
		\nondim{\tau}_{hb} \dv{\nondim{x}_{hb}}{\nondim{t}} &= -\qty(\nondim{F}_{gs} + \nondim{x}_{hb}) \\
		\dv{\nondim{x}_a}{\nondim{t}} &= \nondim{S}_{max}\nondim{S}\qty(\nondim{F}_{gs} - \nondim{x}_a) - \qty(1-\nondim{S}_{max})\nondim{C} \\
		\nondim{\tau}_m \dv{p_m}{\nondim{t}} &= \nondim{C}_m p_T (1-p_m) - p_m \\
		\nondim{\tau}_{gs} \dv{p_{gs}}{\nondim{t}} &= \nondim{C}_{gs} p_T (1-p_{gs}) - p_{gs} \\
		\nondim{\tau}_T \dv{p_T}{\nondim{t}} &= p_T(\infty) - p_T
	\end{split}
	,
\end{equation}
where $\nondim{x}_{hb}$ reflects the position of the hair bundle, $\nondim{x}_{a}$ the position of the myosin adaptation motors; $\nondim{F}_{gs}$ denotes the aggregate restoring force with effects from the gating and extent springs as well as the stereociliary pivot; $p_T$ represents opening probability of the transduction channels; $p_m$ and $p_{gs}$ represent probabilities of calcium binding to the myosin motors and gating springs, respectively; $\nondim{C}_m$ and $\nondim{C}_{gs}$ account for combined calcium and voltage effects near the internal motors and gating springs, respectively; $\nondim{S}$ and $\nondim{C}$ represent slipping and climbing rates, respectively, of the motors, with $\nondim{S}_{max}$ denoting maximum slipping rate. The time constants characterizing various processes of the hair cell are given by $\nondim{\tau}_{hb}$ $\nondim{\tau}_m$, $\nondim{\tau}_{gs}$, and $\nondim{\tau}_T$.

\section{Application of Sensitivity Analysis to Rank the Influence of Free Parameters} \label{sec:sensitivity}
In this section, we rank the importance of each free parameter in the model, quantified by its influence on the overall output of the simulation. We began with the nondimensional model (shown in \cref{eq:nondim}), which produces time-dependent traces of bundle position $\nondim{x}_{hb}$. We selected five prominent features, characterizing our limit cycles, to serve as metrics for assessing the influence of each parameter. We then ranked each free parameter by its influence on bundle dynamics, using statistical techniques from sensitivity analysis. We utilized two sensitivity indices, namely TE \cite{hommaImportanceMeasuresGlobal1996} and PAWN \cite{pianosiSimpleEfficientMethod2015}. We used two indices so as to check our ranking through two independent approaches.

\subsection{Properties of the Numerical Model} \label{sec:properties}
We characterized time-dependent traces of bundle motion with five properties: 1) a boolean quantity indicating whether the trace reflects oscillatory motion, 2-4) mean, amplitude, and frequency, which characterize the oscillation, 5) skewness, which captures the temporal profile of active motility.

\subsubsection{Presence of Active Oscillations}
We define a quantity
\begin{equation} \label{eq:osc}
	\delta\qty{x} \coloneq \begin{cases}
		1, & x \text{ is oscillating} \\
		0, & x \text{ is not oscillating}
	\end{cases}
	.
\end{equation}
The average of this quantity over a given set of traces yields the proportion of traces that exhibit limit cycles. We used $\delta\{\nondim{x}_{hb}\}$ to quantify whether each simulation produced oscillatory behavior.

\subsubsection{Mean}
We define mean $\mean{x}$ of variable $x$ as the arithmetic mean over time $t$. We used $\mean{\nondim{x}_{hb}}$ as the mean for each simulation.

\subsubsection{Amplitude}
We define the amplitude of a time-dependent variable $x(t)$ as half of its peak-to-peak value,
\begin{equation} \label{eq:p2pamp}
	A\{x\} = \frac{1}{2}\qty( \max_t\{x(t)\}-\min_t\{x(t)\} )
	.
\end{equation}
We used $A\{\nondim{x}_{hb}\}$ as the amplitude for each time-dependent trace produced by our model.

\subsubsection{Frequency}
To define the frequency of a time-dependent variable $x(t)$, we compute its analytic function,
\begin{align} \label{eq:asig}
	\begin{split}
		\asig{x}(t) &= \Delta{x}(t) + i\mathcal{H}\{\Delta{x}(t)\} \\
		\Delta{x}(t) &\coloneq x(t)-\mean{x}
	\end{split}
	,
\end{align}
where $\mathcal{H}$ denotes the Hilbert transform. The analytic signal displays an instantaneous frequency $\frac{1}{2\pi} \dv{t}\qty[\arg(\asig{x})]$, where $\arg(x)$ indicates the complex phase of $x$. From this expression, we define frequency of the variable $x$ as the mean instantaneous frequency of $\asig{x}$,
\begin{equation} \label{eq:Hfreq}
	f\{x\} = \frac{1}{2\pi} \mean{\dv{t}\qty[\arg(\asig{x})]}
	.
\end{equation}
We used $f\{\nondim{x}_{hb}\}$ as the frequency characterizing each specific model \cite{justiceAnalyticSignalProcessing1979}.

\subsubsection{Skewness}
We define the skewness of a variable $x$ as the third standardized moment of $x$,
\begin{equation} \label{eq:skew}
	\skewness{x} = \frac{\mean{\qty(x-\mean{x})^3}}{\variance{x}^{3/2}}
	.
\end{equation}
Skewness measures the degree of asymmetry in the shape of one period of oscillation of $x(t)$. In the subsequent analysis, we used $\skewness{\nondim{x}_{hb}}$ as the skewness for each simulation.

\subsection{Total-Effect Index} \label{sec:TEind}
We used the TE index to rank all 15 parameters in the nondimensional model.
TE index is defined mathematically as \cite{hommaImportanceMeasuresGlobal1996,saltelliVarianceBasedSensitivity2010}
\begin{equation} \label{eq:TE}
	S_{T_i} = \frac{\mean[\vec{X}_{\sim i}]{ \variance[X_i]{Y|\vec{X}_{\sim i}} }  }{\variance{Y}}
	,
\end{equation}
where $Y$ represents a random variable corresponding to one property, $X_i$ represents a random variable corresponding to one parameter indexed by $i$, and $\vec{X}_{\sim i}$ represents a vector of random variables corresponding to the set parameters not indexed by $i$. Here $\mean[\vec{X}_{\sim i}]{x}$ denotes the arithmetic mean of $x$ taken over all parameter sets in $\vec{X}_{\sim i}$, and $\variance[X_i]{x}$ denotes the variance of $x$ taken over all parameter values in $X_i$. Conceptually, this index indicates the average variance of one property $x$ produced by varying one parameter indexed by $i$. This index is normalized between 0 and 1, inclusively, where a greater index indicates a parameter with greater influence. An index of 0 indicates that the parameter produces no variance for a given property of the model, whereas an index of 1 indicates that the parameter produces all of the total variance. We ranked parameters as most to least influential from greatest to least TE index, respectively.

We found TE index for each combination of 15 parameters and five properties (shown in \cref{fig:rank}). For each parameter, we set the maximum out of these five TE indices as the final TE index for ranking. To calculate these indices, we simulated $\sim$500,000 instances of the model at a uniform, independent collection of random parameter samples (based on the algorithm by \cite{saltelliVarianceBasedSensitivity2010}). We found TE indices from the collection of oscillating and non-oscillating simulations.

\subsection{PAWN Index} \label{sec:PAWNind}
We next used the PAWN index to rank all 15 parameters in the nondimensional model. PAWN index follows from the Kolmogorov-Smirnov statistic.

The Kolmogorov-Smirnov statistic (KS) is defined mathematically as \cite{kolmogoroffSullaDeterminazioneEmpirica1933,smirnovEstimationDiscrepancyEmpirical1939,shiryayevEmpiricalDeterminationDistribution1992}
\begin{equation} \label{eq:KS}
	\text{KS}(x_i) = \max_{y \in Y} \abs{F_Y(y)-F_{Y|X_i=x_i}(y)}
	,
\end{equation}
where $F_Y(y)$ indicates the cumulative distribution function (CDF) of a random variable $Y$, evaluated at $y\in Y$. Conceptually, KS is the maximum distance between two CDFs. It is normalized between 0 and 1, inclusively, where a greater KS indicates that two CDFs are farther apart.

The PAWN index is defined mathematically as \cite{pianosiSimpleEfficientMethod2015}
\begin{equation} \label{eq:PAWN}
	T_i = \underset{x_i \in X_i}{\text{stat}} \text{KS}(x_i)
	,
\end{equation}
where $\text{stat}$ represents any statistic functional (e.g.\ mean, median, maximum). For the remainder of this study, we chose $\text{stat}=\text{max}$. Conceptually, this index measures the influence of a parameter on a single model output. It is normalized between 0 and 1, inclusively, where a greater index indicates a parameter with greater influence on the simulation. An index of 0 indicates that the parameter produces no influence for some model property. We ranked parameters as most to least influential from greatest to least PAWN index, respectively.

\begin{figure*}
	\includegraphics{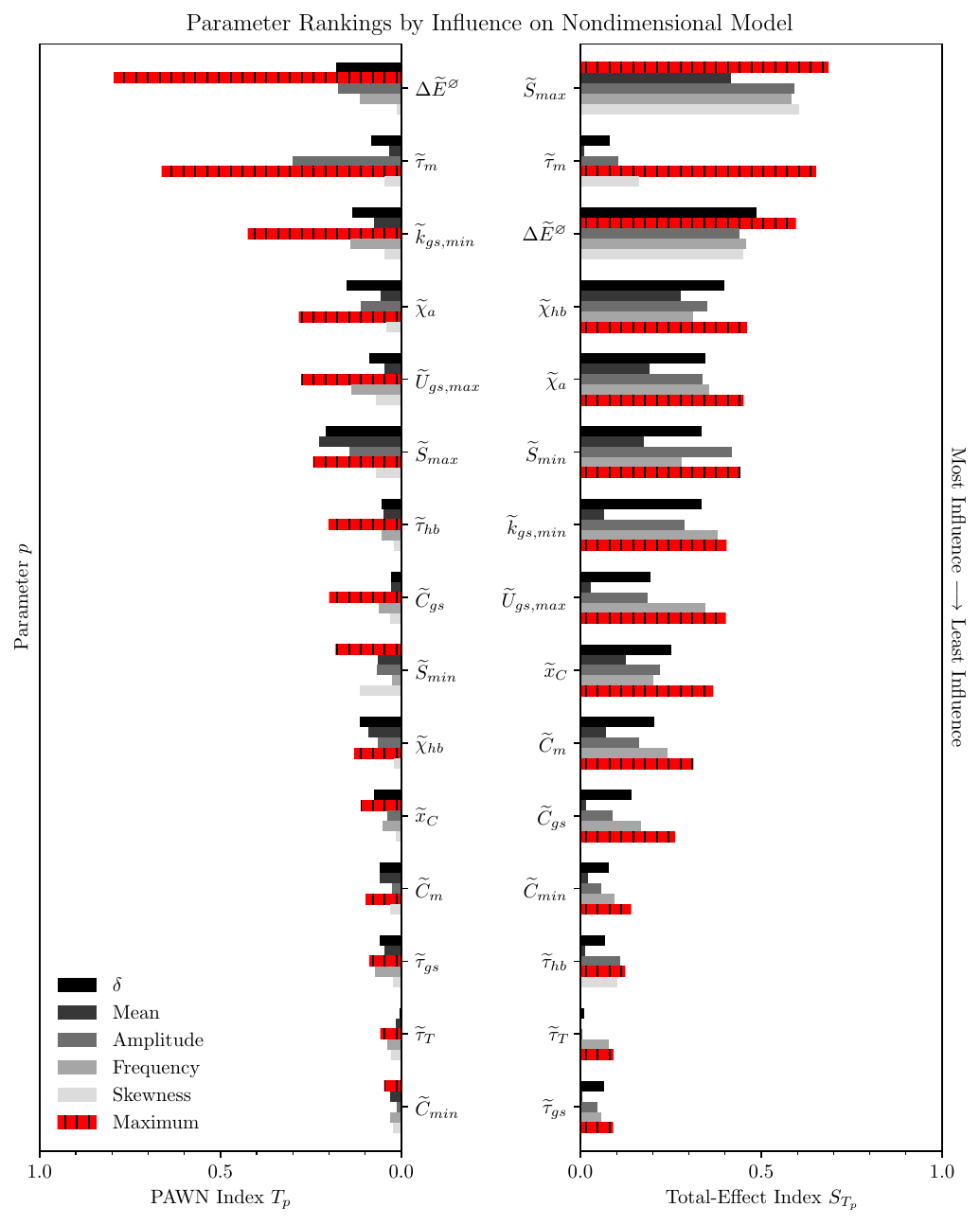}
	\caption{Sensitivity indices for each combination of five properties (described in \cref{sec:properties}) and 15 parameters (shown in \cref{tbl:fixed}). The left and right plots correspond to PAWN (described in \cref{sec:PAWNind}) and total-effect indices (described in \cref{sec:TEind}), respectively. Except for hatched bars, grayscale color indicates the corresponding model property (described in \cref{sec:properties}) for each index. Along their y-axes, each plot orders parameters from most to least influence (top to bottom) by their maximum index (hatched, red bars) of five properties.} \label{fig:rank}
\end{figure*}

We found the PAWN index for each combination of 15 parameters and five model properties. For each parameter, we set the maximum out of these five PAWN indices as the final PAWN index for ranking (shown in \cref{fig:rank}). We found ten KS statistics for each parameter, obtained by binning each parameter at ten distinct values (demonstrated in \cref{fig:KSdemo}). To calculate these indices, we simulated $\sim$500,000 instances of the model at a uniform, independent collection of random parameter samples (same dataset as in \cref{sec:TEind}). Except for $\delta$, we found PAWN indices from the subset of oscillating simulations. To calculate $\delta$, we used all oscillating and non-oscillating simulations.

\subsection{Final Ranking of the Parameters} \label{sec:ranking}
After finding parameter rankings (shown in \cref{fig:rank}), one for each TE and PAWN indices, we compared them. These two ordered lists are roughly $99.5\%$ similar (i.e.\ according to Spearman footrule distribution, see \cref{sec:footrule}), establishing robustness of the two rankings. Hence, we reasonably chose either of the indices, with minor changes to the ranking.

We chose the final parameter ranking that minimizes average correlation between the model properties. TE indices experienced very high correlation across the five properties ($\sim$0.8), while PAWN indices experienced only moderately correlation ($\sim$0.5) across the five properties (shown in \cref{fig:propcorr}). Thus, we chose our final ranking as that obtained from PAWN indices.

\section{Extracting a Reduced Model Limited by Experimental Data} \label{sec:reduction}
\subsection{Using Markov Chain Monte Carlo to Fit a Numerical Model to Data} \label{sec:MCMC}
We used the MCMC method to determine the ML parameter set of a model, given an experimental dataset. MCMC relies on a collection of random walkers to converge toward the ML estimator (MLE). Each walker performs its own fit to the model, attempting to converge toward the MLE, analogous to performing multiple gradient-descent fits simultaneously. However, unlike in gradient descent, these walkers stochastically \enquote{gravitate} toward each other, according to a predefined set of moves, making them mutually dependent. MCMC optimization yielded collections of parameters that accurately modeled the dataset (see \cref{fig:IC-bestfit}).

We calculated the posterior probability for each parameter set, given an empirical dataset. To do so, we first estimated the prior and likelihood probabilities. To estimate the prior probability, we assumed a uniform prior over a constrained region for each parameter (bounds shown in \cref{tbl:nondimparam}). To estimate the likelihood probability, we treated the dataset as a collection of independent, normally distributed observations. For the mean and standard deviation of each distribution, we used the mean and error from the corresponding observation. After finding the prior and likelihood probabilities, we estimated the parameter set at maximum-posterior probability for each dataset. Finally, this maximum-posterior set served as an approximation for the ML parameter set \cite{mcelreathStatisticalRethinking2015,sammutEncyclopediaMachineLearning2017}.

We performed MCMC starting with the full, nondimensional model. We subsequently fixed the least influential parameter. We iterated this procedure by fixing the least influential parameter among the remaining, unfixed parameters, until the MLE failed to reproduce the data sufficiently (illustrated in \cref{fig:ic}).

\subsection{Rescaling the Nondimensional Model}
We rescaled from $\nondim{x}_{hb}$ to $X_{hb}$ by numerically fitting four values $\chr{x}_{hb}$, $\check{x}_{hb}$, $\chr{\tau}_a$, $\check{t}$ such that
\begin{equation} \label{eq:xrescale}
	X_{hb}(t) = \chr{x}_{hb} \qty(\nondim{x}_{hb}\qty(\nondim{t}) - \check{x}_{hb})
	,
\end{equation}
where  $\chr{x}_{hb}$ and $\check{x}_{hb}$ represent multiplicative scaling and constant offset, respectively, of $\nondim{x}_{hb}(\nondim{t})$, while $\chr{\tau}_a$ and $\check{t}$ represent multiplicative scaling and constant offset, respectively, of $\nondim{t}$ (shown in \cref{eq:nondimt} and \cref{tbl:chrparam}). We performed this rescaling operation as the last step in our MCMC fitting procedure, after the MCMC algorithm chose a 15-parameter set.

\subsection{Evaluating an Information Criterion} \label{sec:IC}
Information criteria (ICs) estimate the model prediction error relative to a dataset \cite{mcelreathStatisticalRethinking2015}. They penalize large numbers of degrees of freedom $\nu$, thus favoring fewer free parameters and reducing the chances of overfitting, and they favor large maximum likelihood $\hat{L}$, thus lessening the chances of underfitting. A smaller IC indicates that a model will have a smaller prediction error; hence, models with lesser ICs yield preferable fits to experimental results.

As we sought only a relative comparison of model performance, we required only relative ICs. Relative performance peaks at the greatest possible $\hat{L}$, which occurs when a model exactly matches the most probable values for all independent observations. Equivalently, this greatest likelihood corresponds to the least IC. Accordingly, we report likelihoods and ICs relative to this greatest likelihood (shown in \cref{fig:IC-bestfit,fig:ic}).

\subsubsection{Degrees of Freedom in the Model} \label{sec:dof}
To calculate ICs, we count the total number of degrees of freedom for the model fits.  Each nondimensional model parameter (up to 15 in our system) added one degree of the freedom. We next consider which rescaling variables (described in \cref{eq:xrescale}) add degree of freedoms. Three rescaling variables each added one degree of freedom, i.e.\ $\chr{x}_{hb}$, $\check{x}_{hb}$, $\chr{\tau}_a$. The fourth rescaling variable $\check{t}$ did not add a degree of freedom because the model is invariant to shifts in initial time. In total, our model has
\begin{equation}
	\nu = N_p + 3
\end{equation}
degrees of freedom, where $N_p$ represents number of free parameters in the model.

\subsubsection{Akaike Information Criterion} \label{sec:AIC}
We applied the Akaike information criterion (AIC; \cite{akaikeInformationTheoryExtension1973}) as a measure of model prediction error. AIC is defined mathematically as
\begin{equation} \label{eq:aic}
	\text{AIC} = 2\nu - 2\ln(\hat{L}) + \frac{2\nu(\nu+1)}{N_d-\nu-1}
	,
\end{equation}
where $N_d$ represents the number of observations in a dataset. Notice that AIC increases monotonically as $\nu$ increases, penalizing large numbers of degrees of freedom.

\subsubsection{Bayesian Information Criterion} \label{sec:BIC}
As an independent test, we applied the Bayesian information criterion (BIC; \cite{schwarzEstimatingDimensionModel1978}) as a measure of model prediction error. BIC is defined mathematically as
\begin{equation} \label{eq:bic}
	\text{BIC} = \nu\ln(N_d) - 2\ln(\hat{L})
	.
\end{equation}
Notice that BIC increases monotonically as $\nu$ or $N_d$ increase, penalizing large numbers of degrees of freedom, especially for large datasets.

\subsection{Reducing the Model Based on Fits to Experimental Datasets}
\begin{figure*}
	\centering
	\includegraphics{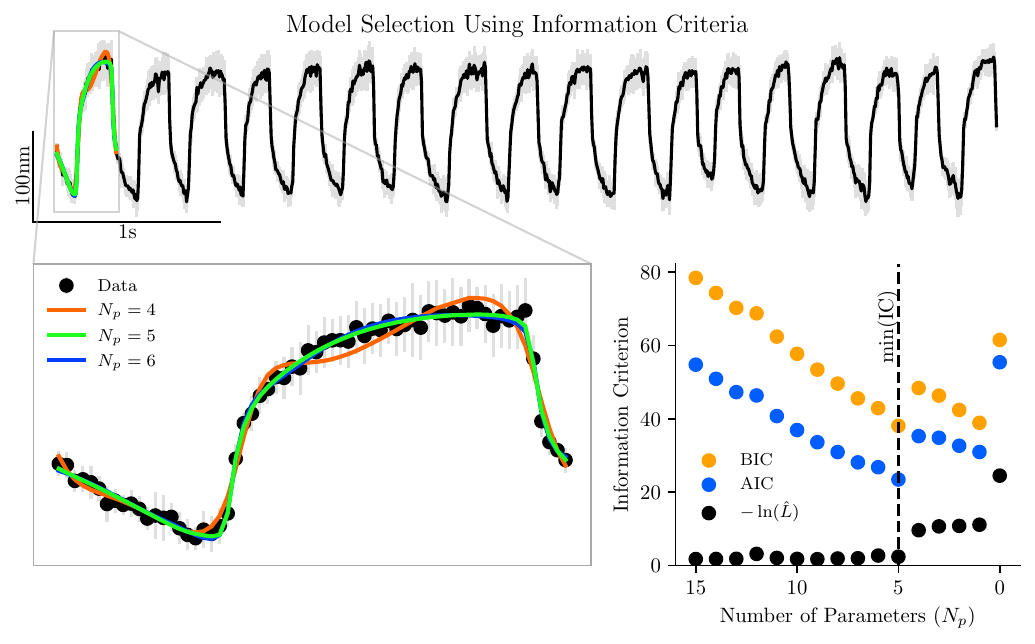}
	\caption{Goodness of fit of the reduced model to a dataset. Top: Five seconds of empirical bundle position (black line with gray error bars) over time. Bottom left: Comparison of best-fit models (described in \cref{eq:nondim} and \cref{tbl:fixed}) relative to dataset between two dashed red lines in full trace (top). Best fits are shown for models with $N_p=6$ parameters (blue line), $N_p=5$ (green line), and $N_p=4$ (orange line) along with truncated dataset (black points with gray error bars). Bottom right: Bayesian (orange points; defined in \cref{eq:bic}) and Akaike (blue points; defined in \cref{eq:aic}) information criteria along with $-\ln(\hat{L})$ (black points; defined in \cref{sec:IC}). These three quantities are reported relative to their corresponding best-possible value. The dashed black line indicates $N_p=5$ parameters for minimum information criteria.} \label{fig:IC-bestfit}
\end{figure*}

We found the models with least AIC and BIC for each dataset. Near (five of the six) minima, we fixed ten parameters at specified values, while five parameters (along with three other degrees of freedom discussed in \cref{sec:dof}) remained varied. Fixed parameter values and unfixed parameters are listed in \cref{tbl:fixed}, along with specific fixed values for the remaining five parameters, which however yielded poorer ICs. This five-parameter model minimizes both ICs for two of the three datasets, and it minimizes BIC for the third dataset (shown in \cref{fig:IC-bestfit,fig:ic}). For this remaining dataset, AIC is minimized by the six-parameter model.

AIC and BIC, in conjunction, have been demonstrated to reduce complex models reliably \cite{burnhamMultimodelInferenceUnderstanding2004,kuhaAICBICComparisons2004,brewerRelativePerformanceAIC2016}. They have been used in a plethora of fields such as astronomy \cite{liddleInformationCriteriaAstrophysical2007,doRelativisticRedshiftStar2019}, ecology \cite{johnsonModelSelectionEcology2004,ahoModelSelectionEcologists2014,albahliDefectPredictionUsing2021,tranEvaluatingPredictivePower2021}, physiology \cite{nevillModelingPhysiologicalAnthropometric2005,gloersenModelingVo2Onkinetics2022,jafariMVARCausalModeling2023}, finance \cite{hartmanModelSelectionAveraging2013,punzoMultivariateTailinflatedNormal2021,shahzadMostConsistentReliable2022}, and machine learning \cite{hossainEstimationARMAModel2020,nguyenComparingPerformanceMachine2021}.

\subsection{Validating the Reduced Model}
We fit one new dataset using the reduced model. To do so, we repeated the previous fit procedure (described in \cref{sec:MCMC}). Visually, the five-parameter model matches this new experimental dataset sufficiently, with no indication of either underfitting or overfitting.

\begin{table*}
	\makegapedcells
	\begin{tabular}{lcrlc}
		\toprule\midrule
		$N_p$ & Parameter & Value & Physical Significance & \\\midrule
		15 & --- & --- & All mechanisms included &  \\\midrule
		14 & $\nondim{C}_{min}$ & 1 & Constant climbing rate & \multirow{10}{*}{Fixed} \\
		13 & $\nondim{\tau}_T$ & 0 & Equilibrium transduction-channel dynamics & \\
		12 & $\nondim{\tau}_{gs}$ & 1 & Moderate calcium-feedback time constant for gating spring & \\
		11 & $\nondim{C}_m$ & 1 & Moderate calcium-feedback strength at motors & \\
		10 & $\nondim{x}_c$ & 0 & Null gating-spring offset & \\
		9 & $\nondim{\chi}_{hb}$ & 1 & Moderate coupling from stereocilia on gating-spring force & \\
		8 & $\nondim{S}_{min}$ & 0 & Maximal variability for slipping rate & \\
		7 & $\nondim{C}_{gs}$ & 1000 & Strong calcium-feedback strength on gating spring & \\
		6 & $\nondim{\tau}_{hb}$ & 1 & Moderate stereocilia time constant & \\
		5 & $\nondim{S}_{max}$ & 0.5 & Equal effective slipping and climbing rates & \\\midrule
		4 & $\nondim{U}_{gs,max}$* & 10 & Elastic potential energy of gating spring & \multirow{5}{*}{Unfixed} \\
		3 & $\nondim{\chi}_a$* & 1 & Coupling from motors on gating-spring force & \\
		2 & $\nondim{k}_{gs,min}$* & 1 & Variability of gating-spring stiffness & \\
		1 & $\nondim{\tau}_{m}$* & 10 & Calcium-feedback time constant for motor & \\
		0 & $\Delta\nondim{E}^\varnothing$* & 1 & Free energy of transduction-channel opening & \\
		\midrule\bottomrule
	\end{tabular}
	\caption{Parameters in the nondimensional model, ranked from least to most influential (top to bottom). $N_p$ represents number of parameters in the model after fixing the adjacent parameter at the corresponding fixed value. Parameters were fixed cumulatively so that all in the above rows remained fixed. The \enquote{Physical Significance} column indicates the importance 1) of fixing the corresponding parameter for fixed parameters or 2) of the corresponding biophysical mechanism for unfixed parameters. * indicates an ultimately unfixed parameter that was fixed at the corresponding value, but fixing produced a lower-quality fit.} \label{tbl:fixed}
\end{table*}

\section{Discussion} \label{sec:discussion}
We applied quantitative methods to assess and rank the importance of parameters, fixing the less influential ones. To demonstrate our methods on a concrete example, we modeled active, innate motility, observed in inner-ear hair cells. We commenced with a complex biophysical model with 27 parameters, finally reducing it to only five. This reduced model reproduced recordings of spontaneous, hair-bundle oscillations adequately, as demonstrated by fits to experimental measurements (shown in \cref{fig:IC-bestfit}). Our robust methods reduce the risk of overfitting and underfitting a computational model to an experimental dataset.

By reducing the biophysical model, we gain insight into the internal cellular processes. Each process and element corresponds to a set of model parameters and explains some observable behavior. Therefore, by identifying the most influential parameters, we can determine which internal elements produce the most impact on observed, bundle movement.

For example, we found that $\Delta{E}^\varnothing$ exerts strong influence on mean bundle position. This finding implies that mechanisms affecting this parameter value (i.e.\ free energy of transduction-channel opening, maximum gating-spring stiffness, and gating swing) determine, in large part, the mean bundle offset. Future experiments could test this prediction by altering some of these mechanisms and measuring their effect on the mean position.

Next, our fits indicate that hair cells exhibit variable-stiffness gating springs. Variability of gating-spring stiffness is encapsulated by the parameter $\nondim{k}_{gs,min}$, where $\nondim{k}_{gs,min}=0$ and $\nondim{k}_{gs,min}=1$ indicate maximum variability and constant stiffness, respectively. In our datasets, this parameter ranged from $\nondim{k}_{gs,min}\approx0.2-0.9$ (see \cref{fig:bestvalues}). Prior experimental studies have indicated the existence of a stiffness-modulating, calcium feedback on the gating spring \cite{martinSpontaneousOscillationHair2003,roongthumskulMultipletimescaleDynamicsUnderlying2011}. These studies analyzed complex behavior such as bursting dynamics in bundle motility and cellular response to electrical stimulation. Here, we demonstrated that this internal element strongly influences even unperturbed, regular oscillations.

The influential parameters therefore illuminate the most important, cellular mechanisms. For spontaneous hair-bundle oscillations, the dominant parameters were those describing gating springs, myosin motors, and their interaction thereof. For more details on these specific parameters, see the unfixed parameters in \cref{tbl:fixed}.

Similarly, the non-influential parameters likewise yield biophysical insight. For example, our model reliably reproduces experimental results while assuming equilibrium dynamics for transduction channels (i.e.\ $\tau_T=0$). This finding is consistent with previous numerical simulations that assumed equilibrium dynamics \cite{martinSpontaneousOscillationHair2003,nadrowskiActiveHairbundleMotility2004}. Other models, based on the weaker assumption of fast-channel dynamics \cite{bormuthTransductionChannelsGating2014,barralFrictionTransductionChannels2018}, likewise reproduced the results and were corroborated by experiments. The same study, however, demonstrated that adjusting other free parameters could compensate for the effects of assuming equilibrium dynamics, specifically by asserting a stronger effective viscous drag on the bundle. In our model, we simultaneously fit at least five parameters, which compensated for equilibrium-channel dynamics sufficiently, yielding consistent results.

Another low-influence parameter, we fixed the timescale of calcium feedback on the gating spring. We note that some of the less influential parameters may be fixed at different values from those applied in this paper (shown in \cref{tbl:fixed}). For example, $\nondim{\tau}_{gs}=0$ also reliably reproduces empirical datasets. While this would reduce the variable count by one (i.e.\ by letting $p_{gs}$ be in instantaneous equilibrium), it greatly increases computational time due to necessary numerical root finding. The optimal balance between analytic simplicity and computing time depends on the specific demands of the numerical simulation or desired interpretability of the equations.

This study focuses on effects from model parameters only on spontaneous hair-bundle oscillations; further, the oscillation mean, amplitude, frequency, and skewness constituted the only properties of the assessed, oscillating traces. Numerical models aimed to describe other phenomena, such as the cell response to an external drive, or phase-locking dynamics, would need to select different properties, likely resulting in a new ranking. Applying the same methods, our model readily extends itself to explore mechanical or electrophysiological drives and/or noise. Future work will entail exploring the crucial mechanisms underlying stimulated bundles (e.g.\ by forcing or voltage), including phenomena such as bundle entrainment to the stimulus as well as its rapid mechanical response to step stimuli \cite{benserRapidActiveHair1996,beurgCalciumBalanceMechanotransduction2010,omaoileidighDiverseEffectsMechanical2012,crawfordElectricalTuningMechanism1981,ricciActiveHairBundle2000}. Furthermore, our reduced model can reliably make novel predictions, motivating subsequent experimental studies.

A prior theoretical analysis examined the effects of noise in a similar model. It found that the mean-field, limit cycle for the stochastic system can differ significantly from that of the deterministic version \cite{shethNoiseinducedDistortionMean2019}. This implies that noise strength greatly influences limit cycle properties. Consequently, we expect that, after introducing sufficiently strong noise into our model, we could fix at least as many or more low-influence parameters as in our deterministic model.

While we performed this study on a specific biophysical system, namely that of active, hair-cell mechanics, our methodology endures in general. Many biological systems exhibit a great degree of complexity, bolstering numerous interacting processes and coupled elements. Furthermore, due to the delicate machinery involved, experiments cannot directly access or fix many parameters. Our techniques therefore provide a fruitful framework for reducing complex numerical models.

\section*{Acknowledgments}
	We thank Dr. Justin Faber for extensive help in graphic presentation of the results as well as the preparation of the manuscript. We also thank Prof. Tuan Do for discussions on model-selection techniques. This work was funded in part by DoD-DA-Army Research Office grant W911NF1910179 and in part by NSF Physics of Living Systems under grant 2210316.

	\setcounter{equation}{0}
	\setcounter{figure}{0}
	\setcounter{table}{0}
	\setcounter{section}{0}
	
\renewcommand{\theequation}{S\arabic{equation}}
\renewcommand{\thefigure}{S\arabic{figure}}
\renewcommand{\thetable}{S\arabic{table}}
\renewcommand{\thesection}{S\arabic{section}}

\title{Supplemental Material:\\ Developing Parameter-Reduction Methods \\ on a Biophysical Model of Auditory Hair Cells}

\author{Joseph M. Marcinik}
\author{Mart\'in A. Toderi}
\affiliation{Department of Physics \& Astronomy, University of California, Los Angeles, California, 90095, USA}
\author{Dolores Bozovic}
\affiliation{Department of Physics \& Astronomy, University of California, Los Angeles, California, 90095, USA}
\affiliation{California NanoSystems Institute, University of California, Los Angeles, California, 90095, USA}

\date{\today}
\maketitle

\section{Derivation of the Nondimensional Model of Hair-Bundle Oscillation} \label{sec:model}
\subsection{Axis of Motion}
When incoming sound waves deflect stereocilia, bundles pivot about their anchor, i.e.\ rootlets connecting them to the cell's cuticular plate \cite{furnessDimensionsCompositionStereociliary2008}. This rotational motion leads to lateral deflections of stereociliary tips, observed in experimental recordings \cite{hudspethIntegratingActiveProcess2014,bozovicActiveBiomechanicsSensory2019}. Hereafter, we use uppercase letters to denote projections of movement onto this direction. Stereocilia stand not perfectly parallel, but angled toward each other \cite{hudspethHowEarWorks1989}. Meanwhile, myosin motors slip and climb along these stereocilia \cite{hudspethPullingSpringsTune1994}, introducing a new direction of movement. We use lowercase letters to denote projections of movement (or force) along gating springs. This prompts a geometric conversion factor
\begin{align} \label{eq:gamma}
	\gamma \coloneq \frac{\beth^*}{\beth}
\end{align}
for corresponding pairs of positions (or forces) $\beth$ and $\beth^*$, where $\beth$ and $\beth^*$ are along the original and projected directions, respectively. Because bundle motion leads to small oscillations, we treat $\gamma$ as constant hereafter \cite{howardMechanoelectricalTransductionHair1988}.

\subsection{Hair-Bundle Position $X_{hb}$}
We derive a deterministic equation for hair-bundle position $X_{hb}$ during spontaneous oscillation, based on prior descriptions of experimental data. We treat a hair bundle as an elastic structure with effective Newtonian mass $m_{hb}$, undergoing laminar flow in a viscous medium. A hair bundle exhibiting spontaneous oscillations is subject to at least two prominent Hookean forces, one exerted each by a stereociliary pivot base \cite{pacentineStereociliaRootletsActinBased2020} and one by a collection of gating springs \cite{assadTiplinkIntegrityMechanical1991,kacharHighresolutionStructureHaircell2000}. A viscous drag force also acts on the bundle, from the surrounding fluid. We assume that only these three forces act on a spontaneously oscillating bundle.

Actin extends along each stereocilium and into the cellular base. It exerts force on a hair bundle through its stereociliary pivot \cite{howardStiffnessSensoryHair1986,jaramilloDisplacementclampMeasurementForces1993,marquisEffectsExtracellularCa21997}. We describe this as Hookean force
\begin{equation} \label{eq:Fsp}
	F_{sp} = -k_{sp} (X_{hb} - X_{sp})
	,
\end{equation}
where $k_{sp}$ and $X_{sp}$ are combined stiffness and equilibrium position, respectively, of the stereociliary pivot \cite{martinSpontaneousOscillationHair2003}.

Transduction elements likewise exert force on a hair bundle \cite{howardComplianceHairBundle1988,martinNegativeHairbundleStiffness2000}. We assume that each transduction element consists of one gating spring attached to one mechano-electrical transduction channel. We treat the collection of $N$ gating springs, each with stiffness $k_{gs}$, as residing in parallel, generating a single, effective spring. Each spring contains an attached adaptation motor with position $x_a$. We assumed that each transduction channel experiences exactly two states, open and closed \cite{markinGatingspringModelsMechanoelectrical1995}, where $p_T$ denotes probability that a channel resides in its open state. When transitioning from closed to open, the channel slackens its corresponding gating spring by distance $d$, termed \enquote{gating-spring swing}. Inversely, when transitioning from open to closed, the channel extends its corresponding gating spring by distance $d$. We treat the effective gating-spring swing as the average compression of all gating springs $p_Td$. Therefore, each transduction element exerts (on average) a force
\begin{equation} \label{eq:Fgs}
	f_{gs} = -k_{gs} (x_{hb}^* - x_a + x_c - p_Td)
\end{equation}
parallel to the gating springs, where $x_c$ represents average equilibrium position of the adaptation motors \cite{martinSpontaneousOscillationHair2003}. We expound upon $k_{gs}$ later as a function of calcium concentration in \cref{eq:kgs}. The full set of transduction elements exerts a force $NF_{gs}^*$ on the bundle.

A hair bundle experiences a drag force from its surrounding, viscous medium. We describe this as Stokean force
\begin{equation} \label{eq:Fdhb}
	F_{d,hb} = -\lambda_{hb}\dot{X}_{hb}
\end{equation}
where $\lambda_{hb}$ represents coefficient of drag for a hair bundle \cite{martinSpontaneousOscillationHair2003}.

By Newton's second law, net force on a hair bundle is, therefore,
\begin{equation} \label{eq:xddot}
	m_{hb}\ddot{X}_{hb} = F_{d,hb} + F_{sp} + NF_{gs}^*
	.
\end{equation}
For unloaded movement, the bundle has negligible mass $m_{hb}\sim0$, so we reduce \cref{eq:xddot} to
\begin{equation}
	-F_{d,hb} = F_{sp} + NF_{gs}^*
	.
\end{equation}

\subsection{Adaptation-Motor Position $x_a$}
We next derive a deterministic equation for adaptation-motor position $x_a$. We introduce their movement along stereocilia \cite{holtChemicalGeneticStrategyImplicates2002,mancevaCalciumRegulationCalmodulin2007} and incorporate calcium-feedback effects. Attached near the gating spring, these motors effectively modulate its tension on the hair bundle.

Myosin motors slip down stereocilia due to two tension forces, namely those of the gating and extent springs. This gating spring exerts a force on these motors with the same strength as it does on stereocilia, shown in \cref{eq:Fgs}. With only a gating spring, models have overestimated myosin slipping rates, prompting others to propose an extent spring in parallel with the gating spring \cite{shepherdExtentAdaptationBullfrog1994,yamoahPhosphateAnalogsBlock1996}. They describe this as Hookean force
\begin{equation} \label{eq:Fes}
	f_{es} = k_{es} (x_a + x_{es})
	,
\end{equation}
where $k_{es}$ and $x_{es}$ are stiffness and equilibrium position, respectively, of an extent spring.
Myosin motors also climb up stereocilia, independent of tension forces. In a moment, we account for these slipping and climbing movements.

Climbing and slipping rates depend on intracellular calcium concentration, so we cannot treat them as constant. As an approximation, we assume linear relationships \cite{martinSpontaneousOscillationHair2003}
\begin{align}
	C & = C_{max} - p_m(C_{max}-C_{min}) \label{eq:C} \\
	S & = S_{min} + p_m(S_{max}-S_{min}) \label{eq:S}
	,
\end{align}
where $C_{min}$ and $C_{max}$ are the minimum and maximum climbing velocities, respectively; and $S_{min}$ and $S_{max}$ are the minimum and maximum slipping rates, respectively.

Overall, we incorporate effects from \cref{eq:Fes,eq:C,eq:S} as myosin velocity
\begin{equation}
	\dot{x}_a = -C + S(f_{gs} - f_{es})
	.
\end{equation}

\subsection{Calcium-Binding Probabilities for Adaptation Motors $p_m$ and Gating Springs $p_{gs}$} \label{sec:Cabind}
We derive a deterministic equation for the probability of calcium binding to adaptation motors $p_m$ and to gating springs $p_{gs}$. By law of mass action \cite{martinSpontaneousOscillationHair2003}, the binding probabilities are
\begin{align}
	\dot{p}_m & = k^+_m[\ce{Ca^2+}]_m(1-p_m) - k^-_mp_m \\
	\dot{p}_{gs} & = k^+_{gs}[\ce{Ca^2+}]_{gs}(1-p_{gs}) - k^-_{gs}p_{gs}
	,
\end{align}
where $k^+_m$ and $k^+_{gs}$ are the association constants for an adaptation motor and gating spring, respectively; $k^-_m$ and $k^-_{gs}$ are the dissociation constants for an adaptation motor and gating spring, respectively; $[\ce{Ca^2+}]_m$ and $[\ce{Ca^2+}]_{gs}$ are calcium-ion concentrations near an adaptation motor and gating spring, respectively. 

We calculate calcium-ion concentration in the vicinity of adaptation motors $[\ce{Ca^2+}]_m$ and that of gating springs $[\ce{Ca^2+}]_{gs}$. To do so, we assert \enquote{fast} calcium diffusion \cite{bergRandomWalksBiology1993} (i.e.\ \enquote{small} time constant $\frac{r^2}{4D}$) from the transduction channels, within an area bounded by $\sim\SI{100}{nm}$ \cite{lumpkinRegulationFreeCa21998}. Hence, we assert equilibrium concentrations \cite{martinSpontaneousOscillationHair2003}:
\begin{align}
	[\ce{Ca^2+}]_m & = -\frac{I_{TCa}}{2\pi z_{Ca}q_eD_{Ca}r_m} \label{eq:Cam} \\
	[\ce{Ca^2+}]_{gs} & = -\frac{I_{TCa}}{2\pi z_{Ca}q_eD_{Ca}r_{gs}} \label{eq:Cags}
	,
\end{align}
where $q_e>0$ represents elementary charge; $D_{Ca}$ and $z_{Ca}$ are diffusion constant and valence number, respectively, for $\ce{Ca^2+}$; $r_m$ and $r_{gs}$ are distances of an adaptation motor and gating spring, respectively, from their corresponding transduction channel. We expound upon $I_{TCa}$ as a function of voltage later in \cref{sec:ITCa}. Finally, we assume that both the internal gating spring and adaptation motor reside sufficiently close to their corresponding transduction channel, warranting fast-diffusion.

Bounded calcium ions modulate gating-spring stiffness, supported by empirical evidence. Specifically, they alter stiffness of an internal element, a spring in series with the tip link and contributing to the overall gating-spring stiffness. To describe these observations, which include multiple timescales seen in hair-bundle oscillations, we assume that gating-spring stiffness $k_{gs}$ varies linearly with $p_{gs}$ \cite{roongthumskulMultipletimescaleDynamicsUnderlying2011}:
\begin{align} \label{eq:kgs}
	k_{gs} = k_{gs,max} - p_{gs}(k_{gs,max}-k_{gs,min})
	,
\end{align}
where $k_{gs,min}$ and $k_{gs,max}$ are minimum and maximum stiffness, respectively, of a gating spring.

\subsection{Transduction-Channel Open Probability $p_T$}
We derive a deterministic equation for open-channel probability $p_T$ of a transduction channel. We assert a two-state, Boltzmann model for a transduction channel, composed of one open and one closed state. At equilibrium, the open-channel probability is \cite{howardMechanoelectricalTransductionHair1988,markinGatingspringModelsMechanoelectrical1995,martinSpontaneousOscillationHair2003}:
\begin{equation} \label{eq:dimtauT}
	\begin{split}
		p_T(\infty) &= \frac{1}{1+A\exp(-\frac{x_{hb}-x_a+x_c-d/2}{\Delta{x_T}})} \\
		A &\coloneq \exp(\frac{\Delta{E}^\varnothing}{k_BT}) \\
		\Delta{x_T} &\coloneq \frac{k_BT}{k_{gs}d}
	\end{split}
	,
\end{equation}
where $p_T(t)$ represents a function of time $t$, and $\Delta{E}^\varnothing$ represents intrinsic energy difference between the transduction channel's two states. We assume that channel-opening dynamics exhibit a finite time constant $\tau_T^\varnothing$ \cite{bormuthTransductionChannelsGating2014, barralFrictionTransductionChannels2018} as follows:
\begin{equation}
	\tau_T^\varnothing\dot{p}_T(t) = p_T(\infty) - p_T(t)
	.
\end{equation}
Channel-opening dynamics effect a hair bundle's viscous drag force.

\subsection{Influx of the Calcium Current $I_{TCa}$} \label{sec:ITCa}
While the influx of $\ce{K^+}$ ions constitute the largest component of transduction current, $\ce{Ca^2+}$ ions carry a fraction of this current. The former determines overall voltage change across the cell membrane, deflecting the stereocilia. The latter, in contrast, exerts a strong feedback on the bundle, as seen in \cref{sec:Cabind}. We derive a deterministic model for the calcium-ion component of transduction current, $I_{TCa}$.

The transduction channels' maximum calcium conductance $g_{TCa,max}$, occurs when all channels are open. This conductance is given by the Goldman-Hodgin-Katz flux equation:
\begin{align}
	\begin{split}
		g_{TCa,max} &= \frac{P_{TCa}z_{Ca}^2q_e^2}{k_BT} \\
		\times& \frac{[\ce{Ca^2+}]_{hb,in}-[\ce{Ca^2+}]_{hb,ex}\exp(-\frac{z_{Ca}q_eV_m}{k_BT})} {1-\exp(-\frac{z_{Ca}q_eV_m}{k_BT})}
	\end{split}
	,
\end{align}
where $P_{TCa}$ represents transduction-channel permeability for $\ce{Ca^2+}$; $[\ce{Ca^2+}]_{hb,in}$; and $[\ce{Ca^2+}]_{hb,ex}$ represent intracellular and extracellular concentrations, respectively, near the transduction channel.

With only a fraction of channels open, a hair bundle's mean conductance varies monotonically with the open-channel probability. We assume a linear relationship, such that $g_{TCa}\propto p_T$. Thus, the conductance for $I_{TCa}$ must be
\begin{equation}
	g_{TCa} = p_T g_{TCa,max}
	.
\end{equation}
Furthermore, by Ohm's law the current is
\begin{equation}
	I_{TCa} = g_{TCa} (V_m-E_{TCa})
	,
\end{equation}
where $E_{Ca}$ represents reversal voltage for calcium.

\subsection{Summary of the model describing hair-bundle motion}
Using the previous derivation, we capture the full biophysical model for hair-bundle dynamics with the following set of differential equations:
\begin{equation} \label{eq:dimmodel}
	\begin{split}
		\lambda_{hb}\dot{X}_{hb} &= -\qty(\gamma Nf_{gs} + k_{sp}\qty(X_{hb}-X_{sp})) \\
		\dot{x}_a &= -C + S(f_{gs} - k_{es}(x_a+x_{es})) \\
		\dot{p}_m &= k_m^+ [\ce{Ca^2+}]_m (1-p_m) - k_m^- p_m \\
		\dot{p}_{gs} &= k_{gs}^+ [\ce{Ca^2+}]_{gs} (1-p_{gs}) - k_{gs}^- p_{gs} \\
		\tau_T^\varnothing \dot{p_T} &= p_T(\infty) - p_T
	\end{split}
	,
\end{equation}
where auxiliary functions are defined in \cref{eq:Fgs,eq:C,eq:S,eq:Cam,eq:Cags,eq:dimtauT}.

\subsection{Conversion to Nondimensional Model}
We fully nondimensionalized the system of ODEs in \cref{eq:dimmodel}. Definitions for all of the nondimensional parameters are summarized in \cref{tbl:nondimparam}, characteristic parameters in \cref{tbl:chrparam}, and nondimensional functions in \cref{tbl:nondimfunc}. We defined the following nondimensional variables
\begin{align}
	\nondim{x}_{hb} &\coloneq \frac{k_{sp}}{k_{gs,max}} \frac{X_{hb}-X_{sp}}{\gamma Nd} \\
	\nondim{x}_a &\coloneq \frac{k_{es}}{k_{gs,max}} \frac{x_a+x_{es}}{d} \\
	\nondim{t} &\coloneq \frac{t+\check{t}}{\chr{\tau}} \label{eq:nondimt}
\end{align}
This leads to the following nondimensional system of ODEs
\begin{equation}
	\begin{split}
		\nondim{\tau}_{hb} \dv{\nondim{x}_{hb}}{\nondim{t}} &= -\qty(\nondim{F}_{gs} + \nondim{x}_{hb}) \\
		\nondim{\tau}_a \dv{\nondim{x}_a}{\nondim{t}} &= \nondim{S}_{max}\nondim{S}\qty(\nondim{F}_{gs} - \nondim{x}_a) - \nondim{C}_{max}\nondim{C} \\
		\nondim{\tau}_m \dv{p_m}{\nondim{t}} &= [\ce{\nondim{Ca}^2+}]_m \nondim{\mathcal{V}}_m p_T (1-p_m) - p_m \\
		\nondim{\tau}_{gs} \dv{p_{gs}}{\nondim{t}} &= [\ce{\nondim{\text{Ca}}^2+}]_{gs} \nondim{\mathcal{V}}_m p_T (1-p_{gs}) - p_{gs} \\
		\nondim{\tau}_T \dv{p_T}{\nondim{t}} &= p_T(\infty) - p_T
	\end{split}
	.
\end{equation}

We make a few further changes to reduce parameter count. After choosing $\chr{\tau}\in\qty{\chr{\tau}_{hb},\chr{\tau}_a,\chr{\tau}_m,\chr{\tau}_{gs},\chr{\tau}_T}$, without loss of generality, we further reduced the parameter count by one, choosing $\chr{\tau}\coloneq\chr{\tau}_a$. Because $V_m$ is constant, we absorb $\nondim{\mathcal{V}}_m$ into $[\ce{\nondim{\text{Ca}}^2+}]_m$ and $[\ce{\nondim{\text{Ca}}^2+}]_{gs}$ to eliminate an additional parameter. This algebraic simplification prompts definitions for parameters $\nondim{C}_m$ and $\nondim{C}_{gs}$, shown in \cref{tbl:nondimparam}. Lastly, note that $\nondim{S}_{max}+\nondim{C}_{max}=1$, implicitly writing $\nondim{C}_{max}$ in terms of $\nondim{S}_{max}$ and reducing parameter count by one more. Afterward, the nondimensional model contains 15 free parameters. It also exhibits all features discussed in previous sections, but converted to simpler notation.

\section{Properties Used for Sensitivity Analysis}
We outline the computational methods to calculate properties of simulated, hair-bundle traces. To determine the five properties outlined in \cref{sec:properties}, we performed calculations until numerical simulations reached steady state. Afterward, we removed the initial, transient part of the simulation.

\subsection{Algorithm for Determining the Presence of Active Oscillations}
We outline the computation used to calculate $\delta\{x\}$ from \cref{eq:osc}. We labeled a total of 357,251 simulations, those that exhibited small, relative standard deviations (i.e.\ less than $10^{-4}$ for $x$ over time), as \enquote{flat}. Of the remaining simulations, we labeled one simulation,  which yielded a large (i.e.\ greater than $0.99$ for $x$ over time) Spearman's rank correlation coefficient \cite{zwillingerCRCStandardProbability2000}, as \enquote{monotonic}. Finally, we labeled the remaining 167,036 simulations, which exhibited neither flat nor monotonic behavior, as \enquote{oscillating}.

With these definitions, we rewrite
\begin{equation}
	\delta\{x\} =
	\begin{cases}
		1, & x \text{ is oscillating} \\
		0, & x \text{ is flat} \\
		0, & x \text{ is monotonic}
	\end{cases}
	.
\end{equation}

\subsection{Comparison of Amplitude Definitions} \label{sec:ampdef}
To select a reliable, amplitude-extracting algorithm for our simulations, we compared four distinct definitions, namely half of the peak-to-peak amplitude, standard deviation, average Hilbert amplitude, and best sine-wave fit amplitude. For the definition of half peak-to-peak amplitude, please refer to \cref{eq:p2pamp}. We defined the average Hilbert amplitude mathematically as
\begin{equation}
	A_H \coloneq \mean{\abs{\asig{\nondim{x}_{hb}}}}
	,
\end{equation}
where $\asig{x}$ represents the analytic signal of $x$ defined in \cref{eq:asig}. We fit a sine wave function
\begin{equation} \label{eq:sinefit}
	A \sin(\omega t + \phi) + C
\end{equation}
to each simulation of $\nondim{x}_{hb}$ and defined the best-fit-sine amplitude mathematically as
\begin{equation}
	A_S \coloneq A
	.
\end{equation}

To yield consistent, reliable results, we compared correlations between each amplitude pair (shown in \cref{fig:defcorr}). We chose half peak-to-peak amplitude, due to its high correlation with amplitudes, in relation to other definitions, and its fast computational speed.

\subsection{Comparison of Frequency Definitions} \label{sec:freqdef}
To construct a reliable, frequency-extracting algorithm, we compared three distinct definitions, namely peak-to-peak frequency, average Hilbert frequency, and best-fit-sine frequency. For the definition of Hilbert frequency, please refer to \cref{eq:Hfreq}. We defined the peak-to-peak frequency mathematically as
\begin{equation}
	f_{P2P} \coloneq \frac{1}{\Delta{t}_{peak}}
	,
\end{equation}
where $\Delta{t}_{peak}$ represents the time difference between two successive, oscillation peaks. We used two arbitrary, successive peaks after $\nondim{x}_{hb}$ settled into equilibrium to calculate the peak-to-peak frequency. We defined best-fit-sine frequency as
\begin{equation}
	f_S \coloneq \frac{\omega}{2\pi}
	,
\end{equation}
where $\omega$ represents angular frequency from \cref{eq:sinefit}.

\begin{figure*}
	\centering
	\includegraphics{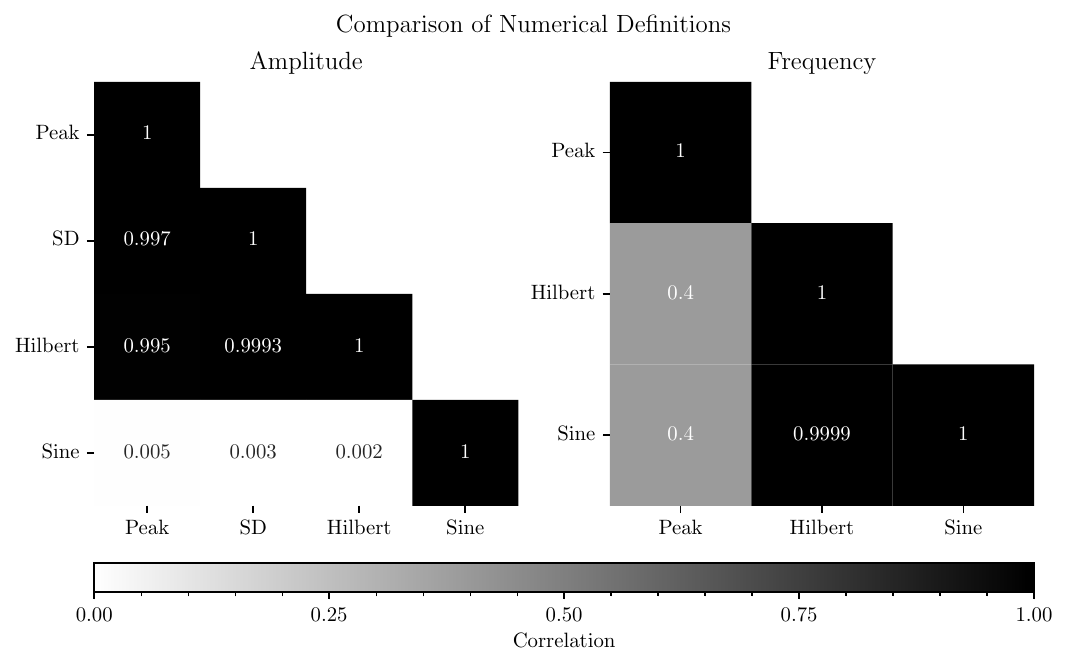}
	\caption{Correlation matrix for four distinct definitions of amplitude (left) and three of frequency (right), taken from a subset of $\sim$150,000 oscillating simulations. Four definitions of amplitude are ordered as follows (from top to bottom, left to right): half of peak-to-peak amplitude (Peak), standard deviation (SD), Hilbert amplitude (Hilbert), and best-fit-sine amplitude (Sine), all defined in \cref{sec:ampdef}. Three definitions of frequency are ordered as follows: peak-to-peak frequency (Peak), Hilbert frequency (Hilbert), and best-fit-sine frequency (Sine), all defined in \cref{sec:freqdef}. Color indicates cross-correlation value.} \label{fig:defcorr}
\end{figure*}

To foster reliable, consistent frequency values, we compared correlations between each frequency pair (shown in \cref{fig:defcorr}). We chose Hilbert frequency, due to its high correlation with frequencies from other definitions, as well as its reliability even when applied to non-oscillating simulations.

\begin{SCfigure*}
	\centering
	\includegraphics{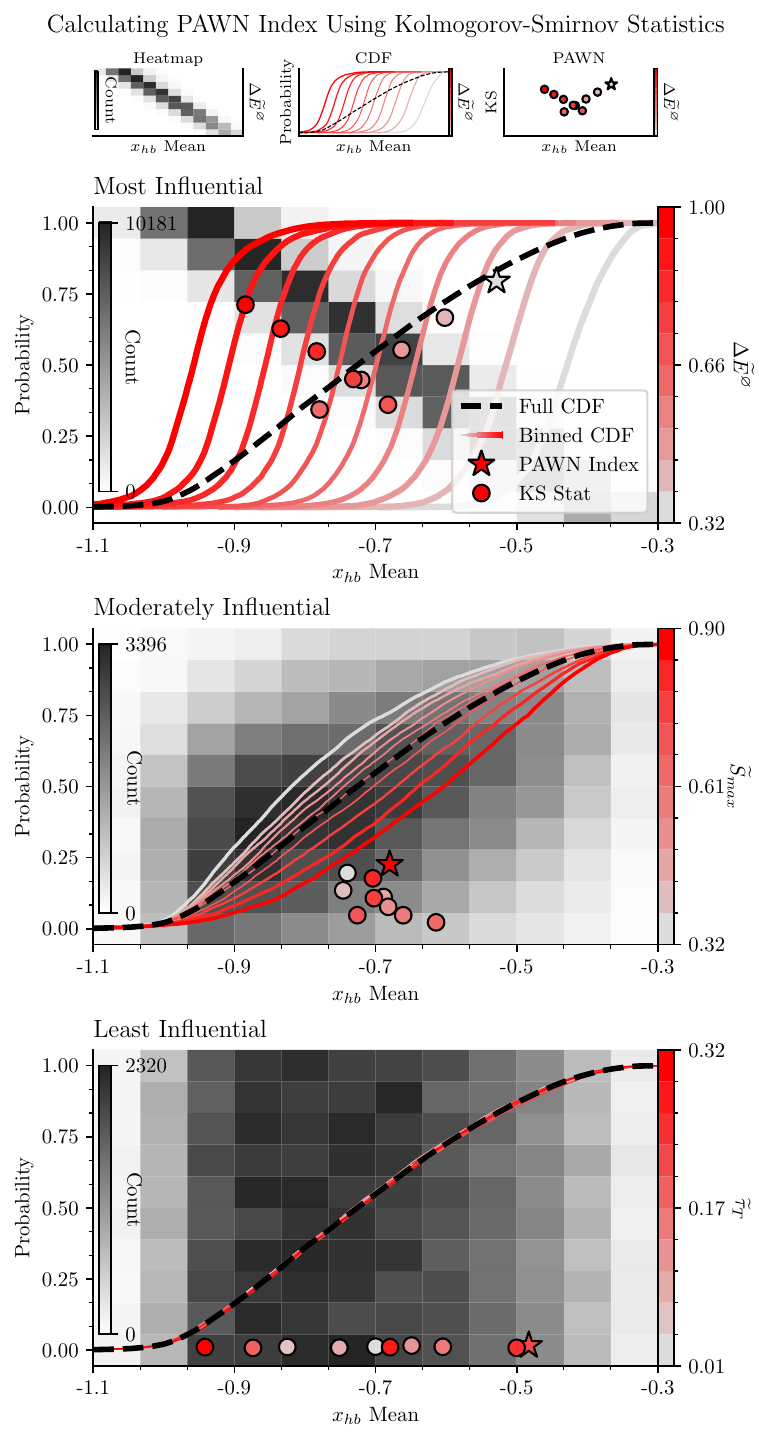}
	\caption{Demonstration on how to calculate PAWN index, taken from the subset of oscillating simulations. Near the top, three miniature plots elucidate three constituent plots superimposed in each large, primary plot. From left to right, they show 1) a heatmap of simulation counts, 2) cumulative distribution functions (CDFs), and 3) Kolmogorov-Smirnov (KS) statistics (defined in \cref{eq:KS}). Details on each constituent plot are as follows. 1) A heatmap shows simulation count. Its axes are as follows: x-axis shows mean $x_{hb}$; y-axis (sharing redscale colorbar ticks) shows parameter value; and left, grayscale colorbar shows counts per bin. 2) A standard line plot of CDFs. Its axes are as follows: x-axis shows mean $x_{hb}$; left y-axis shows probability of CDFs. The black, dashed line shows the full CDF for all oscillating simulations; the solid, redscale lines show CDFs binned by parameter value; and right, redscale colorbar shows parameter value for each bin. The thickness of the solid, red lines increases nonlinearly with the KS statistic for the corresponding binned CDF. 3) A scatter plot of KS statistics. Its axes are as follows: x-axis shows mean $x_{hb}$ where maximum separation occurs between the corresponding binned CDF and full CDF; y-axis shows KS statistic; and right, redscale colorbar shows parameter value, corresponding to binned CDF. Circular points indicate ordinary KS statistics, whereas a star indicates PAWN index (i.e.\ maximum KS statistic). Points are outlined in black to produce contrast with the background colors and convey no extra information. Extra) The order of the primary plots indicates relative parameter rankings, ordered from most to least influential parameter, i.e.\ $\Delta{\nondim{E}}^\varnothing\rightarrow\nondim{S}_{max}\rightarrow\nondim{\tau}_T$ from top to bottom. Note that each horizontal cross-section across the heatmap is equivalent to a probability density function (PDF), which matches with a binned CDF at the corresponding parameter value. The recommended order to parse each primary plot is as follows: heatmap, cross-section PDFs, binned CDFs, KS statistics, PAWN index.} \label{fig:KSdemo}
\end{SCfigure*}

\section{Comparison of Parameter Rankings} \label{sec:footrule}
We used the Spearman footrule metric to judge the distance between two ordered lists. Specifically, we used this metric to determine the distance between the two parameter rankings in \cref{fig:rank}.

We considered a distribution based on a collection of Spearman footrule distances. Let $I_n\coloneq[1,2,...n]$ be an ordered list and $I_n'$ be some random permutation of $I_n$. Let the $n^\text{th}$-order Spearman footrule distribution (SFD) be the distribution of Spearman footrule distances between $I_n$ and $I_n'$. This distribution has an exact mean of $\frac{n^2-1}{3}$, with a standard deviation $\frac{4(n+1)(2n^2+7)}{180}$ \cite{senSpearmanFootruleMarkov1983}\footnote{Errata: Equation (3.11)  in \cite{salamaNoteSpearmanFootrule1990}, for mean and standard deviation, is incorrect due to some algebraic errors. However, the preceding equation (3.10) is correct, so we derived our equations for mean and standard deviation from it. We verified our equations computationally, for various values of $n$, by Monte Carlo sampling.}; the full distribution has been tabulated for small values of $n$ \cite{salamaNoteSpearmanFootrule1990}. This distribution is asymptotically normal as $n\to\infty$ \cite{diaconisSpearmanFootruleMeasure1977}. This property implies that, for $n\gg1$, we can calculate a z-score, which now measures distance between two ordered lists. A negative z-score indicates that these lists are more similar than not.

\begin{SCfigure*}
	\centering
	\includegraphics{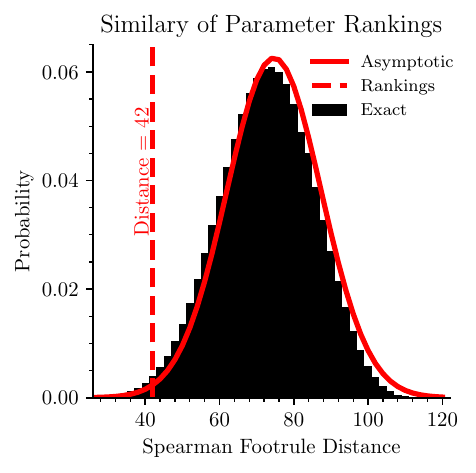}
	\caption{Similarity between two parameter rankings using a $15^\text{th}$-order Spearman footrule distribution. Black bars display the exact distribution. The solid, red line shows the asymptotic Gaussian approximation. At the red, dashed line lies the distance between our two parameter rankings (see \cref{sec:ranking}). This distance resides at z-score $z=-2.6$, about 0.5\% percentile in the asymptotic distribution, and about 0.7\% in the exact distribution.} \label{fig:spearman}
\end{SCfigure*}

We quantify the similarity between our two parameter rankings (shown in \cref{fig:spearman}). They reside distance $42$ from each other with have length $n=15$. Because these two rankings are permutations of each other, we used a $15^\text{th}$-order SFD to convert this distance into z-score $z=-2.6$. We then converted this z-score into a percentile. Because we wanted only similarity of lists, not dissimilarity, we considered the left-tailed distribution, not the two-tailed distribution. For a left-tailed, normal distribution, z-score $z=-2.6$ resides at about the 0.5\% percentile. This implies that our two parameter rankings are about 99.5\% similar.

\begin{figure*}
	\includegraphics{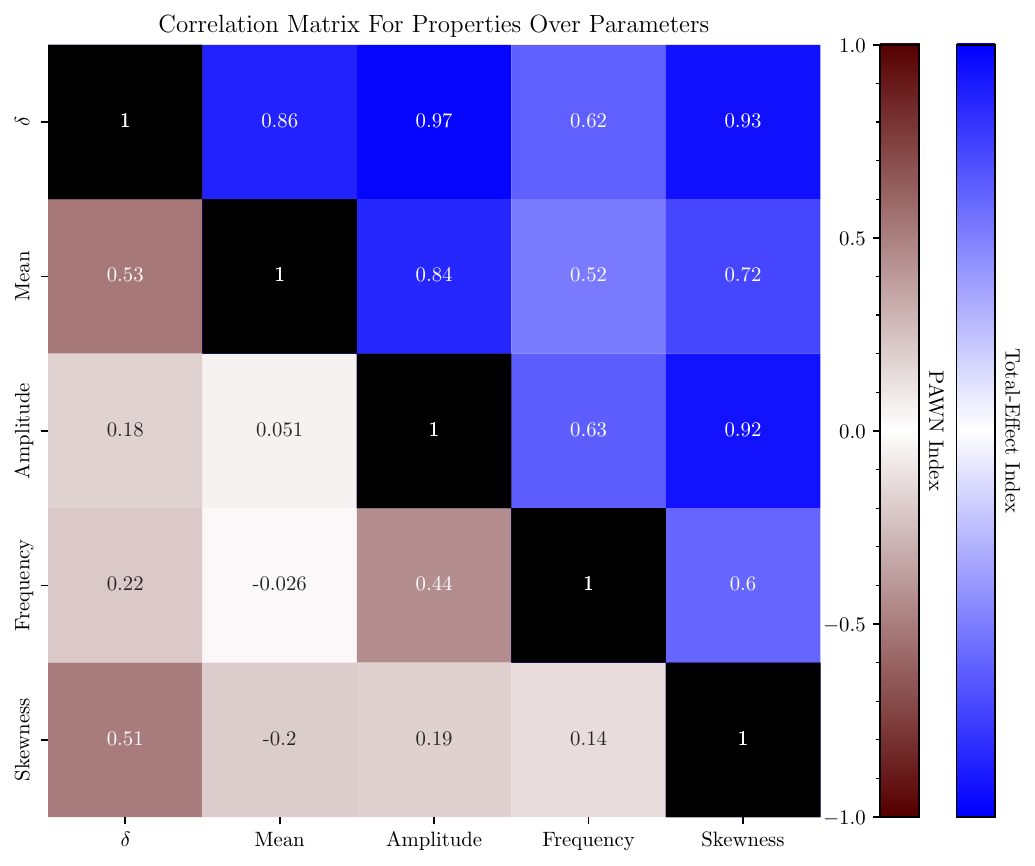}
	\caption{Correlation matrices over parameters for five properties (shown in \cref{fig:rank}). Color and location distinguish between PAWN indices (red, lower left; shown in \cref{eq:PAWN}) and TE indices (blue, upper right; shown in \cref{eq:TE}). Color brightness shows strength of correlation, whether positive or negative, with darker colors indicating stronger correlations. Notice that TE indices have very strong correlation (average magnitude of $\sim$0.8), whereas PAWN indices have only moderate correlation (average magnitude of $\sim$0.5).} \label{fig:propcorr}
\end{figure*}

\section{Numerical Methods and Python Packages}
We present a few important details for our numerical methods. We used 99.5\% stretch moves and 0.5\% kernel-density estimate moves for the MCMC algorithm, along with $N_{walkers}=10N_p$ walkers (where $N_p$ represents the number of free parameters in the respective model). This number of walkers varied from $N_{walkers}=150$ to $N_{walkers}=10$. Due to long computing times for some solutions of the model, we asserted a lower bound on $\chr{\tau}_a$ and an upper bound on $\chr{x}_{hb}$ (shown in \cref{tbl:chrparam}).

Python packages used for these simulations were: \texttt{numpy} \cite{harrisArrayProgrammingNumPy2020}; \texttt{scipy} \cite{virtanenSciPyFundamentalAlgorithms2020}; \texttt{SALib} \cite{hermanSALibOpensourcePython2017,iwanagaSALibAdvancingAccessibility2022}; \texttt{emcee} \cite{foreman-mackeyEmceeMCMCHammer2013}; \texttt{matplotlib} \cite{hunterMatplotlib2DGraphics2007}; \texttt{seaborn} \cite{waskomSeabornStatisticalData2021}; \texttt{pandas} \cite{mckinneyDataStructuresStatistical2010}

\begin{figure*}
	\centering
	\includegraphics{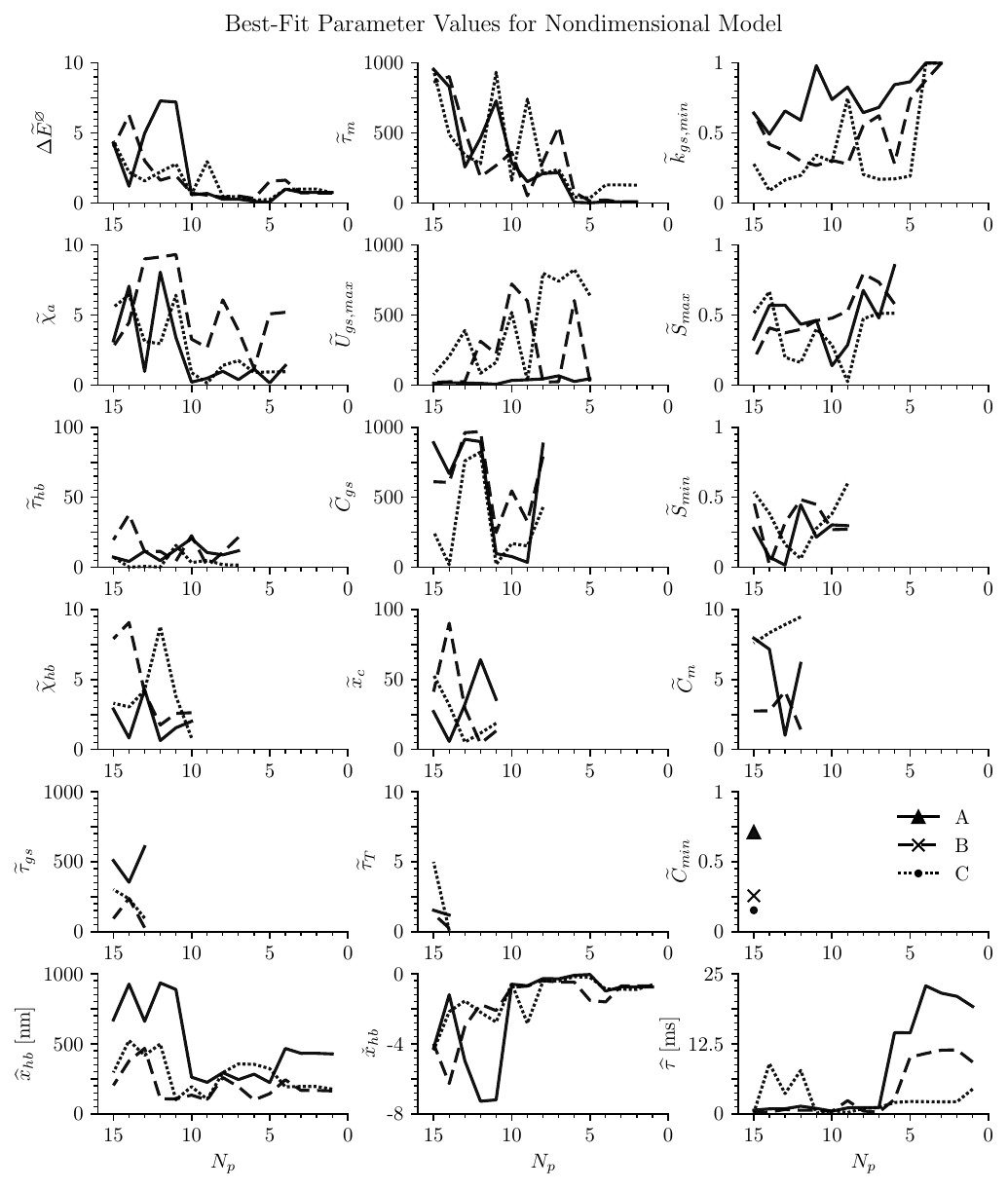}
	\caption{Best-fit parameter values for model fits to three datasets. Each plot corresponds to a different parameter indicated along the y-axis, and number of parameters $N_p$ is indicated along the x-axis. Line style (solid, dashed, dotted) or marker style (triangle, \enquote{x}, dot) indicate which of three datasets was fit. Only fitted values are shown, so lines terminate at the least $N_p$ at which they were still fit and not fixed (see \cref{tbl:fixed}). Rescaling parameters (defined in \cref{eq:xrescale}) are shown in the bottom row of plots, whereas nondimensional parameters (shown in \cref{tbl:nondimparam}) are shown in the above rows. When appropriate, i.e.\ for all nondimensional and some rescaling parameters, y-axis bounds correspond to simulated range of fits (shown in \cref{tbl:nondimparam,tbl:chrparam}).} \label{fig:bestvalues}
\end{figure*}

\begin{figure*}
	\centering
	\includegraphics{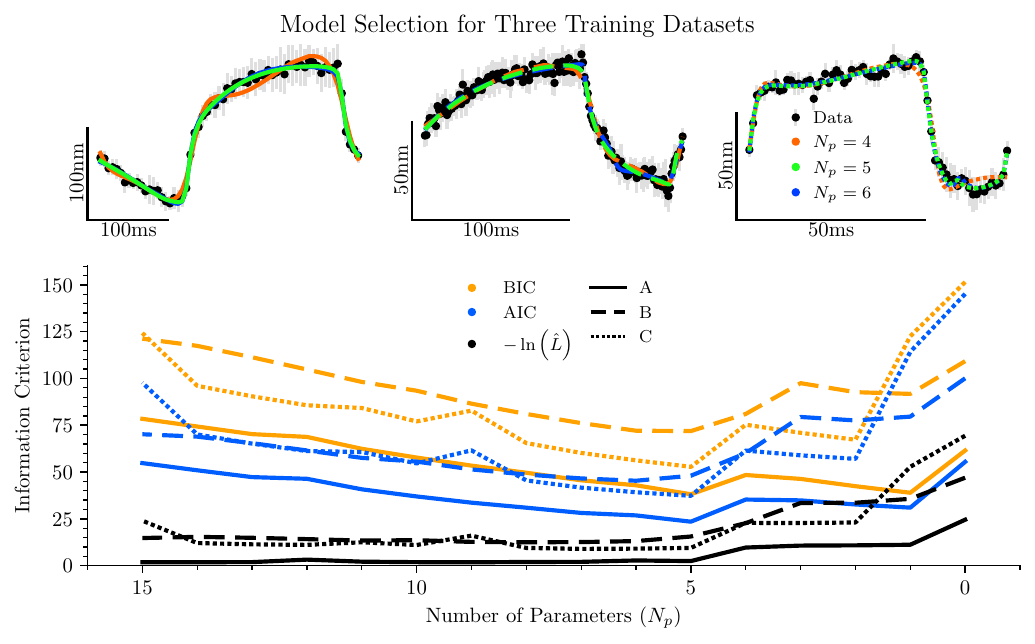}
	\caption{Comparison of best-fit models to three testing datasets along with information criteria. Line style indicates which of three datasets was fit. Top row: Lines show best-fit models, position (y-axis) vs. time (x-axis), with varying numbers of parameters; $N_p=4$ (orange), $N_p=5$ (green), and $N_p=6$ (blue). Each plot corresponds to a different dataset. Points (black) and error bars (light gray) represent measured points for each dataset along with their corresponding errors. Bottom: Information criteria and $-\ln(\hat{L})$ vs. number of parameters $N_p$ in the nondimensional model. Color indicates Bayesian (orange; defined in \cref{eq:bic}) and Akaike (blue; defined in \cref{eq:aic}) information criterion or $-\ln(\hat{L})$ (black; defined in \cref{sec:IC}). Line style indicates dataset, which corresponds to plots in top row.} \label{fig:ic}
\end{figure*}

\clearpage

\begin{table}
	\small
	\centering\makegapedcells
	\begin{tabular}{ll}
		\toprule\midrule
		Function & Definition \\\midrule
		$\nondim{k}_{gs}$* & $\displaystyle 1 - p_{gs}\qty(1-\nondim{k}_{gs,min})$ \\
		$\nondim{x}_{gs}$ & $\displaystyle \nondim{\chi}_{hb}\nondim{x}_{hb} - \nondim{\chi}_a\nondim{x}_a + \nondim{x}_c$ \\
		$\nondim{F}_{gs}$ & $\displaystyle \nondim{k}_{gs} \qty(\nondim{x}_{gs} - p_T)$ \\
		$\nondim{C}$* & $\displaystyle 1 - p_m\qty(1-\nondim{C}_{min})$ \\
		$\nondim{S}$* & $\displaystyle \nondim{S}_{min} + p_m\qty(1-\nondim{S}_{min})$ \\
		$p_T(\infty)$ & $\displaystyle \frac{1}{1+\exp(\nondim{U}_{gs,max} \qty(\Delta{\nondim{E}}^\varnothing - \nondim{k}_{gs}\qty(\nondim{x}_{gs}-\frac{1}{2}) ) )}$ \\
		\midrule\bottomrule
	\end{tabular}
	\caption{Definitions for nondimensional functions in the nondimensional model. * denotes functions that are normalized between 0 and 1, inclusively.} \label{tbl:nondimfunc}
\end{table}

\begin{table}
	\small
	\centering\makegapedcells
	\begin{tabular}{llr}
		\toprule\midrule
		Parameter & Definition & Simulated Range \\\midrule
		$\chr{\tau}_{hb}$ & $\displaystyle \frac{\lambda_{hb}}{k_{sp}}$ & --- \\
		$\chr{\tau}_a$ & $\displaystyle \frac{1}{k_{es}S_{max} \qty(1+\frac{C_{max}}{S_{max}k_{gs,max}d}) }$ & --- \\
		$\chr{\tau}_m$ & $\displaystyle \frac{1}{k_m^-}$ & --- \\
		$\chr{\tau}_{gs}$ & $\displaystyle \frac{1}{k_{gs}^-}$ & --- \\
		$\chr{\tau}_T$ & $\displaystyle \tau_T^\varnothing$ & --- \\
		$\chr{\tau}$ & $\displaystyle \chr{\tau}_a$ & [0.1, $\infty$] \\
		$\chr{x}_{hb}$ & $\displaystyle \frac{k_{gs,max}}{k_{sp}} \, \gamma Nd$ & [0, 1000] \\
		$\check{x}_{hb}$ & $\displaystyle -\frac{k_{sp}}{k_{gs,max}} \frac{X_{sp}}{\gamma Nd}$ & [$-\infty$, $\infty$] \\
		$\check{t}$ & \cref{eq:nondimt} & [0, $\infty$] \\
		\midrule\bottomrule
	\end{tabular}
	\caption{Definitions and simulated ranges for characteristic parameters in nondimensional model. When applicable, an equation reference indicates in which equation this parameter was defined implicitly.} \label{tbl:chrparam}
\end{table}

\begin{table}
	\small
	\centering\makegapedcells
	\begin{tabular}{llr}
		\toprule\midrule
		Parameter & Definition & Simulated Range \\\midrule			
		$\nondim{\tau}_{hb}$ & $\displaystyle \frac{\chr{\tau}_{hb}}{\chr{\tau}}$ & [0.01, 100] \\
		$\nondim{\tau}_m$ & $\displaystyle \frac{\chr{\tau}_m}{\chr{\tau}}$ & [0, 1000] \\
		$\nondim{\tau}_{gs}$ & $\displaystyle \frac{\chr{\tau}_{gs}}{\chr{\tau}}$ & [0, 1000] \\
		$\nondim{\tau}_a$ & $\displaystyle \frac{\chr{\tau}_a}{\chr{\tau}}$ & --- \\
		$\nondim{\tau}_T$ & $\displaystyle \frac{\chr{\tau}_T}{\chr{\tau}}$ & [0, 10] \\
		
		$\nondim{U}_{gs,max}$ & $\displaystyle \frac{k_{gs,max}d^2}{k_BT}$ & [0, 1000] \\
		$\Delta{\nondim{E}^\varnothing}$ & $\displaystyle \frac{\Delta{E}^\varnothing}{k_{gs,max}d^2}$ & [0, 10] \\
		$\nondim{\chi}_{hb}$ & $\displaystyle \frac{k_{gs,max}}{k_{sp}} \frac{\gamma^2N}{\nondim{k}_{sp}}$ & [0, 10] \\
		$\nondim{\chi}_a$ & $\displaystyle \frac{k_{gs,max}}{k_{es}}$ & [0, 10] \\
		$\nondim{x}_c$ & $\displaystyle \frac{\gamma X_{sp} + x_{es} + x_c}{d}$ & [0, 100] \\
		$\nondim{S}_{max}$* & $\displaystyle \frac{S_{max}k_{gs,max}d}{C_{max}+S_{max}k_{gs,max}d}$ & [0, 1] \\
		$\nondim{k}_{gs,min}$* & $\displaystyle \frac{k_{gs,min}}{k_{gs,max}}$ & [0, 1] \\
		$\nondim{C}_{min}$* & $\displaystyle \frac{C_{min}}{C_{max}}$ & [0, 1] \\
		$\nondim{S}_{min}$* & $\displaystyle \frac{S_{min}}{S_{max}}$ & [0, 1] \\
		$\nondim{C}_m$ & $[\ce{\nondim{\text{Ca}}^2+}]_m\nondim{\mathcal{V}}_m$ & [0, 10] \\
		$\nondim{C}_{gs}$ & $[\ce{\nondim{\text{Ca}}^2+}]_{gs}\nondim{\mathcal{V}}_m$ & [0, 1000] \\
		\midrule
		$\nondim{C}_{max}$* & $\displaystyle 1-\nondim{S}_{max}$ & --- \\
		$[\ce{\nondim{\text{Ca}}^2+}]_m$ & $\displaystyle -\frac{k_m^+}{k_m^-r_m} \frac{P_{TCa}[\ce{Ca^2+}]_{hb,ex}}{2\pi D_{Ca}}$ & ---\\
		$[\ce{\nondim{\text{Ca}}^2+}]_{gs}$ & $\displaystyle -\frac{k_{gs}^+}{k_{gs}^-r_{gs}} \frac{P_{TCa}[\ce{Ca^2+}]_{hb,ex}}{2\pi D_{Ca}}$ & --- \\
		$\nondim{V}_m$ & $\displaystyle \frac{z_{Ca}q_eV_m}{k_BT}$ & ---\\
		$\nondim{\mathcal{V}}_m$ & $\displaystyle \nondim{V}_m \frac{\frac{[\ce{Ca^2+}]_{hb,in}}{[\ce{Ca^2+}]_{hb,ex}}-\exp(-\nondim{V}_m)}{1-\exp(-\nondim{V}_m)}$ & --- \\
		\midrule\bottomrule
	\end{tabular}
	\caption{Definitions and simulated ranges for nondimensional parameters in nondimensional model. * denotes parameters that are normalized between 0 and 1, inclusively.} \label{tbl:nondimparam}
\end{table}

	\bibliography{reference}

\begin{thebibliography}{101}%
\makeatletter
\providecommand \@ifxundefined [1]{%
 \@ifx{#1\undefined}
}%
\providecommand \@ifnum [1]{%
 \ifnum #1\expandafter \@firstoftwo
 \else \expandafter \@secondoftwo
 \fi
}%
\providecommand \@ifx [1]{%
 \ifx #1\expandafter \@firstoftwo
 \else \expandafter \@secondoftwo
 \fi
}%
\providecommand \natexlab [1]{#1}%
\providecommand \enquote  [1]{``#1''}%
\providecommand \bibnamefont  [1]{#1}%
\providecommand \bibfnamefont [1]{#1}%
\providecommand \citenamefont [1]{#1}%
\providecommand \href@noop [0]{\@secondoftwo}%
\providecommand \href [0]{\begingroup \@sanitize@url \@href}%
\providecommand \@href[1]{\@@startlink{#1}\@@href}%
\providecommand \@@href[1]{\endgroup#1\@@endlink}%
\providecommand \@sanitize@url [0]{\catcode `\\12\catcode `\$12\catcode
  `\&12\catcode `\#12\catcode `\^12\catcode `\_12\catcode `\%12\relax}%
\providecommand \@@startlink[1]{}%
\providecommand \@@endlink[0]{}%
\providecommand \url  [0]{\begingroup\@sanitize@url \@url }%
\providecommand \@url [1]{\endgroup\@href {#1}{\urlprefix }}%
\providecommand \urlprefix  [0]{URL }%
\providecommand \Eprint [0]{\href }%
\providecommand \doibase [0]{https://doi.org/}%
\providecommand \selectlanguage [0]{\@gobble}%
\providecommand \bibinfo  [0]{\@secondoftwo}%
\providecommand \bibfield  [0]{\@secondoftwo}%
\providecommand \translation [1]{[#1]}%
\providecommand \BibitemOpen [0]{}%
\providecommand \bibitemStop [0]{}%
\providecommand \bibitemNoStop [0]{.\EOS\space}%
\providecommand \EOS [0]{\spacefactor3000\relax}%
\providecommand \BibitemShut  [1]{\csname bibitem#1\endcsname}%
\let\auto@bib@innerbib\@empty
\bibitem [{\citenamefont {Wever}(1949)}]{weverTheoryHearing1949}%
  \BibitemOpen
  \bibfield  {author} {\bibinfo {author} {\bibfnamefont {E.~G.}\ \bibnamefont
  {Wever}},\ }\href@noop {} {\emph {\bibinfo {title} {Theory of {{Hearing}}}}}\
  (\bibinfo  {publisher} {{Wiley}},\ \bibinfo {year} {1949})\BibitemShut
  {NoStop}%
\bibitem [{\citenamefont {Geisler}(1998)}]{geislerSoundSynapsePhysiology1998}%
  \BibitemOpen
  \bibfield  {author} {\bibinfo {author} {\bibfnamefont {C.~D.}\ \bibnamefont
  {Geisler}},\ }\href@noop {} {\emph {\bibinfo {title} {From {{Sound}} to
  {{Synapse}}: {{Physiology}} of the {{Mammalian Ear}}}}}\ (\bibinfo
  {publisher} {{Oxford University Press}},\ \bibinfo {year} {1998})\BibitemShut
  {NoStop}%
\bibitem [{\citenamefont
  {Pickles}(2013)}]{picklesIntroductionPhysiologyHearing2013}%
  \BibitemOpen
  \bibfield  {author} {\bibinfo {author} {\bibfnamefont {J.~O.}\ \bibnamefont
  {Pickles}},\ }\href@noop {} {\emph {\bibinfo {title} {An {{Introduction}} to
  the {{Physiology}} of {{Hearing}}}}}\ (\bibinfo  {publisher} {{Brill}},\
  \bibinfo {year} {2013})\BibitemShut {NoStop}%
\bibitem [{\citenamefont {G{\"o}pfert}\ and\ \citenamefont
  {Robert}(2001)}]{gopfertActiveAuditoryMechanics2001}%
  \BibitemOpen
  \bibfield  {author} {\bibinfo {author} {\bibfnamefont {M.~C.}\ \bibnamefont
  {G{\"o}pfert}}\ and\ \bibinfo {author} {\bibfnamefont {D.}~\bibnamefont
  {Robert}},\ }\bibfield  {title} {\bibinfo {title} {Active auditory mechanics
  in mosquitoes},\ }\href {https://doi.org/10.1098/rspb.2000.1376} {\bibfield
  {journal} {\bibinfo  {journal} {Proc. R. Soc. B}\ }\textbf {\bibinfo {volume}
  {268}},\ \bibinfo {pages} {333} (\bibinfo {year} {2001})}\BibitemShut
  {NoStop}%
\bibitem [{\citenamefont
  {Hudspeth}(2014)}]{hudspethIntegratingActiveProcess2014}%
  \BibitemOpen
  \bibfield  {author} {\bibinfo {author} {\bibfnamefont {A.~J.}\ \bibnamefont
  {Hudspeth}},\ }\bibfield  {title} {\bibinfo {title} {Integrating the active
  process of hair cells with cochlear function},\ }\href
  {https://doi.org/10.1038/nrn3786} {\bibfield  {journal} {\bibinfo  {journal}
  {Nat. Rev. Neurosci.}\ }\textbf {\bibinfo {volume} {15}},\ \bibinfo {pages}
  {600} (\bibinfo {year} {2014})}\BibitemShut {NoStop}%
\bibitem [{\citenamefont
  {Bozovic}(2019)}]{bozovicActiveBiomechanicsSensory2019}%
  \BibitemOpen
  \bibfield  {author} {\bibinfo {author} {\bibfnamefont {D.}~\bibnamefont
  {Bozovic}},\ }\bibfield  {title} {\bibinfo {title} {Active {{Biomechanics}}
  of {{Sensory Hair Bundles}}},\ }\href
  {https://doi.org/10.1101/cshperspect.a035014} {\bibfield  {journal} {\bibinfo
   {journal} {CSH Perspect. Med.}\ }\textbf {\bibinfo {volume} {9}},\ \bibinfo
  {pages} {a035014} (\bibinfo {year} {2019})}\BibitemShut {NoStop}%
\bibitem [{\citenamefont {Hudspeth}(1989)}]{hudspethHowEarWorks1989}%
  \BibitemOpen
  \bibfield  {author} {\bibinfo {author} {\bibfnamefont {A.~J.}\ \bibnamefont
  {Hudspeth}},\ }\bibfield  {title} {\bibinfo {title} {How the ear's works
  work},\ }\href {https://doi.org/10.1038/341397a0} {\bibfield  {journal}
  {\bibinfo  {journal} {Nature}\ }\textbf {\bibinfo {volume} {341}},\ \bibinfo
  {pages} {397} (\bibinfo {year} {1989})}\BibitemShut {NoStop}%
\bibitem [{\citenamefont {Smotherman}\ and\ \citenamefont
  {Narins}(2000)}]{smothermanHearingFrogs2000}%
  \BibitemOpen
  \bibfield  {author} {\bibinfo {author} {\bibfnamefont {M.~S.}\ \bibnamefont
  {Smotherman}}\ and\ \bibinfo {author} {\bibfnamefont {P.~M.}\ \bibnamefont
  {Narins}},\ }\bibfield  {title} {\bibinfo {title} {Hearing in frogs},\ }\href
  {https://doi.org/10.1242/jeb.203.15.2237} {\bibfield  {journal} {\bibinfo
  {journal} {J. Exp. Biol.}\ }\textbf {\bibinfo {volume} {203}},\ \bibinfo
  {pages} {2237} (\bibinfo {year} {2000})}\BibitemShut {NoStop}%
\bibitem [{\citenamefont {Martin}\ \emph {et~al.}(2003)\citenamefont {Martin},
  \citenamefont {Bozovic}, \citenamefont {Choe},\ and\ \citenamefont
  {Hudspeth}}]{martinSpontaneousOscillationHair2003}%
  \BibitemOpen
  \bibfield  {author} {\bibinfo {author} {\bibfnamefont {P.}~\bibnamefont
  {Martin}}, \bibinfo {author} {\bibfnamefont {D.}~\bibnamefont {Bozovic}},
  \bibinfo {author} {\bibfnamefont {Y.}~\bibnamefont {Choe}},\ and\ \bibinfo
  {author} {\bibfnamefont {A.~J.}\ \bibnamefont {Hudspeth}},\ }\bibfield
  {title} {\bibinfo {title} {Spontaneous oscillation by hair bundles of the
  bullfrog's sacculus},\ }\href
  {https://doi.org/10.1523/jneurosci.23-11-04533.2003} {\bibfield  {journal}
  {\bibinfo  {journal} {J. Neurosci.}\ }\textbf {\bibinfo {volume} {23}},\
  \bibinfo {pages} {4533} (\bibinfo {year} {2003})}\BibitemShut {NoStop}%
\bibitem [{\citenamefont {Beurg}\ \emph {et~al.}(2010)\citenamefont {Beurg},
  \citenamefont {Nam}, \citenamefont {Chen},\ and\ \citenamefont
  {Fettiplace}}]{beurgCalciumBalanceMechanotransduction2010}%
  \BibitemOpen
  \bibfield  {author} {\bibinfo {author} {\bibfnamefont {M.}~\bibnamefont
  {Beurg}}, \bibinfo {author} {\bibfnamefont {J.~H.}\ \bibnamefont {Nam}},
  \bibinfo {author} {\bibfnamefont {Q.}~\bibnamefont {Chen}},\ and\ \bibinfo
  {author} {\bibfnamefont {R.}~\bibnamefont {Fettiplace}},\ }\bibfield  {title}
  {\bibinfo {title} {Calcium balance and mechanotransduction in rat cochlear
  hair cells},\ }\href {https://doi.org/10.1152/jn.00019.2010} {\bibfield
  {journal} {\bibinfo  {journal} {J. Neurophysiol.}\ }\textbf {\bibinfo
  {volume} {104}},\ \bibinfo {pages} {18} (\bibinfo {year} {2010})}\BibitemShut
  {NoStop}%
\bibitem [{\citenamefont {Peng}\ \emph {et~al.}(2013)\citenamefont {Peng},
  \citenamefont {Effertz},\ and\ \citenamefont
  {Ricci}}]{pengAdaptationMammalianAuditory2013}%
  \BibitemOpen
  \bibfield  {author} {\bibinfo {author} {\bibfnamefont {A.}~\bibnamefont
  {Peng}}, \bibinfo {author} {\bibfnamefont {T.}~\bibnamefont {Effertz}},\ and\
  \bibinfo {author} {\bibfnamefont {A.}~\bibnamefont {Ricci}},\ }\bibfield
  {title} {\bibinfo {title} {Adaptation of mammalian auditory hair cell
  mechanotransduction is independent of calcium entry},\ }\href
  {https://doi.org/10.1016/j.neuron.2013.08.025} {\bibfield  {journal}
  {\bibinfo  {journal} {Neuron}\ }\textbf {\bibinfo {volume} {80}},\ \bibinfo
  {pages} {960} (\bibinfo {year} {2013})}\BibitemShut {NoStop}%
\bibitem [{\citenamefont {Quino{\~n}es}\ \emph {et~al.}(2015)\citenamefont
  {Quino{\~n}es}, \citenamefont {Meenderink},\ and\ \citenamefont
  {Bozovic}}]{quinonesVoltageandCalciumdependentMotility2015}%
  \BibitemOpen
  \bibfield  {author} {\bibinfo {author} {\bibfnamefont {P.~M.}\ \bibnamefont
  {Quino{\~n}es}}, \bibinfo {author} {\bibfnamefont {S.~W.}\ \bibnamefont
  {Meenderink}},\ and\ \bibinfo {author} {\bibfnamefont {D.}~\bibnamefont
  {Bozovic}},\ }\bibfield  {title} {\bibinfo {title} {Voltage-and
  calcium-dependent motility of saccular hair bundles},\ }\href
  {https://doi.org/10.1063/1.4939321} {\bibfield  {journal} {\bibinfo
  {journal} {AIP Conference Proceedings}\ }\textbf {\bibinfo {volume} {1703}},\
  \bibinfo {pages} {1} (\bibinfo {year} {2015})}\BibitemShut {NoStop}%
\bibitem [{\citenamefont {Meenderink}\ \emph {et~al.}(2015)\citenamefont
  {Meenderink}, \citenamefont {Qui{\~n}ones},\ and\ \citenamefont
  {Bozovic}}]{meenderinkVoltagemediatedControlSpontaneous2015}%
  \BibitemOpen
  \bibfield  {author} {\bibinfo {author} {\bibfnamefont {S.~W.}\ \bibnamefont
  {Meenderink}}, \bibinfo {author} {\bibfnamefont {P.~M.}\ \bibnamefont
  {Qui{\~n}ones}},\ and\ \bibinfo {author} {\bibfnamefont {D.}~\bibnamefont
  {Bozovic}},\ }\bibfield  {title} {\bibinfo {title} {Voltage-mediated control
  of spontaneous bundle oscillations in saccular hair cells},\ }\href
  {https://doi.org/10.1523/JNEUROSCI.1451-15.2015} {\bibfield  {journal}
  {\bibinfo  {journal} {J. Neurosci.}\ }\textbf {\bibinfo {volume} {35}},\
  \bibinfo {pages} {14457} (\bibinfo {year} {2015})}\BibitemShut {NoStop}%
\bibitem [{\citenamefont {Jaramillo}\ and\ \citenamefont
  {Hudspeth}(1993)}]{jaramilloDisplacementclampMeasurementForces1993}%
  \BibitemOpen
  \bibfield  {author} {\bibinfo {author} {\bibfnamefont {F.}~\bibnamefont
  {Jaramillo}}\ and\ \bibinfo {author} {\bibfnamefont {A.~J.}\ \bibnamefont
  {Hudspeth}},\ }\bibfield  {title} {\bibinfo {title} {Displacement-clamp
  measurement of the forces exerted by gating springs in the hair bundle},\
  }\href {https://doi.org/10.1073/pnas.90.4.1330} {\bibfield  {journal}
  {\bibinfo  {journal} {Proc. Natl. Acad. Sci.}\ }\textbf {\bibinfo {volume}
  {90}},\ \bibinfo {pages} {1330} (\bibinfo {year} {1993})}\BibitemShut
  {NoStop}%
\bibitem [{\citenamefont {Strimbu}\ \emph {et~al.}(2012)\citenamefont
  {Strimbu}, \citenamefont {{Fredrickson-Hemsing}},\ and\ \citenamefont
  {Bozovic}}]{strimbuCouplingElasticLoading2012}%
  \BibitemOpen
  \bibfield  {author} {\bibinfo {author} {\bibfnamefont {C.~E.}\ \bibnamefont
  {Strimbu}}, \bibinfo {author} {\bibfnamefont {L.}~\bibnamefont
  {{Fredrickson-Hemsing}}},\ and\ \bibinfo {author} {\bibfnamefont
  {D.}~\bibnamefont {Bozovic}},\ }\bibfield  {title} {\bibinfo {title}
  {Coupling and elastic loading affect the active response by the inner ear
  hair cell bundles},\ }\href {https://doi.org/10.1371/journal.pone.0033862}
  {\bibfield  {journal} {\bibinfo  {journal} {PLoS One}\ }\textbf {\bibinfo
  {volume} {7}},\ \bibinfo {pages} {e33862} (\bibinfo {year}
  {2012})}\BibitemShut {NoStop}%
\bibitem [{\citenamefont {{Ramunno-Johnson}}\ \emph {et~al.}(2010)\citenamefont
  {{Ramunno-Johnson}}, \citenamefont {Strimbu}, \citenamefont {Kao},
  \citenamefont {Fredrickson~Hemsing},\ and\ \citenamefont
  {Bozovic}}]{ramunno-johnsonEffectsSomaticIon2010}%
  \BibitemOpen
  \bibfield  {author} {\bibinfo {author} {\bibfnamefont {D.}~\bibnamefont
  {{Ramunno-Johnson}}}, \bibinfo {author} {\bibfnamefont {C.~E.}\ \bibnamefont
  {Strimbu}}, \bibinfo {author} {\bibfnamefont {A.}~\bibnamefont {Kao}},
  \bibinfo {author} {\bibfnamefont {L.}~\bibnamefont {Fredrickson~Hemsing}},\
  and\ \bibinfo {author} {\bibfnamefont {D.}~\bibnamefont {Bozovic}},\
  }\bibfield  {title} {\bibinfo {title} {Effects of the somatic ion channels
  upon spontaneous mechanical oscillations in hair bundles of the inner ear},\
  }\href {https://doi.org/10.1016/j.heares.2010.05.017} {\bibfield  {journal}
  {\bibinfo  {journal} {Hear. Res.}\ }\textbf {\bibinfo {volume} {268}},\
  \bibinfo {pages} {163} (\bibinfo {year} {2010})}\BibitemShut {NoStop}%
\bibitem [{\citenamefont {Assad}\ \emph {et~al.}(1991)\citenamefont {Assad},
  \citenamefont {Shepherd},\ and\ \citenamefont
  {Corey}}]{assadTiplinkIntegrityMechanical1991}%
  \BibitemOpen
  \bibfield  {author} {\bibinfo {author} {\bibfnamefont {J.~A.}\ \bibnamefont
  {Assad}}, \bibinfo {author} {\bibfnamefont {G.~M.}\ \bibnamefont
  {Shepherd}},\ and\ \bibinfo {author} {\bibfnamefont {D.~P.}\ \bibnamefont
  {Corey}},\ }\bibfield  {title} {\bibinfo {title} {Tip-link integrity and
  mechanical transduction in vertebrate hair cells},\ }\href
  {https://doi.org/10.1016/0896-6273(91)90343-X} {\bibfield  {journal}
  {\bibinfo  {journal} {Neuron}\ }\textbf {\bibinfo {volume} {7}},\ \bibinfo
  {pages} {985} (\bibinfo {year} {1991})}\BibitemShut {NoStop}%
\bibitem [{\citenamefont {Beurg}\ \emph {et~al.}(2008)\citenamefont {Beurg},
  \citenamefont {Nam}, \citenamefont {Crawford},\ and\ \citenamefont
  {Fettiplace}}]{beurgActionsCalciumHair2008}%
  \BibitemOpen
  \bibfield  {author} {\bibinfo {author} {\bibfnamefont {M.}~\bibnamefont
  {Beurg}}, \bibinfo {author} {\bibfnamefont {J.~H.}\ \bibnamefont {Nam}},
  \bibinfo {author} {\bibfnamefont {A.}~\bibnamefont {Crawford}},\ and\
  \bibinfo {author} {\bibfnamefont {R.}~\bibnamefont {Fettiplace}},\ }\bibfield
   {title} {\bibinfo {title} {The actions of calcium on hair bundle mechanics
  in mammalian cochlear hair cells},\ }\href
  {https://doi.org/10.1529/biophysj.107.123257} {\bibfield  {journal} {\bibinfo
   {journal} {Biophys. J.}\ }\textbf {\bibinfo {volume} {94}},\ \bibinfo
  {pages} {2639} (\bibinfo {year} {2008})}\BibitemShut {NoStop}%
\bibitem [{\citenamefont {Hudspeth}\ \emph {et~al.}(2000)\citenamefont
  {Hudspeth}, \citenamefont {Choe}, \citenamefont {Mehta},\ and\ \citenamefont
  {Martin}}]{hudspethPuttingIonChannels2000}%
  \BibitemOpen
  \bibfield  {author} {\bibinfo {author} {\bibfnamefont {A.~J.}\ \bibnamefont
  {Hudspeth}}, \bibinfo {author} {\bibfnamefont {Y.}~\bibnamefont {Choe}},
  \bibinfo {author} {\bibfnamefont {A.~D.}\ \bibnamefont {Mehta}},\ and\
  \bibinfo {author} {\bibfnamefont {P.}~\bibnamefont {Martin}},\ }\bibfield
  {title} {\bibinfo {title} {Putting ion channels to work:
  {{Mechanoelectrical}} transduction, adaptation, and amplification by hair
  cells},\ }\href {https://doi.org/10.1073/pnas.97.22.11765} {\bibfield
  {journal} {\bibinfo  {journal} {Proc. Natl. Acad. Sci.}\ }\textbf {\bibinfo
  {volume} {97}},\ \bibinfo {pages} {11765} (\bibinfo {year}
  {2000})}\BibitemShut {NoStop}%
\bibitem [{\citenamefont {Nadrowski}\ \emph {et~al.}(2004)\citenamefont
  {Nadrowski}, \citenamefont {Martin},\ and\ \citenamefont
  {J{\"u}licher}}]{nadrowskiActiveHairbundleMotility2004}%
  \BibitemOpen
  \bibfield  {author} {\bibinfo {author} {\bibfnamefont {B.}~\bibnamefont
  {Nadrowski}}, \bibinfo {author} {\bibfnamefont {P.}~\bibnamefont {Martin}},\
  and\ \bibinfo {author} {\bibfnamefont {F.}~\bibnamefont {J{\"u}licher}},\
  }\bibfield  {title} {\bibinfo {title} {Active hair-bundle motility harnesses
  noise to operate near an optimum of mechanosensitivity},\ }\href
  {https://doi.org/10.1073/pnas.0403020101} {\bibfield  {journal} {\bibinfo
  {journal} {Proc. Natl. Acad. Sci.}\ }\textbf {\bibinfo {volume} {101}},\
  \bibinfo {pages} {12195} (\bibinfo {year} {2004})}\BibitemShut {NoStop}%
\bibitem [{\citenamefont {Fruth}\ \emph {et~al.}(2014)\citenamefont {Fruth},
  \citenamefont {J{\"u}licher},\ and\ \citenamefont
  {Lindner}}]{fruthActiveOscillatorModel2014}%
  \BibitemOpen
  \bibfield  {author} {\bibinfo {author} {\bibfnamefont {F.}~\bibnamefont
  {Fruth}}, \bibinfo {author} {\bibfnamefont {F.}~\bibnamefont
  {J{\"u}licher}},\ and\ \bibinfo {author} {\bibfnamefont {B.}~\bibnamefont
  {Lindner}},\ }\bibfield  {title} {\bibinfo {title} {An {{Active Oscillator
  Model Describes}} the {{Statistics}} of {{Spontaneous Otoacoustic
  Emissions}}},\ }\href {https://doi.org/10.1016/j.bpj.2014.06.047} {\bibfield
  {journal} {\bibinfo  {journal} {Biophys. J.}\ }\textbf {\bibinfo {volume}
  {107}},\ \bibinfo {pages} {815} (\bibinfo {year} {2014})}\BibitemShut
  {NoStop}%
\bibitem [{\citenamefont {Homma}\ and\ \citenamefont
  {Saltelli}(1996)}]{hommaImportanceMeasuresGlobal1996}%
  \BibitemOpen
  \bibfield  {author} {\bibinfo {author} {\bibfnamefont {T.}~\bibnamefont
  {Homma}}\ and\ \bibinfo {author} {\bibfnamefont {A.}~\bibnamefont
  {Saltelli}},\ }\bibfield  {title} {\bibinfo {title} {Importance measures in
  global sensitivity analysis of nonlinear models},\ }\href
  {https://doi.org/10.1016/0951-8320(96)00002-6} {\bibfield  {journal}
  {\bibinfo  {journal} {Reliability Engineering \& System Safety}\ }\textbf
  {\bibinfo {volume} {52}},\ \bibinfo {pages} {1} (\bibinfo {year}
  {1996})}\BibitemShut {NoStop}%
\bibitem [{\citenamefont
  {Sobol{${'}$}}(2001)}]{sobolGlobalSensitivityIndices2001}%
  \BibitemOpen
  \bibfield  {author} {\bibinfo {author} {\bibfnamefont {I.~M.}\ \bibnamefont
  {Sobol{${'}$}}},\ }\bibfield  {title} {\bibinfo {title} {Global sensitivity
  indices for nonlinear mathematical models and their {{Monte Carlo}}
  estimates},\ }\href {https://doi.org/10.1016/S0378-4754(00)00270-6}
  {\bibfield  {journal} {\bibinfo  {journal} {Math. Comput. Simul.}\ }\bibinfo
  {series} {The {{Second IMACS Seminar}} on {{Monte Carlo Methods}}},\ \textbf
  {\bibinfo {volume} {55}},\ \bibinfo {pages} {271} (\bibinfo {year}
  {2001})}\BibitemShut {NoStop}%
\bibitem [{\citenamefont {Saltelli}\ \emph {et~al.}(2010)\citenamefont
  {Saltelli}, \citenamefont {Annoni}, \citenamefont {Azzini}, \citenamefont
  {Campolongo}, \citenamefont {Ratto},\ and\ \citenamefont
  {Tarantola}}]{saltelliVarianceBasedSensitivity2010}%
  \BibitemOpen
  \bibfield  {author} {\bibinfo {author} {\bibfnamefont {A.}~\bibnamefont
  {Saltelli}}, \bibinfo {author} {\bibfnamefont {P.}~\bibnamefont {Annoni}},
  \bibinfo {author} {\bibfnamefont {I.}~\bibnamefont {Azzini}}, \bibinfo
  {author} {\bibfnamefont {F.}~\bibnamefont {Campolongo}}, \bibinfo {author}
  {\bibfnamefont {M.}~\bibnamefont {Ratto}},\ and\ \bibinfo {author}
  {\bibfnamefont {S.}~\bibnamefont {Tarantola}},\ }\bibfield  {title} {\bibinfo
  {title} {Variance based sensitivity analysis of model output. {{Design}} and
  estimator for the total sensitivity index},\ }\href
  {https://doi.org/10.1016/j.cpc.2009.09.018} {\bibfield  {journal} {\bibinfo
  {journal} {Computer Physics Communications}\ }\textbf {\bibinfo {volume}
  {181}},\ \bibinfo {pages} {259} (\bibinfo {year} {2010})}\BibitemShut
  {NoStop}%
\bibitem [{\citenamefont {Pianosi}\ and\ \citenamefont
  {Wagener}(2015)}]{pianosiSimpleEfficientMethod2015}%
  \BibitemOpen
  \bibfield  {author} {\bibinfo {author} {\bibfnamefont {F.}~\bibnamefont
  {Pianosi}}\ and\ \bibinfo {author} {\bibfnamefont {T.}~\bibnamefont
  {Wagener}},\ }\bibfield  {title} {\bibinfo {title} {A simple and efficient
  method for global sensitivity analysis based on~cumulative distribution
  functions},\ }\href {https://doi.org/10/f677qs} {\bibfield  {journal}
  {\bibinfo  {journal} {Environ. Model. Softw.}\ }\textbf {\bibinfo {volume}
  {67}},\ \bibinfo {pages} {1} (\bibinfo {year} {2015})}\BibitemShut {NoStop}%
\bibitem [{\citenamefont
  {McElreath}(2015)}]{mcelreathStatisticalRethinking2015}%
  \BibitemOpen
  \bibfield  {author} {\bibinfo {author} {\bibfnamefont {R.}~\bibnamefont
  {McElreath}},\ }\href@noop {} {\emph {\bibinfo {title} {Statistical
  {{Rethinking}}}}}\ (\bibinfo {year} {2015})\BibitemShut {NoStop}%
\bibitem [{\citenamefont {Mortlock}\ \emph {et~al.}(2021)\citenamefont
  {Mortlock}, \citenamefont {Georgia},\ and\ \citenamefont
  {Finley}}]{mortlockDynamicRegulationJAKSTAT2021}%
  \BibitemOpen
  \bibfield  {author} {\bibinfo {author} {\bibfnamefont {R.~D.}\ \bibnamefont
  {Mortlock}}, \bibinfo {author} {\bibfnamefont {S.~K.}\ \bibnamefont
  {Georgia}},\ and\ \bibinfo {author} {\bibfnamefont {S.~D.}\ \bibnamefont
  {Finley}},\ }\bibfield  {title} {\bibinfo {title} {Dynamic {{Regulation}} of
  {{JAK-STAT Signaling Through}} the {{Prolactin Receptor Predicted}} by
  {{Computational Modeling}}},\ }\href
  {https://doi.org/10.1007/s12195-020-00647-8} {\bibfield  {journal} {\bibinfo
  {journal} {Cell Mol. Bioeng.}\ }\textbf {\bibinfo {volume} {14}},\ \bibinfo
  {pages} {15} (\bibinfo {year} {2021})}\BibitemShut {NoStop}%
\bibitem [{\citenamefont {Linden}\ \emph {et~al.}(2022)\citenamefont {Linden},
  \citenamefont {Kramer},\ and\ \citenamefont
  {Rangamani}}]{lindenBayesianParameterEstimation2022}%
  \BibitemOpen
  \bibfield  {author} {\bibinfo {author} {\bibfnamefont {N.~J.}\ \bibnamefont
  {Linden}}, \bibinfo {author} {\bibfnamefont {B.}~\bibnamefont {Kramer}},\
  and\ \bibinfo {author} {\bibfnamefont {P.}~\bibnamefont {Rangamani}},\
  }\bibfield  {title} {\bibinfo {title} {Bayesian parameter estimation for
  dynamical models in systems biology},\ }\href {https://doi.org/10/grtqct}
  {\bibfield  {journal} {\bibinfo  {journal} {PLoS Comput. Biol.}\ }\textbf
  {\bibinfo {volume} {18}},\ \bibinfo {pages} {e1010651} (\bibinfo {year}
  {2022})}\BibitemShut {NoStop}%
\bibitem [{\citenamefont {Ospeck}\ \emph {et~al.}(2001)\citenamefont {Ospeck},
  \citenamefont {Egu{\'i}luz},\ and\ \citenamefont
  {Magnasco}}]{ospeckEvidenceHopfBifurcation2001}%
  \BibitemOpen
  \bibfield  {author} {\bibinfo {author} {\bibfnamefont {M.}~\bibnamefont
  {Ospeck}}, \bibinfo {author} {\bibfnamefont {V.~M.}\ \bibnamefont
  {Egu{\'i}luz}},\ and\ \bibinfo {author} {\bibfnamefont {M.~O.}\ \bibnamefont
  {Magnasco}},\ }\bibfield  {title} {\bibinfo {title} {Evidence of a {{Hopf}}
  bifurcation in frog hair cells.},\ }\href
  {https://doi.org/10.1016/S0006-3495(01)76230-3} {\bibfield  {journal}
  {\bibinfo  {journal} {Biophys. J.}\ }\textbf {\bibinfo {volume} {80}},\
  \bibinfo {pages} {2597} (\bibinfo {year} {2001})}\BibitemShut {NoStop}%
\bibitem [{\citenamefont {Hudspeth}\ \emph {et~al.}(2010)\citenamefont
  {Hudspeth}, \citenamefont {J{\"u}licher},\ and\ \citenamefont
  {Martin}}]{hudspethCritiqueCriticalCochlea2010}%
  \BibitemOpen
  \bibfield  {author} {\bibinfo {author} {\bibfnamefont {A.~J.}\ \bibnamefont
  {Hudspeth}}, \bibinfo {author} {\bibfnamefont {F.}~\bibnamefont
  {J{\"u}licher}},\ and\ \bibinfo {author} {\bibfnamefont {P.}~\bibnamefont
  {Martin}},\ }\bibfield  {title} {\bibinfo {title} {A {{Critique}} of the
  {{Critical Cochlea}}: {{Hopf}}{\textemdash}a {{Bifurcation}}{\textemdash}{{Is
  Better Than None}}},\ }\href {https://doi.org/10.1152/jn.00437.2010}
  {\bibfield  {journal} {\bibinfo  {journal} {J. Neurophysiol.}\ }\textbf
  {\bibinfo {volume} {104}},\ \bibinfo {pages} {1219} (\bibinfo {year}
  {2010})}\BibitemShut {NoStop}%
\bibitem [{\citenamefont {Faber}\ and\ \citenamefont
  {Bozovic}(2018)}]{faberChaoticDynamicsInner2018}%
  \BibitemOpen
  \bibfield  {author} {\bibinfo {author} {\bibfnamefont {J.}~\bibnamefont
  {Faber}}\ and\ \bibinfo {author} {\bibfnamefont {D.}~\bibnamefont
  {Bozovic}},\ }\bibfield  {title} {\bibinfo {title} {Chaotic {{Dynamics}} of
  {{Inner Ear Hair Cells}}.},\ }\href
  {https://doi.org/10.1038/s41598-018-21538-z} {\bibfield  {journal} {\bibinfo
  {journal} {Sci. Rep.}\ }\textbf {\bibinfo {volume} {8}},\ \bibinfo {pages}
  {3366} (\bibinfo {year} {2018})},\ \Eprint {https://arxiv.org/abs/1702.02703}
  {arxiv:1702.02703} \BibitemShut {NoStop}%
\bibitem [{\citenamefont {Kachar}\ \emph {et~al.}(2000)\citenamefont {Kachar},
  \citenamefont {Parakkal}, \citenamefont {Kurc}, \citenamefont {Zhao},\ and\
  \citenamefont {Gillespie}}]{kacharHighresolutionStructureHaircell2000}%
  \BibitemOpen
  \bibfield  {author} {\bibinfo {author} {\bibfnamefont {B.}~\bibnamefont
  {Kachar}}, \bibinfo {author} {\bibfnamefont {M.}~\bibnamefont {Parakkal}},
  \bibinfo {author} {\bibfnamefont {M.}~\bibnamefont {Kurc}}, \bibinfo {author}
  {\bibfnamefont {Y.-d.}\ \bibnamefont {Zhao}},\ and\ \bibinfo {author}
  {\bibfnamefont {P.~G.}\ \bibnamefont {Gillespie}},\ }\bibfield  {title}
  {\bibinfo {title} {High-resolution structure of hair-cell tip links},\ }\href
  {https://doi.org/10.1073/pnas.97.24.13336} {\bibfield  {journal} {\bibinfo
  {journal} {Proc. Natl. Acad. Sci.}\ }\textbf {\bibinfo {volume} {97}},\
  \bibinfo {pages} {13336} (\bibinfo {year} {2000})}\BibitemShut {NoStop}%
\bibitem [{\citenamefont {Furness}\ \emph {et~al.}(2008)\citenamefont
  {Furness}, \citenamefont {Mahendrasingam}, \citenamefont {Ohashi},
  \citenamefont {Fettiplace},\ and\ \citenamefont
  {Hackney}}]{furnessDimensionsCompositionStereociliary2008}%
  \BibitemOpen
  \bibfield  {author} {\bibinfo {author} {\bibfnamefont {D.~N.}\ \bibnamefont
  {Furness}}, \bibinfo {author} {\bibfnamefont {S.}~\bibnamefont
  {Mahendrasingam}}, \bibinfo {author} {\bibfnamefont {M.}~\bibnamefont
  {Ohashi}}, \bibinfo {author} {\bibfnamefont {R.}~\bibnamefont {Fettiplace}},\
  and\ \bibinfo {author} {\bibfnamefont {C.~M.}\ \bibnamefont {Hackney}},\
  }\bibfield  {title} {\bibinfo {title} {The {{Dimensions}} and {{Composition}}
  of {{Stereociliary Rootlets}} in {{Mammalian Cochlear Hair Cells}}:
  {{Comparison}} between {{High-}} and {{Low-Frequency Cells}} and {{Evidence}}
  for a {{Connection}} to the {{Lateral Membrane}}},\ }\href
  {https://doi.org/10.1523/JNEUROSCI.1154-08.2008} {\bibfield  {journal}
  {\bibinfo  {journal} {J Neurosci}\ }\textbf {\bibinfo {volume} {28}},\
  \bibinfo {pages} {6342} (\bibinfo {year} {2008})}\BibitemShut {NoStop}%
\bibitem [{\citenamefont {Schwander}\ \emph {et~al.}(2010)\citenamefont
  {Schwander}, \citenamefont {Kachar},\ and\ \citenamefont
  {M{\"u}ller}}]{schwanderCellBiologyHearing2010}%
  \BibitemOpen
  \bibfield  {author} {\bibinfo {author} {\bibfnamefont {M.}~\bibnamefont
  {Schwander}}, \bibinfo {author} {\bibfnamefont {B.}~\bibnamefont {Kachar}},\
  and\ \bibinfo {author} {\bibfnamefont {U.}~\bibnamefont {M{\"u}ller}},\
  }\bibfield  {title} {\bibinfo {title} {The cell biology of hearing},\ }\href
  {https://doi.org/10.1083/jcb.201001138} {\bibfield  {journal} {\bibinfo
  {journal} {J. Cell Biol.}\ }\textbf {\bibinfo {volume} {190}},\ \bibinfo
  {pages} {9} (\bibinfo {year} {2010})}\BibitemShut {NoStop}%
\bibitem [{\citenamefont {Hudspeth}(2019)}]{hudspethAcousticWavesBrain2019}%
  \BibitemOpen
  \bibfield  {author} {\bibinfo {author} {\bibfnamefont {A.~J.}\ \bibnamefont
  {Hudspeth}},\ }\href@noop {} {\bibinfo {title} {Acoustic waves to brain
  signals: {{Identifying}} the gating spring}} (\bibinfo {year}
  {2019})\BibitemShut {NoStop}%
\bibitem [{\citenamefont
  {Ohmori}(1985)}]{ohmoriMechanoElectricalTransduction1985}%
  \BibitemOpen
  \bibfield  {author} {\bibinfo {author} {\bibfnamefont {H.}~\bibnamefont
  {Ohmori}},\ }\bibfield  {title} {\bibinfo {title} {Mechano-electrical
  transduction currents in isolated vestibular hair cells of the chick.},\
  }\href {https://doi.org/10.1113/jphysiol.1985.sp015581} {\bibfield  {journal}
  {\bibinfo  {journal} {J. Physiol.}\ }\textbf {\bibinfo {volume} {359}},\
  \bibinfo {pages} {189} (\bibinfo {year} {1985})}\BibitemShut {NoStop}%
\bibitem [{\citenamefont {Eatock}\ \emph {et~al.}(1987)\citenamefont {Eatock},
  \citenamefont {Corey},\ and\ \citenamefont
  {Hudspeth}}]{eatockAdaptationMechanoelectricalTransduction1987}%
  \BibitemOpen
  \bibfield  {author} {\bibinfo {author} {\bibfnamefont {R.~A.}\ \bibnamefont
  {Eatock}}, \bibinfo {author} {\bibfnamefont {D.~P.}\ \bibnamefont {Corey}},\
  and\ \bibinfo {author} {\bibfnamefont {A.~J.}\ \bibnamefont {Hudspeth}},\
  }\bibfield  {title} {\bibinfo {title} {Adaptation of mechanoelectrical
  transduction in hair cells of the bullfrog's sacculus.},\ }\href
  {https://doi.org/10.1523/jneurosci.07-09-02821.1987} {\bibfield  {journal}
  {\bibinfo  {journal} {J. Neurosci.}\ }\textbf {\bibinfo {volume} {7}},\
  \bibinfo {pages} {2821} (\bibinfo {year} {1987})}\BibitemShut {NoStop}%
\bibitem [{\citenamefont {Howard}\ \emph {et~al.}(1988)\citenamefont {Howard},
  \citenamefont {Roberts},\ and\ \citenamefont
  {Hudspeth}}]{howardMechanoelectricalTransductionHair1988}%
  \BibitemOpen
  \bibfield  {author} {\bibinfo {author} {\bibfnamefont {J.}~\bibnamefont
  {Howard}}, \bibinfo {author} {\bibfnamefont {W.~M.}\ \bibnamefont
  {Roberts}},\ and\ \bibinfo {author} {\bibfnamefont {A.~J.}\ \bibnamefont
  {Hudspeth}},\ }\bibfield  {title} {\bibinfo {title} {Mechanoelectrical
  {{Transduction}} by {{Hair Cells}}},\ }\href
  {https://doi.org/10.1146/annurev.bb.17.060188.000531} {\bibfield  {journal}
  {\bibinfo  {journal} {Annu. Rev. Biophys. Bio.}\ }\textbf {\bibinfo {volume}
  {17}},\ \bibinfo {pages} {99} (\bibinfo {year} {1988})}\BibitemShut {NoStop}%
\bibitem [{\citenamefont {Walker}\ \emph {et~al.}(1993)\citenamefont {Walker},
  \citenamefont {Hudspeth},\ and\ \citenamefont
  {Gillespie}}]{walkerCalmodulinCalmodulinbindingProteins1993}%
  \BibitemOpen
  \bibfield  {author} {\bibinfo {author} {\bibfnamefont {R.~G.}\ \bibnamefont
  {Walker}}, \bibinfo {author} {\bibfnamefont {A.~J.}\ \bibnamefont
  {Hudspeth}},\ and\ \bibinfo {author} {\bibfnamefont {P.~G.}\ \bibnamefont
  {Gillespie}},\ }\bibfield  {title} {\bibinfo {title} {Calmodulin and
  calmodulin-binding proteins in hair bundles},\ }\href@noop {} {\bibfield
  {journal} {\bibinfo  {journal} {Proc. Natl. Acad. Sci.}\ }\textbf {\bibinfo
  {volume} {90}},\ \bibinfo {pages} {2807} (\bibinfo {year}
  {1993})}\BibitemShut {NoStop}%
\bibitem [{\citenamefont {Walker}\ and\ \citenamefont
  {Hudspeth}(1996)}]{walkerCalmodulinControlsAdaptation1996}%
  \BibitemOpen
  \bibfield  {author} {\bibinfo {author} {\bibfnamefont {R.~G.}\ \bibnamefont
  {Walker}}\ and\ \bibinfo {author} {\bibfnamefont {A.~J.}\ \bibnamefont
  {Hudspeth}},\ }\bibfield  {title} {\bibinfo {title} {Calmodulin controls
  adaptation of mechanoelectrical transduction by hair cells of the bullfrog's
  sacculus},\ }\href@noop {} {\bibfield  {journal} {\bibinfo  {journal} {Proc.
  Natl. Acad. Sci.}\ }\textbf {\bibinfo {volume} {93}},\ \bibinfo {pages}
  {2203} (\bibinfo {year} {1996})}\BibitemShut {NoStop}%
\bibitem [{\citenamefont {Cyr}\ \emph {et~al.}(2002)\citenamefont {Cyr},
  \citenamefont {Dumont},\ and\ \citenamefont
  {Gillespie}}]{cyrMyosin1cInteractsHairCell2002}%
  \BibitemOpen
  \bibfield  {author} {\bibinfo {author} {\bibfnamefont {J.~L.}\ \bibnamefont
  {Cyr}}, \bibinfo {author} {\bibfnamefont {R.~A.}\ \bibnamefont {Dumont}},\
  and\ \bibinfo {author} {\bibfnamefont {P.~G.}\ \bibnamefont {Gillespie}},\
  }\bibfield  {title} {\bibinfo {title} {Myosin-1c {{Interacts}} with
  {{Hair-Cell Receptors}} through {{Its Calmodulin-Binding IQ Domains}}},\
  }\href@noop {} {\bibfield  {journal} {\bibinfo  {journal} {J. Neurosci.}\
  }\textbf {\bibinfo {volume} {22}},\ \bibinfo {pages} {2487} (\bibinfo {year}
  {2002})}\BibitemShut {NoStop}%
\bibitem [{\citenamefont {Holt}\ \emph {et~al.}(2002)\citenamefont {Holt},
  \citenamefont {Gillespie}, \citenamefont {William}, \citenamefont {Shah},
  \citenamefont {Shokat}, \citenamefont {Corey}, \citenamefont {Mercer},\ and\
  \citenamefont {Gillespie}}]{holtChemicalGeneticStrategyImplicates2002}%
  \BibitemOpen
  \bibfield  {author} {\bibinfo {author} {\bibfnamefont {J.~R.}\ \bibnamefont
  {Holt}}, \bibinfo {author} {\bibfnamefont {S.~K.~H.}\ \bibnamefont
  {Gillespie}}, \bibinfo {author} {\bibfnamefont {D.}~\bibnamefont {William}},
  \bibinfo {author} {\bibfnamefont {K.}~\bibnamefont {Shah}}, \bibinfo {author}
  {\bibfnamefont {K.~M.}\ \bibnamefont {Shokat}}, \bibinfo {author}
  {\bibfnamefont {D.~P.}\ \bibnamefont {Corey}}, \bibinfo {author}
  {\bibfnamefont {J.~A.}\ \bibnamefont {Mercer}},\ and\ \bibinfo {author}
  {\bibfnamefont {P.~G.}\ \bibnamefont {Gillespie}},\ }\bibfield  {title}
  {\bibinfo {title} {A {{Chemical-Genetic Strategy Implicates Myosin-1c}} in
  {{Adaptation}} by {{Hair Cells}}},\ }\href@noop {} {\bibfield  {journal}
  {\bibinfo  {journal} {Cell}\ }\textbf {\bibinfo {volume} {108}},\ \bibinfo
  {pages} {371} (\bibinfo {year} {2002})}\BibitemShut {NoStop}%
\bibitem [{\citenamefont {Gillespie}\ and\ \citenamefont
  {Cyr}(2004)}]{gillespieMyosin1cHairCell2004}%
  \BibitemOpen
  \bibfield  {author} {\bibinfo {author} {\bibfnamefont {P.~G.}\ \bibnamefont
  {Gillespie}}\ and\ \bibinfo {author} {\bibfnamefont {J.~L.}\ \bibnamefont
  {Cyr}},\ }\bibfield  {title} {\bibinfo {title} {Myosin-1c, the hair cell's
  adaptation motor},\ }\href
  {https://doi.org/10.1146/annurev.physiol.66.032102.112842} {\bibfield
  {journal} {\bibinfo  {journal} {Annu. Rev. Physiol.}\ }\textbf {\bibinfo
  {volume} {66}},\ \bibinfo {pages} {521} (\bibinfo {year} {2004})}\BibitemShut
  {NoStop}%
\bibitem [{\citenamefont {Cheung}\ and\ \citenamefont
  {Corey}(2006)}]{cheungCa2ChangesForce2006}%
  \BibitemOpen
  \bibfield  {author} {\bibinfo {author} {\bibfnamefont {E.~L.}\ \bibnamefont
  {Cheung}}\ and\ \bibinfo {author} {\bibfnamefont {D.~P.}\ \bibnamefont
  {Corey}},\ }\bibfield  {title} {\bibinfo {title} {Ca2+ changes the force
  sensitivity of the hair-cell transduction channel},\ }\href
  {https://doi.org/10.1529/biophysj.105.061226} {\bibfield  {journal} {\bibinfo
   {journal} {Biophys. J.}\ }\textbf {\bibinfo {volume} {90}},\ \bibinfo
  {pages} {124} (\bibinfo {year} {2006})}\BibitemShut {NoStop}%
\bibitem [{\citenamefont {Zhang}\ \emph {et~al.}(2016)\citenamefont {Zhang},
  \citenamefont {He}, \citenamefont {Wong},\ and\ \citenamefont
  {Kindt}}]{zhangFunctionalCalciumImaging2016}%
  \BibitemOpen
  \bibfield  {author} {\bibinfo {author} {\bibfnamefont {Q.~X.}\ \bibnamefont
  {Zhang}}, \bibinfo {author} {\bibfnamefont {X.~J.}\ \bibnamefont {He}},
  \bibinfo {author} {\bibfnamefont {H.~C.}\ \bibnamefont {Wong}},\ and\
  \bibinfo {author} {\bibfnamefont {K.~S.}\ \bibnamefont {Kindt}},\ }\bibfield
  {title} {\bibinfo {title} {Functional calcium imaging in zebrafish
  lateral-line hair cells},\ }\href
  {https://doi.org/10.1016/bs.mcb.2015.12.002} {\bibfield  {journal} {\bibinfo
  {journal} {Method Cell Biol.}\ }\textbf {\bibinfo {volume} {133}},\ \bibinfo
  {pages} {229} (\bibinfo {year} {2016})}\BibitemShut {NoStop}%
\bibitem [{\citenamefont
  {Justice}(1979)}]{justiceAnalyticSignalProcessing1979}%
  \BibitemOpen
  \bibfield  {author} {\bibinfo {author} {\bibfnamefont {J.}~\bibnamefont
  {Justice}},\ }\bibfield  {title} {\bibinfo {title} {Analytic signal
  processing in music computation},\ }\href {https://doi.org/10/b342hx}
  {\bibfield  {journal} {\bibinfo  {journal} {IEEE T. Acoust. Speech}\ }\textbf
  {\bibinfo {volume} {27}},\ \bibinfo {pages} {670} (\bibinfo {year}
  {1979})}\BibitemShut {NoStop}%
\bibitem [{\citenamefont
  {Kolmogoroff}(1933)}]{kolmogoroffSullaDeterminazioneEmpirica1933}%
  \BibitemOpen
  \bibfield  {author} {\bibinfo {author} {\bibfnamefont {A.}~\bibnamefont
  {Kolmogoroff}},\ }\bibfield  {title} {\bibinfo {title} {Sulla determinazione
  empirica di una legge di distribuzione},\ }\href@noop {} {\bibfield
  {journal} {\bibinfo  {journal} {Giorn Dell'inst Ital Degli Att}\ }\textbf
  {\bibinfo {volume} {4}},\ \bibinfo {pages} {83} (\bibinfo {year}
  {1933})}\BibitemShut {NoStop}%
\bibitem [{\citenamefont
  {Smirnov}(1939)}]{smirnovEstimationDiscrepancyEmpirical1939}%
  \BibitemOpen
  \bibfield  {author} {\bibinfo {author} {\bibfnamefont {N.}~\bibnamefont
  {Smirnov}},\ }\bibfield  {title} {\bibinfo {title} {On the {{Estimation}} of
  {{Discrepancy}} between {{Empirical Curves}} of {{Distribution}} for {{Two
  Independent Samples}}},\ }\href@noop {} {\bibfield  {journal} {\bibinfo
  {journal} {Moscow Univ. Math. Bull.}\ }\textbf {\bibinfo {volume} {2}},\
  \bibinfo {pages} {3} (\bibinfo {year} {1939})}\BibitemShut {NoStop}%
\bibitem [{\citenamefont
  {Shiryayev}(1992)}]{shiryayevEmpiricalDeterminationDistribution1992}%
  \BibitemOpen
  \bibfield  {author} {\bibinfo {author} {\bibfnamefont {A.~N.}\ \bibnamefont
  {Shiryayev}},\ }\bibfield  {title} {\bibinfo {title} {On {{The Empirical
  Determination}} of {{A Distribution Law}}},\ }in\ \href@noop {} {\emph
  {\bibinfo {booktitle} {Selected {{Works}} of {{A}}. {{N}}. {{Kolmogorov}}:
  {{Volume II Probability Theory}} and {{Mathematical Statistics}}}}},\
  \bibinfo {series} {Mathematics and {{Its Applications}} ({{Soviet Series}})},
  Vol.~\bibinfo {volume} {2}\ (\bibinfo  {publisher} {{Springer Netherlands}},\
  \bibinfo {address} {{Dordrecht}},\ \bibinfo {year} {1992})\ pp.\ \bibinfo
  {pages} {139--146}\BibitemShut {NoStop}%
\bibitem [{\citenamefont {Sammut}\ and\ \citenamefont
  {Webb}(2017)}]{sammutEncyclopediaMachineLearning2017}%
  \BibitemOpen
  \bibinfo {editor} {\bibfnamefont {C.}~\bibnamefont {Sammut}}\ and\ \bibinfo
  {editor} {\bibfnamefont {G.~I.}\ \bibnamefont {Webb}},\ eds.,\ \href@noop {}
  {\emph {\bibinfo {title} {Encyclopedia of {{Machine Learning}} and {{Data
  Mining}}}}}\ (\bibinfo {year} {2017})\BibitemShut {NoStop}%
\bibitem [{\citenamefont
  {Akaike}(1973)}]{akaikeInformationTheoryExtension1973}%
  \BibitemOpen
  \bibfield  {author} {\bibinfo {author} {\bibfnamefont {H.}~\bibnamefont
  {Akaike}},\ }\bibfield  {title} {\bibinfo {title} {Information {{Theory}} and
  an {{Extension}} of the {{Maximum Likelihood Principle}}},\ }in\ \href@noop
  {} {\emph {\bibinfo {booktitle} {Proceeding of the {{Second International
  Symposium}} on {{Information Theory}}}}},\ \bibinfo {editor} {edited by\
  \bibinfo {editor} {\bibfnamefont {B.~N.}\ \bibnamefont {Petrov}}\ and\
  \bibinfo {editor} {\bibfnamefont {F.}~\bibnamefont {Caski}}}\ (\bibinfo
  {publisher} {{Akademiai Kiado}},\ \bibinfo {address} {{Budapest}},\ \bibinfo
  {year} {1973})\ pp.\ \bibinfo {pages} {267--281}\BibitemShut {NoStop}%
\bibitem [{\citenamefont
  {Schwarz}(1978)}]{schwarzEstimatingDimensionModel1978}%
  \BibitemOpen
  \bibfield  {author} {\bibinfo {author} {\bibfnamefont {G.}~\bibnamefont
  {Schwarz}},\ }\bibfield  {title} {\bibinfo {title} {Estimating the
  {{Dimension}} of a {{Model}}},\ }\href
  {https://doi.org/10.1214/aos/1176344136} {\bibfield  {journal} {\bibinfo
  {journal} {Ann. Stat.}\ }\textbf {\bibinfo {volume} {6}},\ \bibinfo {pages}
  {461} (\bibinfo {year} {1978})}\BibitemShut {NoStop}%
\bibitem [{\citenamefont {Burnham}\ and\ \citenamefont
  {Anderson}(2004)}]{burnhamMultimodelInferenceUnderstanding2004}%
  \BibitemOpen
  \bibfield  {author} {\bibinfo {author} {\bibfnamefont {K.~P.}\ \bibnamefont
  {Burnham}}\ and\ \bibinfo {author} {\bibfnamefont {D.~R.}\ \bibnamefont
  {Anderson}},\ }\bibfield  {title} {\bibinfo {title} {Multimodel
  {{Inference}}: {{Understanding AIC}} and {{BIC}} in {{Model Selection}}},\
  }\href {https://doi.org/10.1177/0049124104268644} {\bibfield  {journal}
  {\bibinfo  {journal} {Sociol. Method Res.}\ }\textbf {\bibinfo {volume}
  {33}},\ \bibinfo {pages} {261} (\bibinfo {year} {2004})}\BibitemShut
  {NoStop}%
\bibitem [{\citenamefont {Kuha}(2004)}]{kuhaAICBICComparisons2004}%
  \BibitemOpen
  \bibfield  {author} {\bibinfo {author} {\bibfnamefont {J.}~\bibnamefont
  {Kuha}},\ }\bibfield  {title} {\bibinfo {title} {{{AIC}} and {{BIC}}:
  {{Comparisons}} of {{Assumptions}} and {{Performance}}},\ }\href
  {https://doi.org/10.1177/0049124103262065} {\bibfield  {journal} {\bibinfo
  {journal} {Sociol. Method Res.}\ }\textbf {\bibinfo {volume} {33}},\ \bibinfo
  {pages} {188} (\bibinfo {year} {2004})}\BibitemShut {NoStop}%
\bibitem [{\citenamefont {Brewer}\ \emph {et~al.}(2016)\citenamefont {Brewer},
  \citenamefont {Butler},\ and\ \citenamefont
  {Cooksley}}]{brewerRelativePerformanceAIC2016}%
  \BibitemOpen
  \bibfield  {author} {\bibinfo {author} {\bibfnamefont {M.~J.}\ \bibnamefont
  {Brewer}}, \bibinfo {author} {\bibfnamefont {A.}~\bibnamefont {Butler}},\
  and\ \bibinfo {author} {\bibfnamefont {S.~L.}\ \bibnamefont {Cooksley}},\
  }\bibfield  {title} {\bibinfo {title} {The relative performance of {{AIC}},
  {{AICC}} and {{BIC}} in the presence of unobserved heterogeneity},\ }\href
  {https://doi.org/10.1111/2041-210X.12541} {\bibfield  {journal} {\bibinfo
  {journal} {Methods Ecol. Evol.}\ }\textbf {\bibinfo {volume} {7}},\ \bibinfo
  {pages} {679} (\bibinfo {year} {2016})}\BibitemShut {NoStop}%
\bibitem [{\citenamefont
  {Liddle}(2007)}]{liddleInformationCriteriaAstrophysical2007}%
  \BibitemOpen
  \bibfield  {author} {\bibinfo {author} {\bibfnamefont {A.~R.}\ \bibnamefont
  {Liddle}},\ }\bibfield  {title} {\bibinfo {title} {Information criteria for
  astrophysical model selection},\ }\href
  {https://doi.org/10.1111/j.1745-3933.2007.00306.x} {\bibfield  {journal}
  {\bibinfo  {journal} {Mon. Not. R. Astron. Soc. Lett.}\ }\textbf {\bibinfo
  {volume} {377}},\ \bibinfo {pages} {L74} (\bibinfo {year}
  {2007})}\BibitemShut {NoStop}%
\bibitem [{\citenamefont {Do}\ \emph {et~al.}(2019)\citenamefont {Do},
  \citenamefont {Hees}, \citenamefont {Ghez}, \citenamefont {Martinez},
  \citenamefont {Chu}, \citenamefont {Jia}, \citenamefont {Sakai},
  \citenamefont {Lu}, \citenamefont {Gautam}, \citenamefont {O'Neil},
  \citenamefont {Becklin}, \citenamefont {Morris}, \citenamefont {Matthews},
  \citenamefont {Nishiyama}, \citenamefont {Campbell}, \citenamefont
  {Chappell}, \citenamefont {Chen}, \citenamefont {Ciurlo}, \citenamefont
  {Dehghanfar}, \citenamefont {{Gallego-Cano}}, \citenamefont {Kerzendorf},
  \citenamefont {Lyke}, \citenamefont {Naoz}, \citenamefont {Saida},
  \citenamefont {Sch{\"o}del}, \citenamefont {Takahashi}, \citenamefont
  {Takamori}, \citenamefont {Witzel},\ and\ \citenamefont
  {Wizinowich}}]{doRelativisticRedshiftStar2019}%
  \BibitemOpen
  \bibfield  {author} {\bibinfo {author} {\bibfnamefont {T.}~\bibnamefont
  {Do}}, \bibinfo {author} {\bibfnamefont {A.}~\bibnamefont {Hees}}, \bibinfo
  {author} {\bibfnamefont {A.}~\bibnamefont {Ghez}}, \bibinfo {author}
  {\bibfnamefont {G.~D.}\ \bibnamefont {Martinez}}, \bibinfo {author}
  {\bibfnamefont {D.~S.}\ \bibnamefont {Chu}}, \bibinfo {author} {\bibfnamefont
  {S.}~\bibnamefont {Jia}}, \bibinfo {author} {\bibfnamefont {S.}~\bibnamefont
  {Sakai}}, \bibinfo {author} {\bibfnamefont {J.~R.}\ \bibnamefont {Lu}},
  \bibinfo {author} {\bibfnamefont {A.~K.}\ \bibnamefont {Gautam}}, \bibinfo
  {author} {\bibfnamefont {K.~K.}\ \bibnamefont {O'Neil}}, \bibinfo {author}
  {\bibfnamefont {E.~E.}\ \bibnamefont {Becklin}}, \bibinfo {author}
  {\bibfnamefont {M.~R.}\ \bibnamefont {Morris}}, \bibinfo {author}
  {\bibfnamefont {K.}~\bibnamefont {Matthews}}, \bibinfo {author}
  {\bibfnamefont {S.}~\bibnamefont {Nishiyama}}, \bibinfo {author}
  {\bibfnamefont {R.}~\bibnamefont {Campbell}}, \bibinfo {author}
  {\bibfnamefont {S.}~\bibnamefont {Chappell}}, \bibinfo {author}
  {\bibfnamefont {Z.}~\bibnamefont {Chen}}, \bibinfo {author} {\bibfnamefont
  {A.}~\bibnamefont {Ciurlo}}, \bibinfo {author} {\bibfnamefont
  {A.}~\bibnamefont {Dehghanfar}}, \bibinfo {author} {\bibfnamefont
  {E.}~\bibnamefont {{Gallego-Cano}}}, \bibinfo {author} {\bibfnamefont
  {W.~E.}\ \bibnamefont {Kerzendorf}}, \bibinfo {author} {\bibfnamefont
  {J.~E.}\ \bibnamefont {Lyke}}, \bibinfo {author} {\bibfnamefont
  {S.}~\bibnamefont {Naoz}}, \bibinfo {author} {\bibfnamefont {H.}~\bibnamefont
  {Saida}}, \bibinfo {author} {\bibfnamefont {R.}~\bibnamefont {Sch{\"o}del}},
  \bibinfo {author} {\bibfnamefont {M.}~\bibnamefont {Takahashi}}, \bibinfo
  {author} {\bibfnamefont {Y.}~\bibnamefont {Takamori}}, \bibinfo {author}
  {\bibfnamefont {G.}~\bibnamefont {Witzel}},\ and\ \bibinfo {author}
  {\bibfnamefont {P.}~\bibnamefont {Wizinowich}},\ }\bibfield  {title}
  {\bibinfo {title} {Relativistic redshift of the star {{S0-2}} orbiting the
  {{Galactic Center}} supermassive black hole},\ }\href
  {https://doi.org/10.1126/science.aav8137} {\bibfield  {journal} {\bibinfo
  {journal} {Science}\ }\textbf {\bibinfo {volume} {365}},\ \bibinfo {pages}
  {664} (\bibinfo {year} {2019})}\BibitemShut {NoStop}%
\bibitem [{\citenamefont {Johnson}\ and\ \citenamefont
  {Omland}(2004)}]{johnsonModelSelectionEcology2004}%
  \BibitemOpen
  \bibfield  {author} {\bibinfo {author} {\bibfnamefont {J.~B.}\ \bibnamefont
  {Johnson}}\ and\ \bibinfo {author} {\bibfnamefont {K.~S.}\ \bibnamefont
  {Omland}},\ }\bibfield  {title} {\bibinfo {title} {Model selection in ecology
  and evolution},\ }\href {https://doi.org/10.1016/j.tree.2003.10.013}
  {\bibfield  {journal} {\bibinfo  {journal} {Trends in Ecology \& Evolution}\
  }\textbf {\bibinfo {volume} {19}},\ \bibinfo {pages} {101} (\bibinfo {year}
  {2004})}\BibitemShut {NoStop}%
\bibitem [{\citenamefont {Aho}\ \emph {et~al.}(2014)\citenamefont {Aho},
  \citenamefont {Derryberry},\ and\ \citenamefont
  {Peterson}}]{ahoModelSelectionEcologists2014}%
  \BibitemOpen
  \bibfield  {author} {\bibinfo {author} {\bibfnamefont {K.}~\bibnamefont
  {Aho}}, \bibinfo {author} {\bibfnamefont {D.}~\bibnamefont {Derryberry}},\
  and\ \bibinfo {author} {\bibfnamefont {T.}~\bibnamefont {Peterson}},\
  }\bibfield  {title} {\bibinfo {title} {Model selection for ecologists: The
  worldviews of {{AIC}} and {{BIC}}},\ }\href@noop {} {\bibfield  {journal}
  {\bibinfo  {journal} {Ecology}\ }\textbf {\bibinfo {volume} {95}},\ \bibinfo
  {pages} {631} (\bibinfo {year} {2014})},\ \Eprint
  {https://arxiv.org/abs/43495189} {43495189} \BibitemShut {NoStop}%
\bibitem [{\citenamefont {Albahli}\ and\ \citenamefont
  {Nabi}(2021)}]{albahliDefectPredictionUsing2021}%
  \BibitemOpen
  \bibfield  {author} {\bibinfo {author} {\bibfnamefont {S.}~\bibnamefont
  {Albahli}}\ and\ \bibinfo {author} {\bibfnamefont {G.}~\bibnamefont {Nabi}},\
  }\bibfield  {title} {\bibinfo {title} {Defect {{Prediction Using Akaike}} and
  {{Bayesian Information Criterion}}},\ }\href
  {https://doi.org/10.32604/csse.2022.021750} {\bibfield  {journal} {\bibinfo
  {journal} {CSSE}\ }\textbf {\bibinfo {volume} {41}},\ \bibinfo {pages} {1117}
  (\bibinfo {year} {2021})}\BibitemShut {NoStop}%
\bibitem [{\citenamefont {Tran}\ \emph {et~al.}(2021)\citenamefont {Tran},
  \citenamefont {Tsujimura}, \citenamefont {Ha}, \citenamefont {Nguyen},
  \citenamefont {Binh}, \citenamefont {Dang}, \citenamefont {Doan},
  \citenamefont {Bui}, \citenamefont {Anh~Ngoc}, \citenamefont {Phu},
  \citenamefont {Thuc},\ and\ \citenamefont
  {Pham}}]{tranEvaluatingPredictivePower2021}%
  \BibitemOpen
  \bibfield  {author} {\bibinfo {author} {\bibfnamefont {D.~A.}\ \bibnamefont
  {Tran}}, \bibinfo {author} {\bibfnamefont {M.}~\bibnamefont {Tsujimura}},
  \bibinfo {author} {\bibfnamefont {N.~T.}\ \bibnamefont {Ha}}, \bibinfo
  {author} {\bibfnamefont {V.~T.}\ \bibnamefont {Nguyen}}, \bibinfo {author}
  {\bibfnamefont {D.~V.}\ \bibnamefont {Binh}}, \bibinfo {author}
  {\bibfnamefont {T.~D.}\ \bibnamefont {Dang}}, \bibinfo {author}
  {\bibfnamefont {Q.-V.}\ \bibnamefont {Doan}}, \bibinfo {author}
  {\bibfnamefont {D.~T.}\ \bibnamefont {Bui}}, \bibinfo {author} {\bibfnamefont
  {T.}~\bibnamefont {Anh~Ngoc}}, \bibinfo {author} {\bibfnamefont {L.~V.}\
  \bibnamefont {Phu}}, \bibinfo {author} {\bibfnamefont {P.~T.~B.}\
  \bibnamefont {Thuc}},\ and\ \bibinfo {author} {\bibfnamefont {T.~D.}\
  \bibnamefont {Pham}},\ }\bibfield  {title} {\bibinfo {title} {Evaluating the
  predictive power of different machine learning algorithms for groundwater
  salinity prediction of multi-layer coastal aquifers in the {{Mekong Delta}},
  {{Vietnam}}},\ }\href {https://doi.org/10.1016/j.ecolind.2021.107790}
  {\bibfield  {journal} {\bibinfo  {journal} {Ecol. Indic.}\ }\textbf {\bibinfo
  {volume} {127}},\ \bibinfo {pages} {107790} (\bibinfo {year}
  {2021})}\BibitemShut {NoStop}%
\bibitem [{\citenamefont {Nevill}\ \emph {et~al.}(2005)\citenamefont {Nevill},
  \citenamefont {Bate},\ and\ \citenamefont
  {Holder}}]{nevillModelingPhysiologicalAnthropometric2005}%
  \BibitemOpen
  \bibfield  {author} {\bibinfo {author} {\bibfnamefont {A.~M.}\ \bibnamefont
  {Nevill}}, \bibinfo {author} {\bibfnamefont {S.}~\bibnamefont {Bate}},\ and\
  \bibinfo {author} {\bibfnamefont {R.~L.}\ \bibnamefont {Holder}},\ }\bibfield
   {title} {\bibinfo {title} {Modeling {{Physiological}} and {{Anthropometric
  Variables Known}} to {{Vary}} with {{Body Size}} and {{Other Confounding
  Variables}}},\ }\href {https://doi.org/10.1002/ajpa.20356} {\bibfield
  {journal} {\bibinfo  {journal} {Am. J. Phys. Anthropol.}\ }\textbf {\bibinfo
  {volume} {128}},\ \bibinfo {pages} {141} (\bibinfo {year}
  {2005})}\BibitemShut {NoStop}%
\bibitem [{\citenamefont {Gl{\o}ersen}\ \emph {et~al.}(2022)\citenamefont
  {Gl{\o}ersen}, \citenamefont {Colosio}, \citenamefont {Boone}, \citenamefont
  {Dysthe}, \citenamefont {{Malthe-S{\o}renssen}}, \citenamefont {Capelli},\
  and\ \citenamefont {Pogliaghi}}]{gloersenModelingVo2Onkinetics2022}%
  \BibitemOpen
  \bibfield  {author} {\bibinfo {author} {\bibfnamefont {{\O}.}~\bibnamefont
  {Gl{\o}ersen}}, \bibinfo {author} {\bibfnamefont {A.~L.}\ \bibnamefont
  {Colosio}}, \bibinfo {author} {\bibfnamefont {J.}~\bibnamefont {Boone}},
  \bibinfo {author} {\bibfnamefont {D.~K.}\ \bibnamefont {Dysthe}}, \bibinfo
  {author} {\bibfnamefont {A.}~\bibnamefont {{Malthe-S{\o}renssen}}}, \bibinfo
  {author} {\bibfnamefont {C.}~\bibnamefont {Capelli}},\ and\ \bibinfo {author}
  {\bibfnamefont {S.}~\bibnamefont {Pogliaghi}},\ }\bibfield  {title} {\bibinfo
  {title} {Modeling {{{\.V}o2}} on-kinetics based on intensity-dependent
  delayed adjustment and loss of efficiency ({{DALE}})},\ }\href
  {https://doi.org/10.1152/japplphysiol.00570.2021} {\bibfield  {journal}
  {\bibinfo  {journal} {J. Appl. Psychol.}\ }\textbf {\bibinfo {volume}
  {132}},\ \bibinfo {pages} {1480} (\bibinfo {year} {2022})}\BibitemShut
  {NoStop}%
\bibitem [{\citenamefont {Jafari}\ \emph {et~al.}(2023)\citenamefont {Jafari},
  \citenamefont {Lai},\ and\ \citenamefont
  {Yanushkevich}}]{jafariMVARCausalModeling2023}%
  \BibitemOpen
  \bibfield  {author} {\bibinfo {author} {\bibfnamefont {B.}~\bibnamefont
  {Jafari}}, \bibinfo {author} {\bibfnamefont {K.}~\bibnamefont {Lai}},\ and\
  \bibinfo {author} {\bibfnamefont {S.}~\bibnamefont {Yanushkevich}},\
  }\bibfield  {title} {\bibinfo {title} {{{MVAR}} and {{Causal Modeling}} of
  {{Relationship}} between {{Physiological Signals}} and {{Affective
  States}}},\ }in\ \href {https://doi.org/10.1109/CAI54212.2023.00065} {\emph
  {\bibinfo {booktitle} {2023 {{IEEE Conference}} on {{Artificial
  Intelligence}} ({{CAI}})}}}\ (\bibinfo {year} {2023})\ pp.\ \bibinfo {pages}
  {134--135}\BibitemShut {NoStop}%
\bibitem [{\citenamefont {Hartman}\ and\ \citenamefont
  {Groendyke}(2013)}]{hartmanModelSelectionAveraging2013}%
  \BibitemOpen
  \bibfield  {author} {\bibinfo {author} {\bibfnamefont {B.~M.}\ \bibnamefont
  {Hartman}}\ and\ \bibinfo {author} {\bibfnamefont {C.}~\bibnamefont
  {Groendyke}},\ }\bibfield  {title} {\bibinfo {title} {Model {{Selection}} and
  {{Averaging}} in {{Financial Risk Management}}},\ }\href
  {https://doi.org/10.1080/10920277.2013.824374} {\bibfield  {journal}
  {\bibinfo  {journal} {N. Am. Actuar. J.}\ }\textbf {\bibinfo {volume} {17}},\
  \bibinfo {pages} {216} (\bibinfo {year} {2013})}\BibitemShut {NoStop}%
\bibitem [{\citenamefont {Punzo}\ and\ \citenamefont
  {Bagnato}(2021)}]{punzoMultivariateTailinflatedNormal2021}%
  \BibitemOpen
  \bibfield  {author} {\bibinfo {author} {\bibfnamefont {A.}~\bibnamefont
  {Punzo}}\ and\ \bibinfo {author} {\bibfnamefont {L.}~\bibnamefont
  {Bagnato}},\ }\bibfield  {title} {\bibinfo {title} {The multivariate
  tail-inflated normal distribution and its application in finance},\ }\href
  {https://doi.org/10.1080/00949655.2020.1805451} {\bibfield  {journal}
  {\bibinfo  {journal} {J. Stat. Comput. Sim.}\ }\textbf {\bibinfo {volume}
  {91}},\ \bibinfo {pages} {1} (\bibinfo {year} {2021})}\BibitemShut {NoStop}%
\bibitem [{\citenamefont {Shahzad}\ \emph {et~al.}(2022)\citenamefont
  {Shahzad}, \citenamefont {Luo}, \citenamefont {Liu}, \citenamefont {Faisal},\
  and\ \citenamefont {Ullah}}]{shahzadMostConsistentReliable2022}%
  \BibitemOpen
  \bibfield  {author} {\bibinfo {author} {\bibfnamefont {U.}~\bibnamefont
  {Shahzad}}, \bibinfo {author} {\bibfnamefont {F.}~\bibnamefont {Luo}},
  \bibinfo {author} {\bibfnamefont {J.}~\bibnamefont {Liu}}, \bibinfo {author}
  {\bibfnamefont {M.}~\bibnamefont {Faisal}},\ and\ \bibinfo {author}
  {\bibfnamefont {H.}~\bibnamefont {Ullah}},\ }\bibfield  {title} {\bibinfo
  {title} {The most consistent and reliable predictors of corporate financial
  choices in {{Pakistan}}: {{New}} evidence using {{BIC}} estimation},\ }\href
  {https://doi.org/10.1002/ijfe.2149} {\bibfield  {journal} {\bibinfo
  {journal} {Int. J. Financ. Econ.}\ }\textbf {\bibinfo {volume} {27}},\
  \bibinfo {pages} {237} (\bibinfo {year} {2022})}\BibitemShut {NoStop}%
\bibitem [{\citenamefont {Hossain}\ \emph {et~al.}(2020)\citenamefont
  {Hossain}, \citenamefont {Moon},\ and\ \citenamefont
  {Chon}}]{hossainEstimationARMAModel2020}%
  \BibitemOpen
  \bibfield  {author} {\bibinfo {author} {\bibfnamefont {M.-B.}\ \bibnamefont
  {Hossain}}, \bibinfo {author} {\bibfnamefont {J.}~\bibnamefont {Moon}},\ and\
  \bibinfo {author} {\bibfnamefont {K.~H.}\ \bibnamefont {Chon}},\ }\bibfield
  {title} {\bibinfo {title} {Estimation of {{ARMA Model Order}} via
  {{Artificial Neural Network}} for {{Modeling Physiological Systems}}},\
  }\href {https://doi.org/10.1109/ACCESS.2020.3029756} {\bibfield  {journal}
  {\bibinfo  {journal} {IEEE Access}\ }\textbf {\bibinfo {volume} {8}},\
  \bibinfo {pages} {186813} (\bibinfo {year} {2020})}\BibitemShut {NoStop}%
\bibitem [{\citenamefont {Nguyen}\ \emph {et~al.}(2021)\citenamefont {Nguyen},
  \citenamefont {Ha}, \citenamefont {{Nguyen-Ngoc}},\ and\ \citenamefont
  {Pham}}]{nguyenComparingPerformanceMachine2021}%
  \BibitemOpen
  \bibfield  {author} {\bibinfo {author} {\bibfnamefont {H.~Q.}\ \bibnamefont
  {Nguyen}}, \bibinfo {author} {\bibfnamefont {N.~T.}\ \bibnamefont {Ha}},
  \bibinfo {author} {\bibfnamefont {L.}~\bibnamefont {{Nguyen-Ngoc}}},\ and\
  \bibinfo {author} {\bibfnamefont {T.~L.}\ \bibnamefont {Pham}},\ }\bibfield
  {title} {\bibinfo {title} {Comparing the performance of machine learning
  algorithms for remote and in situ estimations of chlorophyll-a content: {{A}}
  case study in the {{Tri An Reservoir}}, {{Vietnam}}},\ }\href
  {https://doi.org/10.1002/wer.1643} {\bibfield  {journal} {\bibinfo  {journal}
  {Water Environ. Res.}\ }\textbf {\bibinfo {volume} {93}},\ \bibinfo {pages}
  {2941} (\bibinfo {year} {2021})}\BibitemShut {NoStop}%
\bibitem [{\citenamefont {Roongthumskul}\ \emph {et~al.}(2011)\citenamefont
  {Roongthumskul}, \citenamefont {{Fredrickson-Hemsing}}, \citenamefont {Kao},\
  and\ \citenamefont
  {Bozovic}}]{roongthumskulMultipletimescaleDynamicsUnderlying2011}%
  \BibitemOpen
  \bibfield  {author} {\bibinfo {author} {\bibfnamefont {Y.}~\bibnamefont
  {Roongthumskul}}, \bibinfo {author} {\bibfnamefont {L.}~\bibnamefont
  {{Fredrickson-Hemsing}}}, \bibinfo {author} {\bibfnamefont {A.}~\bibnamefont
  {Kao}},\ and\ \bibinfo {author} {\bibfnamefont {D.}~\bibnamefont {Bozovic}},\
  }\bibfield  {title} {\bibinfo {title} {Multiple-timescale dynamics underlying
  spontaneous oscillations of saccular hair bundles},\ }\href
  {https://doi.org/10.1016/j.bpj.2011.06.027} {\bibfield  {journal} {\bibinfo
  {journal} {Biophys. J.}\ }\textbf {\bibinfo {volume} {101}},\ \bibinfo
  {pages} {603} (\bibinfo {year} {2011})}\BibitemShut {NoStop}%
\bibitem [{\citenamefont {Bormuth}\ \emph {et~al.}(2014)\citenamefont
  {Bormuth}, \citenamefont {Barral}, \citenamefont {Joanny}, \citenamefont
  {J{\"u}licher},\ and\ \citenamefont
  {Martin}}]{bormuthTransductionChannelsGating2014}%
  \BibitemOpen
  \bibfield  {author} {\bibinfo {author} {\bibfnamefont {V.}~\bibnamefont
  {Bormuth}}, \bibinfo {author} {\bibfnamefont {J.}~\bibnamefont {Barral}},
  \bibinfo {author} {\bibfnamefont {J.~F.}\ \bibnamefont {Joanny}}, \bibinfo
  {author} {\bibfnamefont {F.}~\bibnamefont {J{\"u}licher}},\ and\ \bibinfo
  {author} {\bibfnamefont {P.}~\bibnamefont {Martin}},\ }\bibfield  {title}
  {\bibinfo {title} {Transduction channels' gating can control friction on
  vibrating hair-cell bundles in the ear},\ }\href
  {https://doi.org/10.1073/pnas.1402556111} {\bibfield  {journal} {\bibinfo
  {journal} {Proc. Natl. Acad. Sci.}\ }\textbf {\bibinfo {volume} {111}},\
  \bibinfo {pages} {7185} (\bibinfo {year} {2014})}\BibitemShut {NoStop}%
\bibitem [{\citenamefont {Barral}\ \emph {et~al.}(2018)\citenamefont {Barral},
  \citenamefont {J{\"u}licher},\ and\ \citenamefont
  {Martin}}]{barralFrictionTransductionChannels2018}%
  \BibitemOpen
  \bibfield  {author} {\bibinfo {author} {\bibfnamefont {J.}~\bibnamefont
  {Barral}}, \bibinfo {author} {\bibfnamefont {F.}~\bibnamefont
  {J{\"u}licher}},\ and\ \bibinfo {author} {\bibfnamefont {P.}~\bibnamefont
  {Martin}},\ }\bibfield  {title} {\bibinfo {title} {Friction from
  {{Transduction Channels}}' {{Gating Affects Spontaneous Hair-Bundle
  Oscillations}}},\ }\href {https://doi.org/10.1016/j.bpj.2017.11.019}
  {\bibfield  {journal} {\bibinfo  {journal} {Biophys. J.}\ }\textbf {\bibinfo
  {volume} {114}},\ \bibinfo {pages} {425} (\bibinfo {year}
  {2018})}\BibitemShut {NoStop}%
\bibitem [{\citenamefont {Benser}\ \emph {et~al.}(1996)\citenamefont {Benser},
  \citenamefont {Marquis},\ and\ \citenamefont
  {Hudspeth}}]{benserRapidActiveHair1996}%
  \BibitemOpen
  \bibfield  {author} {\bibinfo {author} {\bibfnamefont {M.~E.}\ \bibnamefont
  {Benser}}, \bibinfo {author} {\bibfnamefont {R.~E.}\ \bibnamefont
  {Marquis}},\ and\ \bibinfo {author} {\bibfnamefont {A.~J.}\ \bibnamefont
  {Hudspeth}},\ }\bibfield  {title} {\bibinfo {title} {Rapid, active hair
  bundle movements in hair cells from the bullfrog's sacculus},\ }\href
  {https://doi.org/10.1523/jneurosci.16-18-05629.1996} {\bibfield  {journal}
  {\bibinfo  {journal} {J. Neurosci.}\ }\textbf {\bibinfo {volume} {16}},\
  \bibinfo {pages} {5629} (\bibinfo {year} {1996})}\BibitemShut {NoStop}%
\bibitem [{\citenamefont {{\'O}~Maoil{\'e}idigh}\ \emph
  {et~al.}(2012)\citenamefont {{\'O}~Maoil{\'e}idigh}, \citenamefont {Nicola},\
  and\ \citenamefont {Hudspeth}}]{omaoileidighDiverseEffectsMechanical2012}%
  \BibitemOpen
  \bibfield  {author} {\bibinfo {author} {\bibfnamefont {D.}~\bibnamefont
  {{\'O}~Maoil{\'e}idigh}}, \bibinfo {author} {\bibfnamefont {E.~M.}\
  \bibnamefont {Nicola}},\ and\ \bibinfo {author} {\bibfnamefont {A.~J.}\
  \bibnamefont {Hudspeth}},\ }\bibfield  {title} {\bibinfo {title} {The diverse
  effects of mechanical loading on active hair bundles},\ }\href
  {https://doi.org/10.1073/pnas.1120298109} {\bibfield  {journal} {\bibinfo
  {journal} {Proc. Natl. Acad. Sci.}\ }\textbf {\bibinfo {volume} {109}},\
  \bibinfo {pages} {1943} (\bibinfo {year} {2012})}\BibitemShut {NoStop}%
\bibitem [{\citenamefont {Crawford}\ and\ \citenamefont
  {Fettiplace}(1981)}]{crawfordElectricalTuningMechanism1981}%
  \BibitemOpen
  \bibfield  {author} {\bibinfo {author} {\bibfnamefont {A.~C.}\ \bibnamefont
  {Crawford}}\ and\ \bibinfo {author} {\bibfnamefont {R.}~\bibnamefont
  {Fettiplace}},\ }\bibfield  {title} {\bibinfo {title} {An electrical tuning
  mechanism in turtle cochlear hair cells},\ }\href
  {https://doi.org/10.1113/jphysiol.1981.sp013634} {\bibfield  {journal}
  {\bibinfo  {journal} {J. Physiol.}\ }\textbf {\bibinfo {volume} {312}},\
  \bibinfo {pages} {377} (\bibinfo {year} {1981})}\BibitemShut {NoStop}%
\bibitem [{\citenamefont {Ricci}\ \emph {et~al.}(2000)\citenamefont {Ricci},
  \citenamefont {Crawford},\ and\ \citenamefont
  {Fettiplace}}]{ricciActiveHairBundle2000}%
  \BibitemOpen
  \bibfield  {author} {\bibinfo {author} {\bibfnamefont {A.~J.}\ \bibnamefont
  {Ricci}}, \bibinfo {author} {\bibfnamefont {A.~C.}\ \bibnamefont
  {Crawford}},\ and\ \bibinfo {author} {\bibfnamefont {R.}~\bibnamefont
  {Fettiplace}},\ }\bibfield  {title} {\bibinfo {title} {Active hair bundle
  motion linked to fast transducer adaptation in auditory hair cells},\ }\href
  {https://doi.org/10.1523/jneurosci.20-19-07131.2000} {\bibfield  {journal}
  {\bibinfo  {journal} {J. Neurosci.}\ }\textbf {\bibinfo {volume} {20}},\
  \bibinfo {pages} {7131} (\bibinfo {year} {2000})}\BibitemShut {NoStop}%
\bibitem [{\citenamefont {Sheth}\ \emph {et~al.}(2019)\citenamefont {Sheth},
  \citenamefont {Bozovic},\ and\ \citenamefont
  {Levine}}]{shethNoiseinducedDistortionMean2019}%
  \BibitemOpen
  \bibfield  {author} {\bibinfo {author} {\bibfnamefont {J.}~\bibnamefont
  {Sheth}}, \bibinfo {author} {\bibfnamefont {D.}~\bibnamefont {Bozovic}},\
  and\ \bibinfo {author} {\bibfnamefont {A.~J.}\ \bibnamefont {Levine}},\
  }\bibfield  {title} {\bibinfo {title} {Noise-induced distortion of the mean
  limit cycle of nonlinear oscillators},\ }\href
  {https://doi.org/10.1103/PhysRevE.99.062124} {\bibfield  {journal} {\bibinfo
  {journal} {Phys. Rev. E}\ }\textbf {\bibinfo {volume} {99}},\ \bibinfo
  {pages} {062124} (\bibinfo {year} {2019})}\BibitemShut {NoStop}%
\bibitem [{\citenamefont {Hudspeth}\ and\ \citenamefont
  {Gillespie}(1994)}]{hudspethPullingSpringsTune1994}%
  \BibitemOpen
  \bibfield  {author} {\bibinfo {author} {\bibfnamefont {A.~J.}\ \bibnamefont
  {Hudspeth}}\ and\ \bibinfo {author} {\bibfnamefont {P.~G.}\ \bibnamefont
  {Gillespie}},\ }\bibfield  {title} {\bibinfo {title} {Pulling springs to tune
  transduction: {{Adaptation}} by hair cells},\ }\href
  {https://doi.org/10.1016/0896-6273(94)90147-3} {\bibfield  {journal}
  {\bibinfo  {journal} {Neuron}\ }\textbf {\bibinfo {volume} {12}},\ \bibinfo
  {pages} {1} (\bibinfo {year} {1994})}\BibitemShut {NoStop}%
\bibitem [{\citenamefont {Pacentine}\ \emph {et~al.}(2020)\citenamefont
  {Pacentine}, \citenamefont {Chatterjee},\ and\ \citenamefont
  {{Barr-Gillespie}}}]{pacentineStereociliaRootletsActinBased2020}%
  \BibitemOpen
  \bibfield  {author} {\bibinfo {author} {\bibfnamefont {I.}~\bibnamefont
  {Pacentine}}, \bibinfo {author} {\bibfnamefont {P.}~\bibnamefont
  {Chatterjee}},\ and\ \bibinfo {author} {\bibfnamefont {P.~G.}\ \bibnamefont
  {{Barr-Gillespie}}},\ }\bibfield  {title} {\bibinfo {title} {Stereocilia
  {{Rootlets}}: {{Actin-Based Structures That Are Essential}} for {{Structural
  Stability}} of the {{Hair Bundle}}},\ }\href
  {https://doi.org/10.3390/ijms21010324} {\bibfield  {journal} {\bibinfo
  {journal} {Int. J. Mol. Sci.}\ }\textbf {\bibinfo {volume} {21}},\ \bibinfo
  {pages} {324} (\bibinfo {year} {2020})}\BibitemShut {NoStop}%
\bibitem [{\citenamefont {Howard}\ and\ \citenamefont
  {Ashmore}(1986)}]{howardStiffnessSensoryHair1986}%
  \BibitemOpen
  \bibfield  {author} {\bibinfo {author} {\bibfnamefont {J.}~\bibnamefont
  {Howard}}\ and\ \bibinfo {author} {\bibfnamefont {J.~F.}\ \bibnamefont
  {Ashmore}},\ }\bibfield  {title} {\bibinfo {title} {Stiffness of sensory hair
  bundles in the sacculus of the frog},\ }\href
  {https://doi.org/10.1016/0378-5955(86)90178-4} {\bibfield  {journal}
  {\bibinfo  {journal} {Hear. Res.}\ }\textbf {\bibinfo {volume} {23}},\
  \bibinfo {pages} {93} (\bibinfo {year} {1986})}\BibitemShut {NoStop}%
\bibitem [{\citenamefont {Marquis}\ and\ \citenamefont
  {Hudspeth}(1997)}]{marquisEffectsExtracellularCa21997}%
  \BibitemOpen
  \bibfield  {author} {\bibinfo {author} {\bibfnamefont {R.~E.}\ \bibnamefont
  {Marquis}}\ and\ \bibinfo {author} {\bibfnamefont {A.~J.}\ \bibnamefont
  {Hudspeth}},\ }\bibfield  {title} {\bibinfo {title} {Effects of extracellular
  {{Ca2}}+ concentration on hair-bundle stiffness and gating-spring integrity
  in hair cells},\ }\href {https://doi.org/10.1073/pnas.94.22.11923} {\bibfield
   {journal} {\bibinfo  {journal} {Proc. Natl. Acad. Sci.}\ }\textbf {\bibinfo
  {volume} {94}},\ \bibinfo {pages} {11923} (\bibinfo {year}
  {1997})}\BibitemShut {NoStop}%
\bibitem [{\citenamefont {Howard}\ and\ \citenamefont
  {Hudspeth}(1988)}]{howardComplianceHairBundle1988}%
  \BibitemOpen
  \bibfield  {author} {\bibinfo {author} {\bibfnamefont {J.}~\bibnamefont
  {Howard}}\ and\ \bibinfo {author} {\bibfnamefont {A.~J.}\ \bibnamefont
  {Hudspeth}},\ }\bibfield  {title} {\bibinfo {title} {Compliance of the hair
  bundle associated with gating of mechanoelectrical transduction channels in
  the {{Bullfrog}}'s saccular hair cell},\ }\href
  {https://doi.org/10.1016/0896-6273(88)90139-0} {\bibfield  {journal}
  {\bibinfo  {journal} {Neuron}\ }\textbf {\bibinfo {volume} {1}},\ \bibinfo
  {pages} {189} (\bibinfo {year} {1988})}\BibitemShut {NoStop}%
\bibitem [{\citenamefont {Martin}\ \emph {et~al.}(2000)\citenamefont {Martin},
  \citenamefont {Mehta},\ and\ \citenamefont
  {Hudspeth}}]{martinNegativeHairbundleStiffness2000}%
  \BibitemOpen
  \bibfield  {author} {\bibinfo {author} {\bibfnamefont {P.}~\bibnamefont
  {Martin}}, \bibinfo {author} {\bibfnamefont {A.~D.}\ \bibnamefont {Mehta}},\
  and\ \bibinfo {author} {\bibfnamefont {A.~J.}\ \bibnamefont {Hudspeth}},\
  }\bibfield  {title} {\bibinfo {title} {Negative hair-bundle stiffness betrays
  a mechanism for},\ }\href {https://doi.org/10.1073/pnas.210389497} {\bibfield
   {journal} {\bibinfo  {journal} {Proc. Natl. Acad. Sci.}\ }\textbf {\bibinfo
  {volume} {97}},\ \bibinfo {pages} {12026} (\bibinfo {year}
  {2000})}\BibitemShut {NoStop}%
\bibitem [{\citenamefont {Markin}\ and\ \citenamefont
  {Hudspeth}(1995)}]{markinGatingspringModelsMechanoelectrical1995}%
  \BibitemOpen
  \bibfield  {author} {\bibinfo {author} {\bibfnamefont {V.~S.}\ \bibnamefont
  {Markin}}\ and\ \bibinfo {author} {\bibfnamefont {A.~J.}\ \bibnamefont
  {Hudspeth}},\ }\bibfield  {title} {\bibinfo {title} {Gating-spring models of
  mechanoelectrical transduction by hair cells of the internal ear},\ }\href
  {https://doi.org/10.1146/annurev.bb.24.060195.000423} {\bibfield  {journal}
  {\bibinfo  {journal} {Annu. Rev. Bioph. Biom.}\ }\textbf {\bibinfo {volume}
  {24}},\ \bibinfo {pages} {59} (\bibinfo {year} {1995})}\BibitemShut {NoStop}%
\bibitem [{\citenamefont {Manceva}\ \emph {et~al.}(2007)\citenamefont
  {Manceva}, \citenamefont {Lin}, \citenamefont {Pham}, \citenamefont {Lewis},
  \citenamefont {Goldman},\ and\ \citenamefont
  {Ostap}}]{mancevaCalciumRegulationCalmodulin2007}%
  \BibitemOpen
  \bibfield  {author} {\bibinfo {author} {\bibfnamefont {S.}~\bibnamefont
  {Manceva}}, \bibinfo {author} {\bibfnamefont {T.}~\bibnamefont {Lin}},
  \bibinfo {author} {\bibfnamefont {H.}~\bibnamefont {Pham}}, \bibinfo {author}
  {\bibfnamefont {J.~H.}\ \bibnamefont {Lewis}}, \bibinfo {author}
  {\bibfnamefont {Y.~E.}\ \bibnamefont {Goldman}},\ and\ \bibinfo {author}
  {\bibfnamefont {E.~M.}\ \bibnamefont {Ostap}},\ }\bibfield  {title} {\bibinfo
  {title} {Calcium regulation of calmodulin binding to and dissociation from
  the {{Myo1c}} regulatory domain},\ }\href {https://doi.org/10.1021/bi700894h}
  {\bibfield  {journal} {\bibinfo  {journal} {Biochem.}\ }\textbf {\bibinfo
  {volume} {46}},\ \bibinfo {pages} {11718} (\bibinfo {year}
  {2007})}\BibitemShut {NoStop}%
\bibitem [{\citenamefont {Shepherd}\ and\ \citenamefont
  {Corey}(1994)}]{shepherdExtentAdaptationBullfrog1994}%
  \BibitemOpen
  \bibfield  {author} {\bibinfo {author} {\bibfnamefont {G.~M.}\ \bibnamefont
  {Shepherd}}\ and\ \bibinfo {author} {\bibfnamefont {D.~P.}\ \bibnamefont
  {Corey}},\ }\bibfield  {title} {\bibinfo {title} {The extent of adaptation in
  bullfrog saccular hair cells},\ }\href
  {https://doi.org/10.1523/JNEUROSCI.14-10-06217.1994} {\bibfield  {journal}
  {\bibinfo  {journal} {J. Neurosci.}\ }\textbf {\bibinfo {volume} {14}},\
  \bibinfo {pages} {6217} (\bibinfo {year} {1994})}\BibitemShut {NoStop}%
\bibitem [{\citenamefont {Yamoah}\ and\ \citenamefont
  {Gillespie}(1996)}]{yamoahPhosphateAnalogsBlock1996}%
  \BibitemOpen
  \bibfield  {author} {\bibinfo {author} {\bibfnamefont {E.~N.}\ \bibnamefont
  {Yamoah}}\ and\ \bibinfo {author} {\bibfnamefont {P.~G.}\ \bibnamefont
  {Gillespie}},\ }\bibfield  {title} {\bibinfo {title} {Phosphate analogs block
  adaptation in hair cells by inhibiting adaptation-motor force production},\
  }\href {https://doi.org/10.1016/S0896-6273(00)80184-1} {\bibfield  {journal}
  {\bibinfo  {journal} {Neuron}\ }\textbf {\bibinfo {volume} {17}},\ \bibinfo
  {pages} {523} (\bibinfo {year} {1996})}\BibitemShut {NoStop}%
\bibitem [{\citenamefont {Berg}(1993)}]{bergRandomWalksBiology1993}%
  \BibitemOpen
  \bibfield  {author} {\bibinfo {author} {\bibfnamefont {H.~C.}\ \bibnamefont
  {Berg}},\ }\href@noop {} {\emph {\bibinfo {title} {Random {{Walks}} in
  {{Biology}}}}}\ (\bibinfo  {publisher} {{Princeton University Press}},\
  \bibinfo {address} {{Princeton, New Jersey}},\ \bibinfo {year}
  {1993})\BibitemShut {NoStop}%
\bibitem [{\citenamefont {Lumpkin}\ and\ \citenamefont
  {Hudspeth}(1998)}]{lumpkinRegulationFreeCa21998}%
  \BibitemOpen
  \bibfield  {author} {\bibinfo {author} {\bibfnamefont {E.~A.}\ \bibnamefont
  {Lumpkin}}\ and\ \bibinfo {author} {\bibfnamefont {A.~J.}\ \bibnamefont
  {Hudspeth}},\ }\bibfield  {title} {\bibinfo {title} {Regulation of free
  {{Ca2}}+ concentration in hair-cell stereocilia},\ }\href
  {https://doi.org/10.1523/jneurosci.18-16-06300.1998} {\bibfield  {journal}
  {\bibinfo  {journal} {J. Neurosci.}\ }\textbf {\bibinfo {volume} {18}},\
  \bibinfo {pages} {6300} (\bibinfo {year} {1998})}\BibitemShut {NoStop}%
\bibitem [{\citenamefont {Zwillinger}\ and\ \citenamefont
  {Kokoska}(2000)}]{zwillingerCRCStandardProbability2000}%
  \BibitemOpen
  \bibfield  {author} {\bibinfo {author} {\bibfnamefont {D.}~\bibnamefont
  {Zwillinger}}\ and\ \bibinfo {author} {\bibfnamefont {S.}~\bibnamefont
  {Kokoska}},\ }\href@noop {} {\emph {\bibinfo {title} {{{CRC Standard
  Probability}} and {{Statistics Tables}} and {{Formulae}}}}}\ (\bibinfo
  {publisher} {{Chapman \& Hall/CRC}},\ \bibinfo {address} {{Boca Raton, FL}},\
  \bibinfo {year} {2000})\BibitemShut {NoStop}%
\bibitem [{\citenamefont {Sen}\ and\ \citenamefont
  {Salama}(1983)}]{senSpearmanFootruleMarkov1983}%
  \BibitemOpen
  \bibfield  {author} {\bibinfo {author} {\bibfnamefont {P.~K.}\ \bibnamefont
  {Sen}}\ and\ \bibinfo {author} {\bibfnamefont {I.~A.}\ \bibnamefont
  {Salama}},\ }\bibfield  {title} {\bibinfo {title} {The {{Spearman}} footrule
  and a {{Markov}} chain property},\ }\href
  {https://doi.org/10.1016/0167-7152(83)90046-9} {\bibfield  {journal}
  {\bibinfo  {journal} {Stat. Probabil. Lett.}\ }\textbf {\bibinfo {volume}
  {1}},\ \bibinfo {pages} {285} (\bibinfo {year} {1983})}\BibitemShut {NoStop}%
\bibitem [{\citenamefont {Salama}\ and\ \citenamefont
  {Quade}(1990)}]{salamaNoteSpearmanFootrule1990}%
  \BibitemOpen
  \bibfield  {author} {\bibinfo {author} {\bibfnamefont {I.~A.}\ \bibnamefont
  {Salama}}\ and\ \bibinfo {author} {\bibfnamefont {D.}~\bibnamefont {Quade}},\
  }\bibfield  {title} {\bibinfo {title} {A note on spearman's footrule},\
  }\href {https://doi.org/10.1080/03610919008812876} {\bibfield  {journal}
  {\bibinfo  {journal} {Commun. Stat. Simul. Comput.}\ }\textbf {\bibinfo
  {volume} {19}},\ \bibinfo {pages} {591} (\bibinfo {year} {1990})}\BibitemShut
  {NoStop}%
\bibitem [{\citenamefont {Diaconis}\ and\ \citenamefont
  {Graham}(1977)}]{diaconisSpearmanFootruleMeasure1977}%
  \BibitemOpen
  \bibfield  {author} {\bibinfo {author} {\bibfnamefont {P.}~\bibnamefont
  {Diaconis}}\ and\ \bibinfo {author} {\bibfnamefont {R.~L.}\ \bibnamefont
  {Graham}},\ }\bibfield  {title} {\bibinfo {title} {Spearman's {{Footrule}} as
  a {{Measure}} of {{Disarray}}},\ }\href
  {https://doi.org/10.1111/j.2517-6161.1977.tb01624.x} {\bibfield  {journal}
  {\bibinfo  {journal} {J. R. Stat. Soc. Series B Methodol.}\ }\textbf
  {\bibinfo {volume} {39}},\ \bibinfo {pages} {262} (\bibinfo {year} {1977})},\
  \Eprint {https://arxiv.org/abs/2984804} {2984804} \BibitemShut {NoStop}%
\bibitem [{\citenamefont {Harris}\ \emph {et~al.}(2020)\citenamefont {Harris},
  \citenamefont {Millman}, \citenamefont {{van der Walt}}, \citenamefont
  {Gommers}, \citenamefont {Virtanen}, \citenamefont {Cournapeau},
  \citenamefont {Wieser}, \citenamefont {Taylor}, \citenamefont {Berg},
  \citenamefont {Smith}, \citenamefont {Kern}, \citenamefont {Picus},
  \citenamefont {Hoyer}, \citenamefont {{van Kerkwijk}}, \citenamefont {Brett},
  \citenamefont {Haldane}, \citenamefont {{del R{\'i}o}}, \citenamefont
  {Wiebe}, \citenamefont {Peterson}, \citenamefont {{G{\'e}rard-Marchant}},
  \citenamefont {Sheppard}, \citenamefont {Reddy}, \citenamefont {Weckesser},
  \citenamefont {Abbasi}, \citenamefont {Gohlke},\ and\ \citenamefont
  {Oliphant}}]{harrisArrayProgrammingNumPy2020}%
  \BibitemOpen
  \bibfield  {author} {\bibinfo {author} {\bibfnamefont {C.~R.}\ \bibnamefont
  {Harris}}, \bibinfo {author} {\bibfnamefont {K.~J.}\ \bibnamefont {Millman}},
  \bibinfo {author} {\bibfnamefont {S.~J.}\ \bibnamefont {{van der Walt}}},
  \bibinfo {author} {\bibfnamefont {R.}~\bibnamefont {Gommers}}, \bibinfo
  {author} {\bibfnamefont {P.}~\bibnamefont {Virtanen}}, \bibinfo {author}
  {\bibfnamefont {D.}~\bibnamefont {Cournapeau}}, \bibinfo {author}
  {\bibfnamefont {E.}~\bibnamefont {Wieser}}, \bibinfo {author} {\bibfnamefont
  {J.}~\bibnamefont {Taylor}}, \bibinfo {author} {\bibfnamefont
  {S.}~\bibnamefont {Berg}}, \bibinfo {author} {\bibfnamefont {N.~J.}\
  \bibnamefont {Smith}}, \bibinfo {author} {\bibfnamefont {R.}~\bibnamefont
  {Kern}}, \bibinfo {author} {\bibfnamefont {M.}~\bibnamefont {Picus}},
  \bibinfo {author} {\bibfnamefont {S.}~\bibnamefont {Hoyer}}, \bibinfo
  {author} {\bibfnamefont {M.~H.}\ \bibnamefont {{van Kerkwijk}}}, \bibinfo
  {author} {\bibfnamefont {M.}~\bibnamefont {Brett}}, \bibinfo {author}
  {\bibfnamefont {A.}~\bibnamefont {Haldane}}, \bibinfo {author} {\bibfnamefont
  {J.~F.}\ \bibnamefont {{del R{\'i}o}}}, \bibinfo {author} {\bibfnamefont
  {M.}~\bibnamefont {Wiebe}}, \bibinfo {author} {\bibfnamefont
  {P.}~\bibnamefont {Peterson}}, \bibinfo {author} {\bibfnamefont
  {P.}~\bibnamefont {{G{\'e}rard-Marchant}}}, \bibinfo {author} {\bibfnamefont
  {K.}~\bibnamefont {Sheppard}}, \bibinfo {author} {\bibfnamefont
  {T.}~\bibnamefont {Reddy}}, \bibinfo {author} {\bibfnamefont
  {W.}~\bibnamefont {Weckesser}}, \bibinfo {author} {\bibfnamefont
  {H.}~\bibnamefont {Abbasi}}, \bibinfo {author} {\bibfnamefont
  {C.}~\bibnamefont {Gohlke}},\ and\ \bibinfo {author} {\bibfnamefont {T.~E.}\
  \bibnamefont {Oliphant}},\ }\bibfield  {title} {\bibinfo {title} {Array
  programming with {{NumPy}}},\ }\href
  {https://doi.org/10.1038/s41586-020-2649-2} {\bibfield  {journal} {\bibinfo
  {journal} {Nature}\ }\textbf {\bibinfo {volume} {585}},\ \bibinfo {pages}
  {357} (\bibinfo {year} {2020})}\BibitemShut {NoStop}%
\bibitem [{\citenamefont {Virtanen}\ \emph {et~al.}(2020)\citenamefont
  {Virtanen}, \citenamefont {Gommers}, \citenamefont {Oliphant}, \citenamefont
  {Haberland}, \citenamefont {Reddy}, \citenamefont {Cournapeau}, \citenamefont
  {Burovski}, \citenamefont {Peterson}, \citenamefont {Weckesser},
  \citenamefont {Bright}, \citenamefont {{van der Walt}}, \citenamefont
  {Brett}, \citenamefont {Wilson}, \citenamefont {Millman}, \citenamefont
  {Mayorov}, \citenamefont {Nelson}, \citenamefont {Jones}, \citenamefont
  {Kern}, \citenamefont {Larson}, \citenamefont {Carey}, \citenamefont {Polat},
  \citenamefont {Feng}, \citenamefont {Moore}, \citenamefont {VanderPlas},
  \citenamefont {Laxalde}, \citenamefont {Perktold}, \citenamefont {Cimrman},
  \citenamefont {Henriksen}, \citenamefont {Quintero}, \citenamefont {Harris},
  \citenamefont {Archibald}, \citenamefont {Ribeiro}, \citenamefont
  {Pedregosa},\ and\ \citenamefont {{van
  Mulbregt}}}]{virtanenSciPyFundamentalAlgorithms2020}%
  \BibitemOpen
  \bibfield  {author} {\bibinfo {author} {\bibfnamefont {P.}~\bibnamefont
  {Virtanen}}, \bibinfo {author} {\bibfnamefont {R.}~\bibnamefont {Gommers}},
  \bibinfo {author} {\bibfnamefont {T.~E.}\ \bibnamefont {Oliphant}}, \bibinfo
  {author} {\bibfnamefont {M.}~\bibnamefont {Haberland}}, \bibinfo {author}
  {\bibfnamefont {T.}~\bibnamefont {Reddy}}, \bibinfo {author} {\bibfnamefont
  {D.}~\bibnamefont {Cournapeau}}, \bibinfo {author} {\bibfnamefont
  {E.}~\bibnamefont {Burovski}}, \bibinfo {author} {\bibfnamefont
  {P.}~\bibnamefont {Peterson}}, \bibinfo {author} {\bibfnamefont
  {W.}~\bibnamefont {Weckesser}}, \bibinfo {author} {\bibfnamefont
  {J.}~\bibnamefont {Bright}}, \bibinfo {author} {\bibfnamefont {S.~J.}\
  \bibnamefont {{van der Walt}}}, \bibinfo {author} {\bibfnamefont
  {M.}~\bibnamefont {Brett}}, \bibinfo {author} {\bibfnamefont
  {J.}~\bibnamefont {Wilson}}, \bibinfo {author} {\bibfnamefont {K.~J.}\
  \bibnamefont {Millman}}, \bibinfo {author} {\bibfnamefont {N.}~\bibnamefont
  {Mayorov}}, \bibinfo {author} {\bibfnamefont {A.~R.~J.}\ \bibnamefont
  {Nelson}}, \bibinfo {author} {\bibfnamefont {E.}~\bibnamefont {Jones}},
  \bibinfo {author} {\bibfnamefont {R.}~\bibnamefont {Kern}}, \bibinfo {author}
  {\bibfnamefont {E.}~\bibnamefont {Larson}}, \bibinfo {author} {\bibfnamefont
  {C.~J.}\ \bibnamefont {Carey}}, \bibinfo {author} {\bibfnamefont
  {{\.I}.}~\bibnamefont {Polat}}, \bibinfo {author} {\bibfnamefont
  {Y.}~\bibnamefont {Feng}}, \bibinfo {author} {\bibfnamefont {E.~W.}\
  \bibnamefont {Moore}}, \bibinfo {author} {\bibfnamefont {J.}~\bibnamefont
  {VanderPlas}}, \bibinfo {author} {\bibfnamefont {D.}~\bibnamefont {Laxalde}},
  \bibinfo {author} {\bibfnamefont {J.}~\bibnamefont {Perktold}}, \bibinfo
  {author} {\bibfnamefont {R.}~\bibnamefont {Cimrman}}, \bibinfo {author}
  {\bibfnamefont {I.}~\bibnamefont {Henriksen}}, \bibinfo {author}
  {\bibfnamefont {E.~A.}\ \bibnamefont {Quintero}}, \bibinfo {author}
  {\bibfnamefont {C.~R.}\ \bibnamefont {Harris}}, \bibinfo {author}
  {\bibfnamefont {A.~M.}\ \bibnamefont {Archibald}}, \bibinfo {author}
  {\bibfnamefont {A.~H.}\ \bibnamefont {Ribeiro}}, \bibinfo {author}
  {\bibfnamefont {F.}~\bibnamefont {Pedregosa}},\ and\ \bibinfo {author}
  {\bibfnamefont {P.}~\bibnamefont {{van Mulbregt}}},\ }\bibfield  {title}
  {\bibinfo {title} {{{SciPy}} 1.0: Fundamental algorithms for scientific
  computing in {{Python}}},\ }\href {https://doi.org/10.1038/s41592-019-0686-2}
  {\bibfield  {journal} {\bibinfo  {journal} {Nat. Methods}\ }\textbf {\bibinfo
  {volume} {17}},\ \bibinfo {pages} {261} (\bibinfo {year} {2020})}\BibitemShut
  {NoStop}%
\bibitem [{\citenamefont {Herman}\ and\ \citenamefont
  {Usher}(2017)}]{hermanSALibOpensourcePython2017}%
  \BibitemOpen
  \bibfield  {author} {\bibinfo {author} {\bibfnamefont {J.}~\bibnamefont
  {Herman}}\ and\ \bibinfo {author} {\bibfnamefont {W.}~\bibnamefont {Usher}},\
  }\bibfield  {title} {\bibinfo {title} {{{SALib}}: {{An}} open-source
  {{Python}} library for {{Sensitivity Analysis}}},\ }\href
  {https://doi.org/10.21105/joss.00097} {\bibfield  {journal} {\bibinfo
  {journal} {J. Open Source Softw.}\ }\textbf {\bibinfo {volume} {2}},\
  \bibinfo {pages} {97} (\bibinfo {year} {2017})}\BibitemShut {NoStop}%
\bibitem [{\citenamefont {Iwanaga}\ \emph {et~al.}(2022)\citenamefont
  {Iwanaga}, \citenamefont {Usher},\ and\ \citenamefont
  {Herman}}]{iwanagaSALibAdvancingAccessibility2022}%
  \BibitemOpen
  \bibfield  {author} {\bibinfo {author} {\bibfnamefont {T.}~\bibnamefont
  {Iwanaga}}, \bibinfo {author} {\bibfnamefont {W.}~\bibnamefont {Usher}},\
  and\ \bibinfo {author} {\bibfnamefont {J.}~\bibnamefont {Herman}},\
  }\bibfield  {title} {\bibinfo {title} {Toward {{SALib}} 2.0: {{Advancing}}
  the accessibility and interpretability of global sensitivity analyses},\
  }\href {https://doi.org/10.18174/sesmo.18155} {\bibfield  {journal} {\bibinfo
   {journal} {Socioenviron. Syst. Model.}\ }\textbf {\bibinfo {volume} {4}},\
  \bibinfo {pages} {18155} (\bibinfo {year} {2022})}\BibitemShut {NoStop}%
\bibitem [{\citenamefont {{Foreman-Mackey}}\ \emph {et~al.}(2013)\citenamefont
  {{Foreman-Mackey}}, \citenamefont {Hogg}, \citenamefont {Lang},\ and\
  \citenamefont {Goodman}}]{foreman-mackeyEmceeMCMCHammer2013}%
  \BibitemOpen
  \bibfield  {author} {\bibinfo {author} {\bibfnamefont {D.}~\bibnamefont
  {{Foreman-Mackey}}}, \bibinfo {author} {\bibfnamefont {D.~W.}\ \bibnamefont
  {Hogg}}, \bibinfo {author} {\bibfnamefont {D.}~\bibnamefont {Lang}},\ and\
  \bibinfo {author} {\bibfnamefont {J.}~\bibnamefont {Goodman}},\ }\bibfield
  {title} {\bibinfo {title} {Emcee: {{The MCMC Hammer}}},\ }\href
  {https://doi.org/10.1086/670067} {\bibfield  {journal} {\bibinfo  {journal}
  {Publ. Astron. Soc. Pac.}\ }\textbf {\bibinfo {volume} {125}},\ \bibinfo
  {pages} {306} (\bibinfo {year} {2013})}\BibitemShut {NoStop}%
\bibitem [{\citenamefont {Hunter}(2007)}]{hunterMatplotlib2DGraphics2007}%
  \BibitemOpen
  \bibfield  {author} {\bibinfo {author} {\bibfnamefont {J.~D.}\ \bibnamefont
  {Hunter}},\ }\bibfield  {title} {\bibinfo {title} {Matplotlib: {{A 2D
  Graphics Environment}}},\ }\href {https://doi.org/10.1109/MCSE.2007.55}
  {\bibfield  {journal} {\bibinfo  {journal} {Comput. Sci. Eng.}\ }\textbf
  {\bibinfo {volume} {9}},\ \bibinfo {pages} {90} (\bibinfo {year}
  {2007})}\BibitemShut {NoStop}%
\bibitem [{\citenamefont {Waskom}(2021)}]{waskomSeabornStatisticalData2021}%
  \BibitemOpen
  \bibfield  {author} {\bibinfo {author} {\bibfnamefont {M.~L.}\ \bibnamefont
  {Waskom}},\ }\bibfield  {title} {\bibinfo {title} {Seaborn: Statistical data
  visualization},\ }\href {https://doi.org/10.21105/joss.03021} {\bibfield
  {journal} {\bibinfo  {journal} {J. Open Source Softw.}\ }\textbf {\bibinfo
  {volume} {6}},\ \bibinfo {pages} {3021} (\bibinfo {year} {2021})}\BibitemShut
  {NoStop}%
\bibitem [{\citenamefont
  {McKinney}(2010)}]{mckinneyDataStructuresStatistical2010}%
  \BibitemOpen
  \bibfield  {author} {\bibinfo {author} {\bibfnamefont {W.}~\bibnamefont
  {McKinney}},\ }\bibfield  {title} {\bibinfo {title} {Data {{Structures}} for
  {{Statistical Computing}} in {{Python}}},\ }in\ \href
  {https://doi.org/10.25080/Majora-92bf1922-00a} {\emph {\bibinfo {booktitle}
  {Proceedings of the 9th {{Python}} in {{Science Conference}}}}},\ \bibinfo
  {editor} {edited by\ \bibinfo {editor} {\bibfnamefont {S.}~\bibnamefont {{van
  der Walt}}}\ and\ \bibinfo {editor} {\bibfnamefont {J.}~\bibnamefont
  {Millman}}}\ (\bibinfo {year} {2010})\ pp.\ \bibinfo {pages}
  {56--61}\BibitemShut {NoStop}%
\end{thebibliography}%
\end{document}